\documentclass[
 reprint,
 amsmath,amssymb,
 aps,superscriptaddress
]{revtex4-2}

\usepackage{bm}
\usepackage{dcolumn}
\usepackage{graphicx}
\usepackage{mathtools}
\usepackage[T1]{fontenc}

\usepackage{color}

\newcommand{\varA}[1]{{\operatorname{#1}}}

\begin{document}
\title{Multifractal-Spectral Features Enhance Classification of Anomalous Diffusion}
\author{Henrik Seckler}
\affiliation{Institute for Physics \& Astronomy, University of Potsdam, 14476 Potsdam-Golm, Germany}

\author{Ralf Metzler}\email{rmetzler@uni-potsdam.de}
\affiliation{Institute for Physics \& Astronomy, University of Potsdam, 14476 Potsdam-Golm, Germany}
\affiliation{{Asia Pacific Center for Theoretical Physics, Pohang 37673,
Republic of Korea}}

\author{Damian G. Kelty-Stephen}
\affiliation{Department of Psychology, State University of New York at New Paltz, New Paltz, New York 12561, USA}

\author{Madhur Mangalam}\email{mmangalam@unomaha.edu}
\affiliation{Department of Biomechanics and Center for Research in Human Movement Variability, University of Nebraska at Omaha, Omaha, Nebraska 68182, USA}

\date{\today}

\begin{abstract}
    Anomalous diffusion processes, characterized by their nonstandard scaling of the mean squared displacement, pose a unique challenge in classification and characterization. In a previous study (Mangalam \textit{et al.}, 2023, \textit{Physical Review Research} \textbf{5}, 023144), we established a comprehensive framework for understanding anomalous diffusion using multifractal formalism. The present study delves into the potential of multifractal spectral features for effectively distinguishing anomalous diffusion trajectories from five widely used models: fractional Brownian motion, scaled Brownian motion, continuous time random walk, annealed transient time motion, and Lévy walk. To accomplish this, we generate extensive datasets comprising $10^6$ trajectories from these five anomalous diffusion models and extract multiple multifractal spectra from each trajectory. Our investigation entails a thorough analysis of neural network performance, encompassing features derived from varying numbers of spectra. Furthermore, we explore the integration of multifractal spectra into traditional feature datasets, enabling us to assess their impact comprehensively. To ensure a statistically meaningful comparison, we categorize features into concept groups and train neural networks using features from each designated group. Notably, several feature groups demonstrate similar levels of accuracy, with the highest performance observed in groups utilizing moving-window characteristics and $p$-variation features. Multifractal spectral features, particularly those derived from three spectra involving different timescales and cutoffs, closely follow, highlighting their robust discriminatory potential. Remarkably, a neural network exclusively trained on features from a single multifractal spectrum exhibits commendable performance, surpassing other feature groups. In summary, our findings underscore the diverse and potent efficacy of multifractal spectral features in enhancing the predictive capacity of machine learning to classify anomalous diffusion processes.
\\[0.1cm]

\textit{Keywords:} Anomalous diffusion, Feature extraction, Multifractal spectra, Predictive modeling, Spatial dependencies, Temporal dependencies
\end{abstract}

\maketitle

\section{Introduction}

Anomalous diffusion is a ubiquitous phenomenon found in diverse natural settings, including atoms confined in magneto-optical traps \cite{sagi2012observation, zhao2014direct}, the behavior of various biological entities such as DNA, lipids, and proteins \cite{banks2005anomalous, barkai2012single, guigas2008sampling, hofling2013anomalous, jeon2011vivo, jeon2012anomalous, krapf2019strange, metzler2000random, ritchie2005detection, tolic2004anomalous}, microorganisms like bacteria, cells, and parasites \cite{angelini2011glass, dieterich2008anomalous, dieterich2022anomalous, golding2006physical, hapca2009anomalous, lagarde2020colloidal, mukherjee2021anomalous, petrovskii2009dispersal}, as well as in the behavior of foraging wild animals \cite{benhamou2007many, james2011assessing, reynolds2009levy}, and even among human hunter-gatherer societies \cite{brown2007levy, raichlen2014evidence}. Anomalous diffusion manifests itself in economic markets \cite{cherstvy2021scaled, plerou2000economic, vazquez2006modeling} and a wide array of other processes \cite{oliveira2019anomalous, sokolov2005diffusion, timashev2010anomalous, vilk2022unravelling, weigel2011ergodic}, and it exhibits characteristics that span multiple temporal and spatial scales. In these processes, there is a distinctive, erratic evolution of an observable property (e.g., position, temperature, or stock price) over time. The term ``anomalous'' signifies that the mean squared displacement ($\varA{MSD}$) of the observable $x$ does not follow a linear growth with time $t$ as predicted by Fick's theory of diffusion but rather adheres to the more general power-law behavior 
\begin{equation}
    \langle x^{2}(t)\rangle\propto t^{\beta}, \tag{1}\label{eq: 1}
\end{equation}
where $\beta\neq1$. Frequently, one observes $\beta<1$, indicating subdiffusion \cite{bancaud2009molecular, bouchaud1990anomalous, bronstein2009transient, caspi2000enhanced, fernandez2020diffusion, golding2006physical, hofling2011anomalous, hofling2013anomalous, jeon2016protein, saxton2001anomalous, seisenberger2001real, smith1999anomalous, weber2010bacterial, weiss2013single}, while superdiffusion, characterized by $\beta>1$, is less commonly observed but can be found in active physical and biological systems \cite{arcizet2008temporal, caspi2002diffusion, de2011levy, duits2009mapping, gonzalez2008understanding, leoni2014structural, mashanova2010evidence, meyer2023directedeness, nathan2008movement, ott1990anomalous}.

One of the primary motivations for investigating anomalous diffusion models is identifying and categorizing specific anomalous diffusion phenomena in real-world data. Nevertheless, the complexities outlined above render this classification a formidable challenge. As a result, recent endeavors have encompassed Bayesian methodologies \cite{krog2017bayesian, krog2018bayesian, park2021bayesian, thapa2018bayesian, thapa2022bayesian}, as well as machine-learning (ML) strategies \cite{bo2019measurement, cichos2020machine, gajowczyk2021detection, gentili2021characterization, granik2019single, janczura2020classification, kowalek2022boosting, munoz2020single, munoz2021objective, seckler2022bayesian, seckler2023machine}, and even unsupervised techniques \cite{pineda2022geometric, munoz2021unsupervised}. However, these approaches often rely on atheoretical features, which may not correspond to plausible generative mechanisms \cite{kowalek2019classification, loch2020impact, seckler2023machine}. A more theoretically grounded feature set can enhance the ML-based characterization of anomalous diffusion processes within empirical data.

In a previous study \cite{mangalam2023ergodic}, we took an initial stride towards establishing a comprehensive framework rooted in understanding anomalous diffusion using the multifractal formalism \cite{ihlen2012introduction, kelty2013tutorial, kelty2023multifractaltest} aimed at re-establishing ergodicity within the description of anomalous-diffusion phenomena \cite{ritschel2021universality, vinod2022nonergodicity, wang2022restoring}. We harnessed synthetic data that mirror a wide spectrum of anomalous-diffusion processes, spanning various values of the anomalous exponent $\beta$. These processes, encompassing both ergodic and nonergodic behaviors, were approximated through five distinct mathematical models: fractional Brownian motion (FBM, ergodic), scaled Brownian motion (SBM, weakly nonergodic), continuous-time random walk (CTRW, weakly nonergodic), annealed transient time motion (ATTM, weakly nonergodic), and Lévy walk (LW, ultra-weakly nonergodic). Our investigation revealed that descriptors linked to the time-averaged and ensemble-averaged mean-squared displacement ($\varA{TAMSD}$ and $\varA{MSD}$), including linear metrics such as standard deviation, coefficient of variation, and root mean square, exhibit a disruption of ergodicity. In stark contrast, descriptors addressing the temporal structure and potential nonlinearity, such as multifractality and, to a certain extent, fractality, display time-independent behavior, functioning as ergodic descriptors insensitive to the minor ergodicity deviations inherent to these processes. Consequently, these descriptors provide consistent information across various diffusion processes and anomalous exponents $\beta$. Further analysis traced back these patterns to the multiplicative cascades underpinning these diffusion phenomena, as the multifractal spectrum's shape and symmetry, in conjunction with those of corresponding surrogate series, distinguish these processes.

This previous work \cite{mangalam2023ergodic} has opened the door to the potential use of multifractal spectral (MFS) features in improving the classification of anomalous diffusion. Multifractal geometry provides a formalism explicitly tailored to address the intermittent, nonergodic fluctuations that manifest themselves across multiple space- and timescales, encompassing the intricate interplay between short-range events and large-scale contextual factors \cite{mandelbrot1974intermittent, schertzer1997multifractal}. This perspective does not imply that the underlying models generating these diverse forms of anomalous diffusion are inherently multifractal. Instead, it recognizes multifractal geometry as a versatile modeling framework with a long-standing history of explaining how these modes of anomalous diffusion evolve, occasionally transitioning between different regimes \cite{shlesinger1987levy}. Given the successful application of multifractal geometrical estimations as ergodic descriptors in previous work for various anomalous diffusion processes, incorporating MFS features alongside traditional feature sets in ML models holds promise for enhancing the classification of anomalous diffusion.

In this investigation, we delve into the potential of MFS features for discerning anomalous-diffusion patterns. Our approach begins by generating datasets comprising trajectories derived from five distinct anomalous diffusion models---FBM, SBM, CTRW, ATTM, and LW. From each trajectory, we extract multiple multifractal spectra. Our analysis encompasses assessing a neural network's performance when trained on features derived from varying numbers of spectra. Furthermore, we explore the augmentation of datasets containing traditional features, as documented in previous works \cite{kowalek2019classification, kowalek2022boosting}, with multifractal spectra. To culminate the study, we categorize features into distinct concept groups and gauge the performance of each group; this categorization enables a meaningful comparison against the novel concept introduced herein---MFS (multifractal spectral) features.

The paper follows a structured sequence. Beginning with a concise introduction to the employed dataset, traditional features, multifractal analysis, and the machine learning model in Section \ref{Section: Methods}, we assess the performance of MFS features in Section \ref{Section: Results}. This encompasses the outcomes achieved by utilizing MFS features independently, alongside, or in contrast to traditional feature sets. The paper concludes with a comprehensive discussion and a glimpse into avenues for future research in Section \ref{Section: Discussion}.

\section{Methods}\label{Section: Methods}

\subsection{Diffusion models and dataset}

\begin{figure*}
    \includegraphics{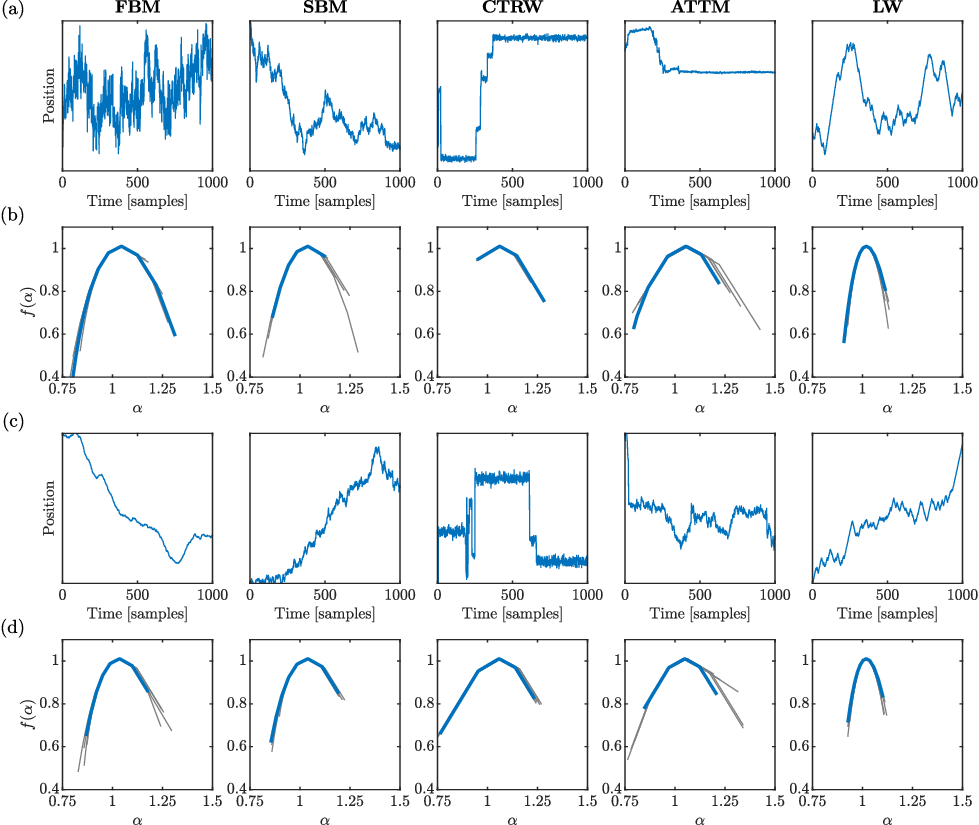}
    \caption{Representative trajectories of the five anomalous diffusion processes for various anomalous exponents (a, c) and the respective multifractal spectrum (b, d). The spectrum in blue corresponds to the original time series, while the spectra in gray correspond to five IAAFT surrogates.}
    \label{fig: Trajectories}
\end{figure*}

To ensure comparability, our dataset generation process closely aligns with that employed in the \emph{Anomalous-Diffusion-(AnDi-)Challenge} \cite{munoz2020single, munoz2021objective, munoz2023quantitative}. Nevertheless, it is noteworthy that, in contrast to the \emph{AnDi-Challenge} dataset, we focus solely on trajectories within the $250\leq N<1000$ range, where $N$ is the number of datapoints. This omission of shorter trajectories is necessitated by the specific constraints associated with the features utilized in our study. Each trajectory is randomly assigned to one of five distinct diffusion models; all yield anomalous diffusion patterns conforming to Eq~(\ref{eq: 1}). We present example trajectories for each in Fig.~\ref{fig: Trajectories}a,c to visually represent these models.

These diffusion processes show differences in how increments are generated, corresponding to distinct statistical mechanisms of anomalous diffusion (see Appendix \ref{TheoreticalModels}).

We generated a dataset comprising a total of $10^6$ trajectories using the \texttt{andi-datasets} \cite{munoz2021unsupervised} Python package. These trajectories have a range of randomly selected anomalous exponents $\beta$, with values $\beta\in\{0.05,0.10,\dots,1.95,2\}$, albeit with some variations based on the specific model. Particularly, the CTRW and ATTM models exhibit only sub- and normal-diffusive behaviors ($\beta\leq1$), while the LW model is exclusively superdiffusive ($\beta>1$) and even ballistic ($\beta=2$). Additionally, this dataset does not consider ballistic ($\beta=2$) FBM. To simulate conditions more akin to experimental data, we introduced additive white Gaussian noise $\xi_{n}$ to all trajectories at varying levels, resulting in signal-to-noise ratios (snr) of either $0.1,0.5,\;\mathrm{or}\;1$. Given a trajectory $\tilde{x}_{n}$, we obtain the noisy trajectory $x_{n}=\tilde{x}_{n}+\xi_{n}$ with the superimposed noise
\begin{equation*}
    \xi_{n}\sim\frac{\sigma_{\Delta\tilde{x}}}{\mathrm{snr}}\mathcal{N}(0,1)\text{,}\tag{2}\label{eq: 2}
\end{equation*}
where $\sigma_{\Delta\tilde{x}}$ is the standard deviation of the unperturbed increment process $\Delta\tilde{x}_{n}=\tilde{x}_{n+1}-\tilde{x}_{n}$.

Of the $10^6$ trajectories, $9\cdot10^5$ ($90\%$) were allocated for training the ML algorithms. The remaining $10^5$ trajectories ($10\%$) were evenly divided into a validation set to refine the training parameters and a separate test set to evaluate the performance metrics reported in this study.

\subsection{Traditional features}\label{Section: TraditionalFeatures}

The traditional features considered for anomalous diffusion classification encompassed many parameters based on Ref. \cite{kowalek2022boosting}, as listed in Table \ref{Table: Table1}.
Appendix \ref{TraditionalFeatures} provides definitions and mathematical details for each feature (see also Ref. \cite{kowalek2022boosting}).

\begin{table}
	\begin{center}
		\begin{tabular}{c}
            Traditional features \\
            \hline
            Anomalous exponent \\
            Diffusion coefficient \\
            Efficiency \\
            Empirical velocity autocorrelation function \\
            Fractal dimension \\
            Maximal excursion \\
            Mean maximal excursion \\
            Kurtosis \\
            Mean Gaussianity \\
            Mean-squared displacement ratio \\
            Statistics based on $p$-variation \\
            Straightness \\
            Trappedness \\
            D’Agostino--Pearson test statistic \\
            Kolmogorov--Smirnov (KS) statistic against $\chi^{2}$ distribution \\
            Joseph exponent \\
            Noah exponent \\
            Moses exponent \\
            Detrending moving average (DMA) \\
            Average moving window characteristics \\
            Maximum standard deviation \\
		\end{tabular}
	\end{center}
    \caption{Traditional features considered for anomalous diffusion classification based on Ref. \cite{kowalek2022boosting}. See Appendix \ref{TraditionalFeatures} or Ref. \cite{kowalek2022boosting} for detailed definitions.}
    \label{Table: Table1}
\end{table}

\subsection{Multifractal spectral (MFS) features}

\subsubsection{Multifractal analysis}

We used Chhabra and Jensen’s \cite{chhabra1989direct} direct method for all analyses of this section. This method estimates the multifractal spectrum width $\Delta \alpha$ by sampling a series $x_{k}$ at progressively longer scales using the proportion $P_{v}(n)$ of the signal falling within the $v$th bin of length $n$ as
\begin{equation*}
    P_{v}(n)=\frac{\sum\limits_{k=(v-1)\,n+1}^{v\cdot n}x_{k}}{\sum\limits_{k=1}^{N}{x_{k}}},\quad n=\{4,8,16,\dots\}<N_\mathrm{max}. \tag{3}\label{eq: 3}
\end{equation*}
As $n$ increases, $P_{v}(n)$ represents a progressively larger proportion of $x(t)$,
\begin{equation*}
    P(n)\propto n^{\alpha}, \tag{4}\label{eq: 4}
\end{equation*}
suggesting a growth of the proportion according to one ``singularity'' strength $\alpha$ \cite{mandelbrot1982fractal}. $P(n)$ exhibits multifractal dynamics when it grows heterogeneously across time scales $n$ according to multiple singularity strengths, such that
\begin{equation*}
    P(n_{v})\propto n^{\alpha_{v}}, \tag{5}\label{eq: 5}
\end{equation*}
whereby each $v$th bin may show a distinct relationship of $P(n)$ with $n$. The width of this singularity spectrum, $\Delta\alpha=(\alpha_\mathrm{max}-\alpha_\mathrm{min})$, indicates the heterogeneity of these relationships \cite{halsey1986fractal, mandelbrot2013fractals}.

Chhabra and Jensen's \cite{chhabra1989direct} method estimates $P(n)$ for $N_{n}$ nonoverlapping windows of size $n$ and transforms them into a ``mass'' $\mu(q)$ using a parameter $q$ emphasizing higher or lower $P(n)$ for $q>1$ and $q<1$, respectively, in the form
\begin{equation*}
    \mu_{v}(q,n)=\frac{\bigl[P_{v}(n)\bigl]^{q}}{\sum\limits_{j=1}^{N_{n}}\bigl[ P_{j}(n)\bigl]^{q}}. \tag{6}\label{eq: 6}
\end{equation*}
Then, $\alpha(q)$ is the singularity for mass $\mu$-weighted $P(n)$ estimated as
\begin{equation*}
    \alpha(q)=-\lim_{N_{n}\to\infty}\frac{1}{\log{N_{n}}}\sum_{v=1}^{N_{n}}\mu_{v} (q,n)\log P_{v}(n)
\end{equation*}
\begin{equation*}
    = \lim_{n\to 0}\frac{1}{\log{n}}\sum_{v=1}^{N_{n}}\mu_{v}(q,n)\log{P_{v}(n)}. \tag{7}\label{eq: 7}
\end{equation*}
Each estimated value of $\alpha(q)$ belongs to the multifractal spectrum only when the Shannon entropy of $\mu(q,n)$ scales with $n$ according to the Hausdorff dimension $f(q)$ \cite{chhabra1989direct}, where
\begin{equation*}
    f(q)=-\lim_{N_{n}\to\infty}\frac{1}{\log N_{n}}\sum_{v=1}^{N_{n}}\mu_{v}(q,n)\log\mu_{v}(q,n)
\end{equation*}
\begin{equation*}
    = \lim_{v\to0}\frac{1}{\log{n}}\sum_{v=1}^{N_{n}}\mu_{v}(q,n)\log{\mu_{v}(q,n)}. \tag{8}\label{eq: 8}
\end{equation*}

For values of $q$ yielding a strong relationship between Eqs.~(\ref{eq: 7}) and (\ref{eq: 8}), as constituted by a minimum value $r$ for the correlation coefficient, the parametric curve $\{\alpha(q),f(q)\}$ or $\{\alpha,f(\alpha)\}$ constitutes the multifractal spectrum and $\Delta \alpha$ (i.e., $\alpha_\mathrm{max}-\alpha_\mathrm{min}$) constitutes the multifractal spectrum width. The cutoff $r$ determines that only scaling relationships of comparable strength can support the estimation of the multifractal spectrum, whether generated as cascades or surrogates. Using a correlation benchmark aims to operationalize previously raised concerns about mis-specifications of the multifractal spectrum \cite{zamir2003critique}. For each trajectory, we compute $9$ multifractal spectra $\{\alpha,f(\alpha)\}$, corresponding to all combinations of the scaling ranges of $N_\mathrm{max}\in\{N/4, N/8, N/16$\}, where $N$ is the trajectory length, and the cutoff $r\in\{0.92, 0.95, 0.97$\}. 

Our next objective was to discern whether a nonzero $\Delta\alpha$ truly signified multifractality arising from nonlinear interactions across various timescales. We compared $\Delta \alpha$ values between the original series and $32$ IAAFT (iterated amplitude adjusted Fourier transform) surrogates \cite{ihlen2012introduction, schreiber1996improved} for each simulated series across generations $9$ through $15$. IAAFT stands out as a method capable of symmetrically reshuffling the original values around their autoregressive structure. Consequently, it generates surrogates that disentangle the phase ordering of spectral amplitudes within the series while preserving the linear temporal correlations. The one-sample $\mathcal{T}$-statistic, $\mathcal{T}_{MF}$, comes into play by computing the difference between $\Delta \alpha$ for the original series and the corresponding values for the $32$ surrogates, which is then divided by the standard error of the spectrum width for these surrogates, facilitating a robust statistical assessment of multifractal nonlinearity.

\begin{figure*}
    \includegraphics{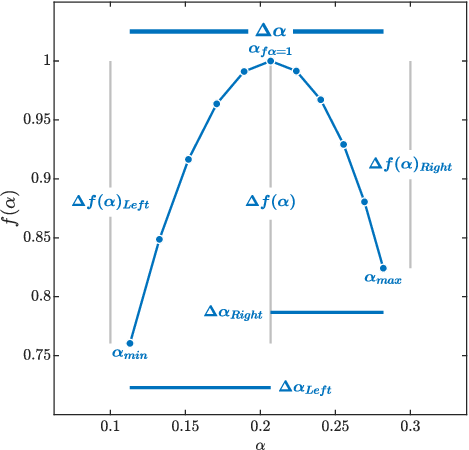}
    \caption{Determining MFS features of anomalous diffusion trajectories. The multifractal spectrum of each trajectory was created by plotting the parametric curve $\{\alpha(q),f(q)\}$. $\alpha(q)$ is the singularity exponent and $f(q)$ the corresponding singularity dimension as defined in Eqs. (\ref{eq: 7}, \ref{eq: 8}).}
    \label{fig: MultifractalSpectrum}
\end{figure*}

We extracted $9$ features from multifractal spectra obtained for each trajectory, as listed in Table \ref{Table: Table2} and depicted in Fig.~\ref{fig: MultifractalSpectrum}.

\begin{table*}
	\begin{center}
		\begin{tabular}{c}
            Multifractal spectral (MFS) features \\
            \hline
            MFS width of the original time series, $\Delta\alpha$ \\
            MFS width of the IAAFT surrogate time series, $\Delta\alpha_\mathrm{Surr}$ \\
            Multifractal nonlinearity, $\mathcal{T}_{MF}$ \\
            Left-side width of the original spectrum, $\Delta\alpha_\mathrm{Left}$ \\
            Right-side width of the original spectrum, $\Delta\alpha_\mathrm{Right}$ \\
            Horizontal location of the singularity, $\alpha_{f(\alpha)=1}$ \\  
            Height of the original spectrum, $\Delta f(\alpha)$ \\
            Left-side height of the original spectrum, $\Delta f(\alpha)_\mathrm{Left}$ \\
            Right-size height of the original spectrum, $\Delta f(\alpha)_\mathrm{Right}$ \\
            Difference in the left- and right-side height of the original spectrum, $\Delta f(\alpha)_{|\mathrm{Left}-\mathrm{Right}|}$ \\
            Mean of $\alpha$ values, $\overline{\alpha}$ \\
            Mean of $f(\alpha)$ values, $\overline{f(\alpha)}$ \\
            Number of points in the original spectrum, $N_\mathrm{Spec}$ \\
		\end{tabular}
	\end{center}
    \caption{MFS features utilized for anomalous diffusion classification. See also Fig.~\ref{fig: MultifractalSpectrum} to visualize the features.}
    \label{Table: Table2}
\end{table*}

\subsection{ML classifiers}

A neural network can essentially be described as a sophisticated function approximator. It aims to align its outputs, denoted as $f_{\theta}(X_{i})$, with the actual target values, represented as $\hat{Y_{i}}$, based on the corresponding input data $X_{i}$ \cite{fausett1994fundamentals,alpaydin2020introduction}. In its simplest form, a neural network comprises multiple layers of neurons. Within each layer, the value of a neuron denoted as $O_{k,l}$ in layer $l$, is determined by the weighted sum of all neurons in the preceding layer, $O_{k,l-1}$, which is then passed through an activation function $h$,
\begin{equation}
    O_{k',l}=h\Biggl(\sum_k\theta_{k',k}^{(l)}\cdot O_{k,l-1}+\theta_{k',0}^{(l)}\Biggl), \tag{9}\label{eq: 9}
\end{equation}
where $\theta_{k',k}^{(l)}$ is the weight between neuron $k'$ in the $l$th layer and neuron $k$ in the $(l-1)$-the layer with $\theta_{k',0}^{(l)}$ as an additional offset.
The output $f_\theta(X_{i})$ of the neural network corresponds to the values of the neurons in the final layer, whereas the input $X_{i}$ corresponds to the values of the neurons in the first layer.

The neural network's weights, collectively represented as $\theta$, are derived by minimizing a loss function applied to a training dataset. Frequently, this loss function is the negative-log-likelihood loss \cite{alpaydin2020introduction},
\begin{equation}
    \mathcal{L}_\mathrm{nll}=-\sum_{i}\log p(\hat{Y}_{i}|f_{\theta}(X_{i})), \tag{10}\label{eq: 10}
\end{equation}
where $p(\hat{Y}_{i}|f_{\theta}(X_{i}))$ is the probability the neural network assigned to the true target $\hat{Y_{i}}$ for input $X_{i}$.
In classification tasks, we usually aim to predict discrete probabilities, denoted as $p_{i,k}$, about each class $k$ as the true label for input $X_{i}$. In this context, the negative-log-likelihood is transformed into the well-known cross-entropy loss \cite{zhang2018generalized},
\begin{equation}
    \mathcal{L}_\mathrm{cel}=-\sum_{i,k}\hat{Y}_{i,k}\log(p_{i,k}),
     \tag{11}\label{eq: 11}
\end{equation}
where $\hat{Y}_{i,k}=\delta_{j_ik}$ is a binary indicator of the true label $j_{i}$ of input $X_{i}$.

The optimization of this loss function is accomplished through the utilization of stochastic gradient descent \cite{bottou2010large}. This study employed an advanced variant of stochastic gradient descent known as ``Adaptive Moment Estimation'' (Adam) \cite{kingma2014adam}. In addition to Adam, we incorporated ``Stochastic Weight Averaging Gaussian'' (SWAG)---which captures the uncertainty of the neural network's weight parameters, $\theta$---toward the conclusion of the training process. This is achieved by fitting an approximate Gaussian distribution to the observed changes of $\theta$ during the gradient descent process. For in-depth insights into SWAG, we refer readers to Ref. \cite{maddox2019simple}, and for an application to anomalous diffusion to Ref. \cite{seckler2022bayesian}.

\begin{figure*}
    \includegraphics{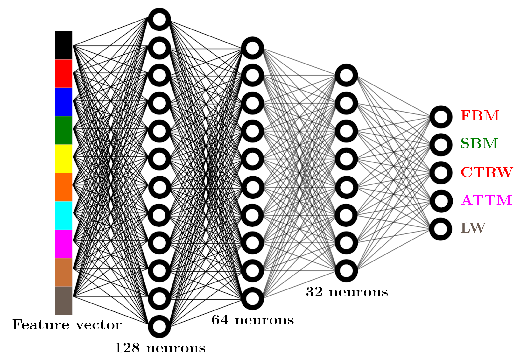}
    \caption{Neural network architecture used for anomalous diffusion classification. A fully connected neural network was used with three hidden layers of size $128$, $64$, and $32$ (or $256$, $128$, and $64$ when using all additional features). The input layer comprised the (normalized) feature vector, with its dimension determined by the number of spectra used ($13$ features per spectrum). It incorporated an additional $26$ or $39$ features for the original or extended sets, respectively. The network then generated model scores for the five diffusion models examined.}
    \label{fig: NeuralNetwork}
\end{figure*}

In the latter approach, a recurrent neural network was employed to classify anomalous diffusion models directly from raw positional data \cite{seckler2022bayesian}. In the present study, however, we took a distinct approach by working with extracted features, which enabled us to adopt a simpler neural network architecture. The used neural network comprised three hidden layers with dimensions $128$, $64$, and $32$, utilizing the rectified linear unit (ReLU) \cite{hahnloser2000digital} as the activation function, as visually represented in Fig.~\ref{fig: NeuralNetwork}. Notably, when working with an extended feature set encompassing all the features outlined in Ref. \cite{kowalek2022boosting}, we observed that a more expansive network configuration yields benefits, thus opting for larger hidden-layer sizes of $256$, $128$, and $64$. The network generates membership scores for each of the five classes as a ``logit vector,'' denoted as $Z_{i}=f_{\theta}(X_{i})$, with values subsequently related to model probabilities $p_{i,k}$ through a normalized exponential (softmax) function \cite{gao2017properties},
\begin{equation}
    p_{i,k}=\frac{\exp{(z_{i,k}})}{\sum_k\exp{(z_{i,k})}}. \tag{12}\label{eq: 12}
\end{equation}

To train the neural network, we conducted $300$ epochs, during which we systematically shuffled the training dataset and divided it into batches of size $512$. Each epoch constitutes one pass through the whole training dataset. The network weights are iteratively updated with each batch to minimize the loss function [as expressed in Eq.~(\ref{eq: 11})], employing the \textit{Adam} optimizer with a learning rate that decays from $10^{-3}$ to $10^{-4}$. The final $20$ epochs estimated SWAG, which entails determining a Gaussian probability density function on the network weights $\theta$.

\section{Results}\label{Section: Results}

Our analysis began with assessing their standalone performance to evaluate the effectiveness of the newly introduced MFS features in ML-based classification. We gauged the achieved accuracy across varying numbers of spectra and delved into the significance of each individual feature. Considering that these MFS features may not individually rival state-of-the-art techniques, which often employ a wide array of features or operate directly on trajectory data, we explored an additional dimension. We investigated how the MFS features might enhance established feature sets by incorporating supplementary features inspired by Kowalek \textit{et al.}'s work in Refs. \cite{kowalek2019classification, kowalek2022boosting} into our dataset. This encompassed the original feature set introduced in Ref. \cite{kowalek2019classification} and the enhanced extended feature set outlined in Ref. \cite{kowalek2022boosting}. Furthermore, we organized the features into groups based on similar conceptual foundations for the extended feature set. This categorization enabled us to make meaningful performance comparisons by training ML models on individual feature groups as an extension to the MFS features.

\subsection{Classification performance with MFS features exclusively}

\begin{figure*}
    \includegraphics{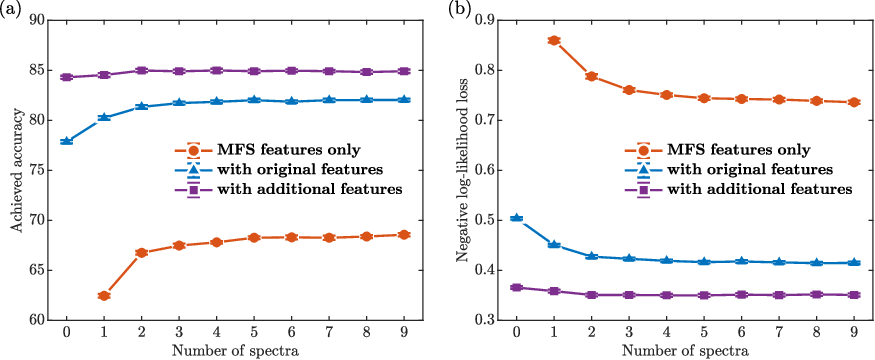}
    \caption{Achieved accuracy (a) and loss (b) for anomalous diffusion classification using features from different numbers of multifractal spectra and feature combinations. The depicted error bars are obtained via subsampling on the test dataset.}
    \label{fig: AccuracyLoss}
\end{figure*}

We trained a neural network to predict the anomalous diffusion model exclusively based on the features extracted from multifractal spectra. To scrutinize the influence of spectrum selection, we employed various spectra, each comprising $13$ distinct features, and documented the outcomes for the most promising combination of spectra. Fig.~\ref{fig: AccuracyLoss} presents the attained accuracy and loss on the test dataset in relation to the number of spectra utilized; we only consider the graphs labeled as ``MFS features only.'' In Fig.~\ref{fig: AccuracyLoss}a, the accuracy ranged from $62.5\%$ to $68.6\%$ based on the number of incorporated spectra. Notably, we observed a substantial surge in accuracy when transitioning from a single spectrum ($62.5\%$) to two spectra ($66.7\%$), but this improvement diminished as more spectra were included. A comparable pattern was observed in loss in Fig.~\ref{fig: AccuracyLoss}b, spanning from $0.860$ to $0.736$. It is worth noting that the loss function considers not only the predicted class but also the assigned probabilities for all five classes, as elucidated in Eq.~(\ref{eq: 11}).

\begin{figure*}
    \includegraphics{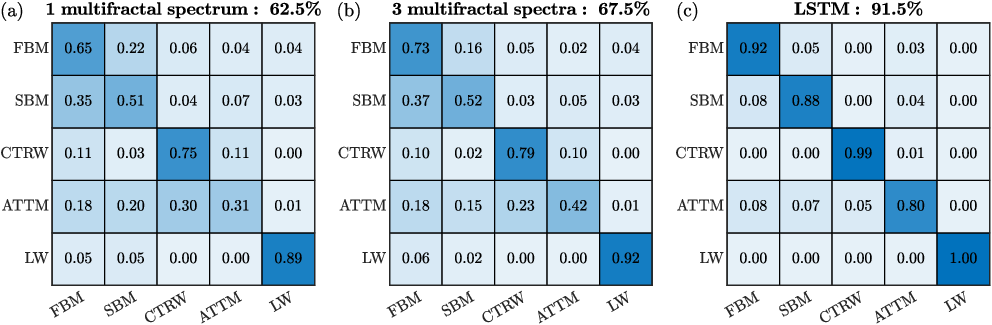}
    \caption{Confusion matrices showing the accuracy of the anomalous diffusion classification using only the MFS features from just $1$ spectrum (a) and $3$ spectra (b), as well as for a state-of-the-art LSTM neural network trained on raw trajectories (c). The matrices show the probability of a ground truth model on the vertical axis to be predicted as one of the models on the horizontal axis.}
    \label{fig: ConfusionMatrices1}
\end{figure*}

Figures~\ref{fig: ConfusionMatrices1}, a and b, respectively, present the confusion matrices for the models trained on features extracted from a single multifractal spectrum and features extracted from three multifractal spectra. These matrices illustrate the likelihood of the respective neural networks predicting each true class (rows) as one of the five classes (columns); therefore, the probabilities for correct predictions may be recovered from the diagonal entries. For comparison, Fig.~\ref{fig: ConfusionMatrices1}c also shows the confusion matrix obtained when employing a state-of-the-art LSTM neural network, as introduced in Ref. \cite{seckler2022bayesian}, on the same dataset. The confusion matrices reveal that a network trained on a single spectrum (Fig.~\ref{fig: ConfusionMatrices1}a) demonstrates proficiency in accurately identifying LW trajectories ($89\%$) and CTRW trajectories ($75\%$), but faces challenges in distinguishing between FBM ($66\%$), SBM ($50\%$), and particularly ATTM ($31\%$). However, incorporating features from multiple spectra (Fig.~\ref{fig: ConfusionMatrices1}b) significantly enhances performance for ATTM ($42\%$) and FBM ($73\%$), with noticeable, though less pronounced, improvements observed for all other models.

A notable advantage of feature-based ML, in contrast to non-feature-based approaches such as Bayesian deep learning \cite{seckler2022bayesian}, lies in its heightened interpretability. Specifically, these methods empower the identification of each feature's influence on the neural network's performance. For instance, one can selectively choose a feature, permute its position among the inputs, and observe the resulting accuracy decrease attributable to this particular feature's misallocation. Tables~\ref{Table: Table3} and \ref{Table: Table4} present the permutation feature importance for models trained on singular or dual multifractal spectra, respectively. Notably, when utilizing a solitary spectrum (Table~\ref{Table: Table3}), we discern the utmost significance for the right end of the spectrum, denoted as $\Delta\alpha_\mathrm{Right}$ and $\Delta f(\alpha)_\mathrm{Right}$. A consistent pattern emerges across various scenarios, wherein the spatial orientation of the spectrum, as indicated by the singularity exponent $\alpha$, outweighs the precise singularity dimension $f(\alpha)$ at those specific locations in terms of importance. The significance of features persists in the context of two spectra (Table~\ref{Table: Table4}), aligning with the observed behavior in the singular spectrum scenario. Slight enhancements in importance across most features are evident, indicative of an overall improved performance. Only $\Delta\alpha_\mathrm{Left}$ ascends from the fourth position in Table~\ref{Table: Table3} to become the most crucial feature in Table~\ref{Table: Table4}---a shift suggesting that heterogeneity in $\Delta\alpha_\mathrm{Left}$ between them may assume heightened importance when leveraging multiple spectra.

\begin{table}
	\begin{center}
		\begin{tabular}{c|c}
			Spectral feature & Permutation importance \\
			\hline
			$\Delta\alpha_\mathrm{Right}$ & $0.202$ \\
			$\Delta f(\alpha)_\mathrm{Right}$ & $0.185$ \\
			$\Delta\alpha$ & $0.174$ \\
			$\Delta\alpha_\mathrm{Left}$ & $0.173$ \\
			$\alpha_{f(\alpha)=1}$ & $0.139$ \\
			$\Delta f(\alpha)_{|\mathrm{Left}-\mathrm{Right}|}$ & $0.136$ \\
			$N_\mathrm{Spec}$ & $0.133$ \\
			$\Delta f(\alpha)_\mathrm{Left}$ & $0.125$ \\
			$\Delta \alpha_{Surr}$ & $0.112$ \\
			$\mathcal{T}_{MF}$ & $0.059$ \\
			$\overline{f(\alpha)}$ & $0.016$ \\
			$\Delta f(\alpha)$ & $0.013$ \\
			$\overline{\alpha}$ & $0.005$ \\
		\end{tabular}
	\end{center}
    \caption{Permutation importance of the various MFS features obtained using only one multifractal spectrum.}
    \label{Table: Table3}
\end{table}

\begin{table}
	\begin{center}
		\begin{tabular}{c|c}
			Spectral feature & Permutation importance \\
			\hline
			$\Delta\alpha_\mathrm{Left}$ & $0.215$ \\
			$\Delta\alpha_\mathrm{Right}$ & $0.212$ \\
			$\Delta f(\alpha)_\mathrm{Right}$ & $0.203$ \\
			$\Delta\alpha$  & $0.185$ \\
			$\alpha_{f(\alpha)=1}$ & $0.180$ \\
			$\Delta f(\alpha)_{|\mathrm{Left}-\mathrm{Right}|}$ & $0.164$ \\
			$N_\mathrm{Spec}$ & $0.159$ \\
			$\Delta f(\alpha)_\mathrm{Left}$ & $0.147$ \\
			$\Delta \alpha_{Surr}$ & $0.141$ \\
			$\mathcal{T}_{MF}$ & $0.078$ \\
			$\overline{f(\alpha)}$ & $0.021$ \\
			$\Delta f(\alpha)$ & $0.015$ \\
			$\overline{\alpha}$ & $0.005$ \\
		\end{tabular}
	\end{center}
    \caption{Permutation importance of the various MFS features obtained using two multifractal spectra. The ranking of the MFS features stays mostly the same, except that the feature $\Delta\alpha_\mathrm{Left}$ moves up from $4$th to the most important position.}
    \label{Table: Table4}
\end{table}

Although achieving an accuracy from $62.5\%$ to $68.6\%$, as we saw for the MFS features only in Fig.~\ref{fig: AccuracyLoss}a, is a notable improvement over random predictions ($20\%$ for predicting one out of five models), it does not reach the levels of performance attainable with state-of-the-art techniques developed during and after the \emph{AnDi-Challenge} (e.g., \cite{munoz2021objective, seckler2022bayesian, seckler2023machine}). This outcome aligns with expectations, given that our model relies solely on a single category of features. Notably, a model leveraging the features introduced in Refs. \cite{kowalek2019classification, kowalek2022boosting} demonstrated the ability to achieve an accuracy of $77.8\%$, which further increased to $84.3\%$ when utilizing the extended feature set. Additionally, when employing the LSTM neural network from \cite{seckler2022bayesian}, which primarily operates on minimally preprocessed raw trajectories, we accomplished an accuracy of $91.7\%$ using the same dataset as depicted in Fig.~\ref{fig: ConfusionMatrices1}c. Consequently, we proceed with our investigation to determine whether incorporating MFS features into established feature sets can improve classification performance.

\subsection{Classification performance after adding MFS features to established feature sets}

\begin{figure*}
    \includegraphics{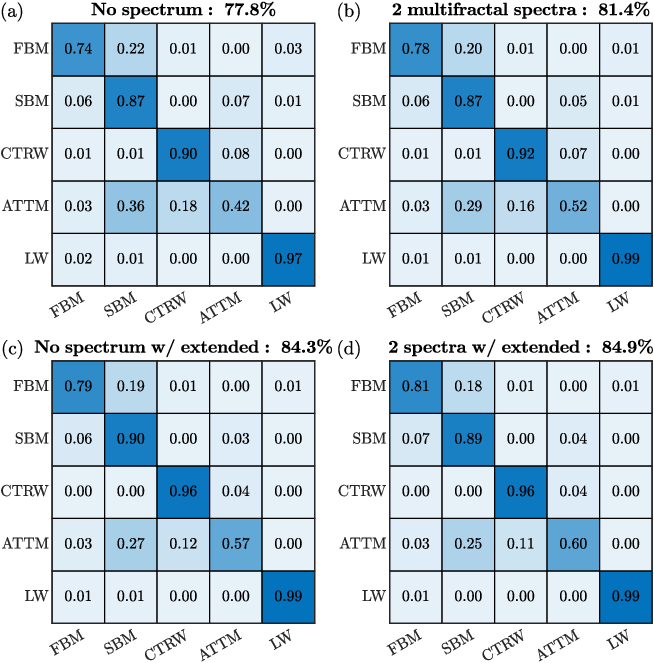}
    \caption{Confusion matrices showing the accuracy of the anomalous diffusion classification using MFS features and traditional features. The Confusion Matrices show the probability of a ground truth model on the vertical axis to be predicted as one of the models on the horizontal axis. (a) Smaller feature set without any spectrum. (b) Smaller feature set with $2$ multifractal spectra. (c) Extended feature set without any spectrum. (d) Extended feature set with $2$ multifractal spectra.}
    \label{fig: ConfusionMatrices2}
\end{figure*}

We next assessed the implications of augmenting the traditional feature sets, detailed in Section \ref{Section: TraditionalFeatures} of Ref. \cite{kowalek2022boosting}, for classification performance, together with the recently introduced MFS features. Fig.~\ref{fig: AccuracyLoss} showcases the attained accuracies and losses. The initial datapoint, representing no spectrum, delineates the classification performance without MFS features. In addition, we present the confusion matrices for the standalone traditional feature sets and their integration with MFS features from $2$ spectra in Fig.~\ref{fig: ConfusionMatrices2}. This comprehensive visualization provides insights into these distinct feature sets' comparative performance and interactions.

In the case of the smaller feature set---so-called ``original'' feature set, initially employed in Ref. \cite{kowalek2019classification}, a discernible enhancement in accuracy is evident---from the initial $77.8\%$ without MFS features to an elevated $82.0\%$ with the inclusion of all nine spectra, as visible in the corresponding graph in Fig.~\ref{fig: AccuracyLoss}a. Notably, the influence of additional spectra diminishes rapidly, exhibiting no discernible changes surpassing random fluctuations beyond the inclusion of five spectra. A parallel pattern emerges for the loss (Fig.~\ref{fig: AccuracyLoss}b), where a notable improvement---from $0.504$ without spectra to $0.415$ with all $9$ spectra---is observed. Although the impact of additional spectra diminishes, the decline is not as abrupt as witnessed in the accuracy domain. Examining the confusion matrices in Fig.~\ref{fig: ConfusionMatrices2}a, b unveils some intriguing insights. For instance, despite the models trained solely on MFS features exhibiting the weakest performance for ATTM, the most significant improvement is witnessed in the detection of ATTM, escalating from $42\%$ (in Fig.~\ref{fig: ConfusionMatrices2}a) to $52\%$ (in Fig.~\ref{fig: ConfusionMatrices2}b). Subsequent enhancements are noted for FBM (from $74\%$ to $78\%$), marginal gains for CTRW and LW (by $2\%$), and no measurable improvement for SBM.

Regarding the ``extended'' feature set \cite{kowalek2022boosting}, we observed a marginal yet perceptible uptick in accuracy---from an initial $84.3\%$ to a refined $84.9\%$ from the corresponding graph in Fig.~\ref{fig: AccuracyLoss}a. This suggests that most information encapsulated in the multifractal spectra may already exist within the extended feature set. Likewise, the loss in Fig.~\ref{fig: AccuracyLoss}b exhibits a slight reduction, transitioning from $0.366$ to $0.351$. Including two-plus multifractal spectra had a minimal to negligible impact on loss and accuracy. Mirroring the trends observed in the smaller feature set, the confusion matrices presented in Figs.~\ref{fig: ConfusionMatrices2}c, d showcase modest improvements for ATTM (ascending from $57\%$ in Fig.~\ref{fig: ConfusionMatrices2}c to $60\%$ in Fig.~\ref{fig: ConfusionMatrices2}d) and FBM (progressing from $79\%$ to $81\%$), while the performance for the other three models remained relatively stable. This underscores the nuanced interplay between the extended feature set and MFS features in influencing model performance.

Tabs.~\ref{Table: Table5} and \ref{Table: Table6} present feature-importance scores, largely aligning with the previously elucidated trends. These tables enumerate the top $10$ features ranked by permutation importance for neural networks trained on the original or extended feature sets, incorporating MFS features from two spectra. It is pertinent to note that certain features, such as $p$-variation, encompass multiple values collectively permuted for conciseness and are singularly represented with unified importance. Noteworthy prominence is accorded to the MFS features and the smaller feature set in Tab.~\ref{Table: Table5}, with most of the top ten comprising these novel elements. Interestingly, the original features involving $p$-variation and the anomalous exponent $\beta$ claim the top two positions. Notable inclusions within the top $10$ are the previously underrated features $\Delta\alpha_\mathrm{Surr}$, derived from surrogates, and $\Delta f(\alpha)$. This dynamic underscores the nuanced landscape of feature importance, shedding light on the distinctive contributions of the traditional and newly introduced features. 
Consistent with the diminished performance noted in the extended feature set, the MFS features assume a relatively lower level of importance in Tab.~\ref{Table: Table6}, signaling that a substantial portion of the information encapsulated in these new features might already be retrievable from the traditional feature set. Nevertheless, two specific MFS features---$\alpha_{f(\alpha)=1}$ and $\Delta f(\alpha)_\mathrm{Right}$---hold positions towards the tail end of the top ten features, implying a heightened significance compared to several traditional features within the dataset.

\begin{table}
	\begin{center}
		\begin{tabular}{c|c}
			Feature & Permutation importance \\
			\hline
			$p$-variation & $0.294$ \\
			$\alpha$ & $0.195$ \\
			\boldmath$\Delta f(\alpha)_\mathrm{Right}$ & \boldmath$0.117$ \\
			\boldmath$\Delta\alpha_\mathrm{Right}$ & \boldmath$0.109$ \\
            \boldmath$\alpha_{f(\alpha)=1}$ & \boldmath$0.098$ \\
			\boldmath$\Delta\alpha_\mathrm{Left}$ & \boldmath$0.094$ \\
			\boldmath$\Delta\alpha_\mathrm{Surr}$ & \boldmath$0.090$ \\
			Mean Gaussianity & $0.080$ \\
			\boldmath$\Delta f(\alpha)$ & \boldmath$0.076$ \\
			\boldmath$\Delta\alpha$ & \boldmath$0.075$ \\
		\end{tabular}
	\end{center}
    \caption{Permutation importance for the top $10$ features, combining two multifractal spectra with the original features as presented in Refs. \cite{kowalek2019classification, kowalek2022boosting}. Although the original features, primarily involving $p$-variation and the anomalous exponent $\beta$, still exhibit the highest importance, they are closely followed by the new MFS features. Seven of the top $10$ features have been derived from this newly introduced multifractal framework (highlighted in boldface).}
    \label{Table: Table5}
\end{table}

\begin{table}
	\begin{center}
		\begin{tabular}{c|c}
			Feature & Permutation importance \\
			\hline
			Moving window & $0.248$ \\
			$p$-variation & $0.184$ \\
			Anomalous exponent & $0.143$ \\
			KS statistic against $\chi^{2}$ & $0.127$ \\
			Direct statistics & $0.104$ \\
			Maximum standard deviation & $0.085$ \\
			Joseph exponent & $0.082$ \\
			Mean Gaussianity & $0.060$ \\
			\boldmath$\alpha_{f(\alpha)=1}$ & \boldmath$0.058$ \\
			\boldmath$\Delta f(\alpha)_\mathrm{Right}$ & \boldmath$0.054$ \\
		\end{tabular}
	\end{center}
    \caption{Permutation importance for the top $10$ features, combining two multifractal spectra with the original and extended features as presented in Ref. \cite{kowalek2022boosting}. These results reveal that most of the top $10$ features are the additional features introduced in Ref. \cite{kowalek2022boosting}. Notably, two MFS features, $\alpha_{f(\alpha)=1}$ and $\Delta f(\alpha)_\mathrm{Right}$ (highlighted in boldface), occupy the $9$th and $10$th positions in terms of importance, suggesting that while the newly introduced features may not be the most critical, they surpass several other traditional features present in anomalous diffusion in terms of importance.}
    \label{Table: Table6}
\end{table}

To conclude our investigation, we systematically categorized the features within the extended feature set based on shared conceptual foundations. This organization enables a direct and meaningful comparison with the MFS feature group. The features were systematically grouped as listed in table \ref{Table: Table7}.

\begin{table*}
	\begin{center}
		\begin{tabular}{c|c}
            Concept group & Features \\
            \hline
            \begin{tabular}{c}Features extracted from the time-averaged \\ Mean squared displacement (TAMSD) \end{tabular} &
            \begin{tabular}{c}
                Anomalous exponent \\
                Diffusion coefficient \\
                Mean-squared displacement ratio \\
                Trappedness \\
            \end{tabular} \\
            \hline
            Statistics based on $p$-variation & \begin{tabular}{c}
                Scaling powers $\gamma_{p}$ \\
                Monotonicity statistic $P$
            \end{tabular} \\
            \hline
            Joseph, Noah, and Moses effect &
            \begin{tabular}{c}
                 Joseph Exponent \\
                 Noah Exponent \\
                 Moses Exponent
            \end{tabular} \\
            \hline
            Statistics based on comparison against the normal distribution & 
            \begin{tabular}{c}
                Kurtosis \\
                Mean Gaussianity \\
                D’Agostino--Pearson test statistic \\
                KS statistic against $\chi^{2}$ distribution
            \end{tabular} \\
            \hline
            Measures for the directedness of motion &
            \begin{tabular}{c}
                Empirical velocity autocorrelation function \\
                Straightness \\
                Efficiency \\
                Fractal dimension
            \end{tabular} \\
            \hline
            Detrending Moving Average & DMA \\
            \hline
            Moving window measures &
            \begin{tabular}{c}
                 Average moving window characteristics \\
                 Maximum standard deviation
            \end{tabular} \\
            \hline
            Measures to detect large jumps (excursions) &
            \begin{tabular}{c}
                 Maximal excursion \\
                 Mean maximal excursion
            \end{tabular}
		\end{tabular}
	\end{center}
    \caption{Traditional features organized into groups based on shared conceptual foundations. See Appendix \ref{TraditionalFeatures} for detailed definitions of the features.}
    \label{Table: Table7}
\end{table*}

Finally, we trained an additional neural network with only the features within each designated group. We recorded the ensuing accuracies on the test set, as delineated in Table~\ref{Table: Table8}. Except for the DMA, which failed to surpass the random-prediction accuracy of $20\%$, the remaining groups demonstrated classification performances within a comparable accuracy range of approximately $50$--$70\%$. Prominent among these groups were those harnessing the moving window characteristics ($72.2\%$) and $p$-variation ($71.7\%$), securing the highest performance accuracies. Notably, the network fashioned from the MFS features extracted from three spectra closely trailed with an accuracy of $68.2\%$. Even the network trained solely on features from a singular multifractal spectrum showcased commendable performance, outpacing most other feature groups with an accuracy of $63\%$, except for the directedness measures, which achieved an accuracy of $65.4\%$. These findings underscore the diverse efficacy of MFS features in contributing to the overall predictive capacity of the neural network in anomalous diffusion classification.

\begin{table}
	\begin{center}
		\begin{tabular}{c|c}
			Feature group & Achieved accuracy \\
			\hline
			Moving window & $72.2\%$ \\
			$p$-variation & $71.7\%$ \\
			\boldmath$3$ \textbf{spectra} & \boldmath$68.2\%$ \\
			Directedness measures & $65.4\%$ \\
			\boldmath$1$ \textbf{spectrum} & \boldmath$63.0\%$ \\
			vs. normal & $59.6\%$ \\
			Joseph, Noah, and Moses & $58.1\%$ \\
			TAMSD & $52.9\%$ \\
            Excursions & $49.8\%$\\
			DMA & $20.0\%$ \\
		\end{tabular}
    \end{center}
    \caption{Achieved performance accuracies across various feature groupings reveal noteworthy distinctions. Feature sets derived from one or three multifractal spectra (highlighted in boldface) emerged as among the most proficient, exhibiting only marginal performance differentials compared to the groups centered around $p$-variation or moving window statistics (by $3.5$-$4\%$). These findings underscore the competitive performance of MFS features, positioning them close to other robust feature groupings in predictive efficacy.}
    \label{Table: Table8}
\end{table}

\section{Discussion}\label{Section: Discussion}

We explored the potential of MFS features to discern effectively anomalous-diffusion trajectories originating from five prevalent models---FBM, SBM, CTRW, ATTM, and LW. To achieve this objective, we generated a dataset of $10^6$ trajectories from these models, each yielding multiple---up to nine---multifractal spectra. Our study entailed a thorough analysis of neural network performance, incorporating features derived from varying numbers of spectra. Additionally, we systematically explored the augmentation of traditional feature datasets with multifractal spectra, enabling a comprehensive assessment of their impact. To facilitate a definitive comparison, we categorized features into concept groups, and neural networks were trained to utilize features from each designated group. The principal outcomes of these investigations are illustrated in Fig.\ref{fig: AccuracyLoss} and tabulated in Tab.\ref{Table: Table8}.

Concretely, the subdivision of traditional features into two distinct sets---an older and a more comprehensive contemporary set has discerned a notable impact stemming from including new Multifractal Spectrum (MFS) features. This impact is particularly pronounced for the former set and remains discernibly measurable for the latter subset of our quantifiers, albeit of a lesser magnitude. Notably, methodologies based on the \emph{AnDi-Challenge}, whether leveraging raw particle trajectories or a more exhaustive feature set, consistently outperform approaches reliant solely on MFS features. Nevertheless, upon further investigation into the segmentation of traditional features into smaller conceptual groups, MFS features closely trailed the two top-performing quantifiers. These quantifiers, derived from moving-average and p-variation statistics analyses, exhibited superior performance. Specifically, our analysis revealed a marginal reduction in accuracy for features derived from three spectra. At the same time, a comparatively more substantial decline in performance was observed for the less sensitive single-spectrum features (see highlighted in bold entries in Table~\ref{Table: Table8}). Notably, the latter still outperformed several traditional features, exemplified by quantifiers comprising the Joseph, Moses, and Noah exponents. In summation, our findings underscore the diverse and potent efficacy of MFS features in augmenting the predictive capabilities of neural networks for classifying anomalous diffusion processes.

The multifractal formalism provides a set of parameters well suited to the ergodic causal modeling of anomalous diffusion processes. While traditional features, often employed in linear causal modeling \cite{kian2017relationship, vink2017eeg, vink2020eeg}, tend to disrupt ergodicity, multifractal descriptors, in contrast, exhibit ergodicity \cite{kelty2022fractal, kelty2023multifractal, mangalam2022ergodic, mangalam2023ergodic, mangalam2023multifractal}, thereby offering a dependable and consistent set of causal predictors \cite{bloomfield2021perceiving, booth2018expectations, carver2017multifractal, dixon2012multifractal, jacobson2021multifractality, kelty2014interwoven, kelty2021multifractal, mangalam2020bodywide, mangalam2020global, mangalam2020multifractal, mangalam2020multiplicative, wallot2018interaction}. The present finding that MFS features improved anomalous diffusion classification supports the growing interest in multifractal modeling within various fields of active matter. For example, in areas such as the dynamics of biomolecules within cells \cite{cardenas2012dynamics, chaieb2008wrinkling, rezania2021multifractality, wawrzkiewicz2020multifractal}, the foraging behavior of wild animals \cite{gutierrez2015neural, ikeda2020c, schmitt2001multifractal, seuront2014anomalous}, and the study of collective dynamics, including swarming and milling behaviors \cite{balaban2018quantifying, carver2017multifractal, koorehdavoudi2017multi}, multifractal formalisms have gained traction. The present finding underscores the significance of such approaches in these fields.

Notably, the frequently observed disruption of ergodicity in traditional features can serve as an advantage in model classification. Unlike the current focus on individual trajectories, this approach necessitates access to a set of trajectories stemming from the same motion. By capturing the interplay between ensemble and time averages, such as through an amplitude scatter function, distinctions may arise based on the underlying diffusion model \cite{jeon2010analysis, jeon2014scaled}.

An intriguing avenue for further research lies in exploring alternative single-particle models that have not been explicitly addressed in the present investigation. Complex systems frequently manifest multiple dynamics in practical scenarios, presenting various facets of heterogeneities. These may include combinations of temporal variations in diffusivity, as evidenced in SBM, spatial heterogeneities, and correlated increments, as observed in FBM \cite{cherstvy2015ergodicity, wang2020anomalous, wang2022anomalous}. An investigation within this context would scrutinize the applicability of multifractal modeling to elucidate and characterize such intricate systems.

Interactions spanning diverse spatiotemporal scales can fundamentally reshape the backdrop for subsequent fluctuations. For instance, cascade-dynamical instabilities can generate turbulent structures, intricate patterns of motion where formerly parallel currents transform into a complex array of vortices and eddies, with intermittent surges and recessions throughout space and time \cite{lovejoy1998diffusion, mandelbrot1974intermittent, shlesinger1987levy} (see also \cite{mangalam2023multifractalnonlinearity, mangalam2024multifractal}). Significantly, anomalous diffusion generating sometimes Lévy-like distributions distinguish active turbulence from its inertial counterpart \cite{cherstvy2021inertia, mukherjee2021anomalous}. The present finding that MFS features improved anomalous diffusion classification suggests a relationship between different diffusion coefficients and the specific geometries of the underlying fluctuations characterizing the observed data series. The intricate connection between multifractality and various models of anomalous diffusion is becoming increasingly evident, both from a theoretical \cite{afanasiev1991chaotic, chen2010anomalous, de2004anomalous, gmachowski2015fractal, lim2002self} and empirical \cite{bickel1999simple, lovejoy2018weather, menu2020anomalous, schmitt2001multifractal, seuront2004random, seuront2014anomalous, sharifi2012investigation} perspective. Multifractal formalisms and anomalous diffusion processes thus appear to be intricately intertwined, with their connection rooted in the far-from-equilibrium capacity to break time-reversal symmetry and to yield instead the multiscaled, nonlinear and multi-body interactions characteristic of evolving living systems \cite{shaebani2020computational}.

While the primary focus of the statistical physics community has predominantly centered on the study of anomalous diffusion at the level of single-particle trajectories \cite{metzler2000random, metzler2004restaurant, metzler2014anomalous}, a comprehensive investigation of biological processes implies the potential existence of multiple ``particles,'' each exhibiting their distinct trajectories. Specifically, examining particle trajectories may have inspired existing models. Implicit in these models may have been their couching of particle behavior within shared, aggregate behavior in which the particle dynamics are sometimes only downstream effects of larger scale, ensemble behaviors. Hence, it is essential to recognize that shared constraints could influence the seemingly independent trajectories of these entities in ways that may not be readily discernible in linear spatiotemporal analysis. In this sense, the particle models reflect only a reduction of the behavior of a multiscaled architecture to a single point mass, that is, a lower-dimensional projection of a higher-dimensional system. The particle models thus carry this signature of high dimensionality.

Future research could elaborate single-particle-trajectory models towards network modeling of the ensemble to make the cascade-dynamical relationships more explicit (cf. \cite{kelty2021multifractal, mangalam2020global, mangalam2020multifractal, stephen2012multifractal}). This endeavor would aim to reconcile qualitative disparities among the particle models (i.e., among FBM, SBM, CTRW, ATTM, and LW) with the pervasive appearance \cite{mangalam2023ergodic} and emerging predictive capacity of multifractal structures within these models, suggesting a shared ancestry within a cascade-dynamical framework. The robust predictive capabilities of multifractal parameters in these models underscore the potential similarity in the underlying cascade dynamics, transcending the diversity of particle models. As such, multifractal modeling promises to unveil causal interactions among these trajectories. The proficiency in categorizing anomalous processes based on cascade-related features implies that ML algorithms may be able to classify trajectories stemming from a spectrum of cascade dynamics. This attempt could serve precise classification of higher-dimensional biological structures but also extend to biomedical applications, affording novel theoretical traction on the cascade-dynamical character of healthy and diseased biological and psychological systems \cite{humeau2008multifractality, goldberger2002fractal, suckling2008endogenous, west2003multifractality}, extending beyond single trajectories.

\appendix

\section{Theoretical models} \label{TheoreticalModels}

\subsection{Fractional Brownian motion}

In fractional Brownian motion (FBM), $x(t)$ is a Gaussian process with stationary increments; it is symmetric, $\langle x(t)\rangle=0$, and importantly, its $\varA{MSD}$ scales as $\langle x(t)\rangle=2K_{H}t^{2H}$, where $H$ is the Hurst exponent and is related to the anomalous exponent $\beta$ as $H=\beta/2$ \cite{jeon2010fractional, kolmogorov1940wiener, mandelbrot1968fractional}. The two-time correlation for FBM is $\langle x(t_{1})x(t_{2})\rangle=K_{H}(t_{1}^{2H}+t_{2}^{2H}-|t_{1}-t_{2}|^{2H})$. FBM can also be defined as a process that arises from a generalized Langevin equation with nonwhite noise (or fractional Gaussian noise, FGN). The FGN has a standard normal distribution with zero mean and power-law correlation:
\begin{multline}
    <\xi_{fGn}(t_{1})\xi_{fGn}(t_{2})>=2K_{H}H(2H-1)|t_{1}-t_{2}|^{2H-2}+\\4K_{H}H|t_{1}-t_{2}|^{2H-1}\delta(t_{1}-t_{2}).
\end{multline}

The FBM features two regimes: one in which the noise is positively correlated ($1/2<H<1$, i.e., $1<\beta<2$, superdiffusive) and the other in which the noise is negatively correlated ($0<H<2$, i.e., $0<\beta<1$, subdiffusive). For $H=1/2$ ($\beta=1$), the noise is uncorrelated. Hence, the FBM converges to Brownian motion.

Various numerical approaches have been proposed to solve the FBM generalized Langevin equation. We use the method by Davies and Harte \cite{davies1987tests} (or Hosking \cite{hosking1984modeling} for $H$ close to $1$) via the \textrm{fbm} python package \cite{wood1994simulation}. Details about the numerical implementations can be found in the associated reference.

\subsection{Scaled Brownian motion}

The scaled Brownian motion (SBM) is a process described by the Langevin equation with a time-dependent diffusivity
\begin{equation}
    \frac{dx(t)}{dt}=\sqrt{2K(t)}\xi(t),
\end{equation}
where $\xi(t)$ is white Gaussian noise \cite{munoz2022stochastic}. In the case when $K(t)$ has a power-law dependence on to $t$ such that $K(t)=\beta K_{\beta}t^{\beta-1}$, $\varA{MSD}$ follows $<x^{2}(t)>_{N}=2K_{\beta}t^{\beta}$. The numerical implementation of SBM is presented in \textit{Algorithm 1}.\\

\setlength{\parindent}{0pt}
\textbf{Algorithm 1:} Generate SBM trajectory

\setlength{\parindent}{10pt}
\textbf{Input:}

\setlength{\parindent}{20pt}
length of the trajectory $T=N\Delta t$

anomalous exponent $\beta$

\setlength{\parindent}{10pt}
\textbf{Define:}

\setlength{\parindent}{20pt}
\texttt{erfcinv}($\vec a$) $\rightarrow$ Inverse complementary error function of $\vec a$

$U(N)\rightarrow$ returns $N$ uniform random numbers $\in[0,1]$

\setlength{\parindent}{10pt}
\textbf{Calculate:}

\setlength{\parindent}{20pt}
$\vec{\Delta x}\leftarrow(t_{2}^{\beta},t_{3}^{\beta},\dots, t_{N}^{\beta})-(t_{1}^{\beta},t_{2}^{\beta},\dots,t_{N-1}^{\beta})$

$\vec{\Delta x}\leftarrow\sqrt{2\vec{\Delta x}}\texttt{erfcinv}(2-2U(N-1))$

$\vec x\leftarrow$ cumsum ($\vec{\Delta x}$)

\setlength{\parindent}{10pt}
\textbf{Return:} \textbf{$\vec x$}

\subsection{Continuous time random walk}

The continuous time random walk (CTRW) is a family of random walks with arbitrary displacement density. The waiting time between subsequent steps is a stochastic variable \cite{scher1975anomalous}. We considered a specific case of CTRW with waiting times following a power-law distribution $\psi(t)\propto t^{-\sigma}$ and displacements following a Gaussian distribution with variance $D$ and zero means. In such case, the anomalous exponent is $\beta=\sigma-1$ ($\varA{MSD}=\langle x(t)^{2} \rangle\propto t^{\beta}$). As the waiting times follow a power-law distribution, for $\sigma=2$, $\varA{MSD}$ features Brownian motion with logarithmic corrections \cite{klafter2011first}.

\setlength{\parindent}{10pt}The numerical implementation of CTRW is presented in \textbf{Algorithm 2}. Notice that the variable $\tau$ represents the total time at $i$-th iteration. The output vector $\vec x$ corresponds to the position of the particle at the irregular times given by $\vec t$.\\

\setlength{\parindent}{0pt}
\textbf{Algorithm 2:} Generate CTRW trajectory

\setlength{\parindent}{10pt}
\textbf{Input:}

\setlength{\parindent}{20pt}
length of the trajectory $T$

anomalous exponent $\beta$

diffusion coefficient $D$

\setlength{\parindent}{10pt}
\textbf{Define:}

\setlength{\parindent}{20pt}
$\vec x\rightarrow$ empty vector

$\vec t\rightarrow$ empty vector

$N(\mu,S)\rightarrow$ Gaussian random number generator with mean $\mu$ and standard deviation $S$

$i=0$; $\tau=0$

\setlength{\parindent}{10pt}
\textbf{While} $\tau<T$ \textbf{do}

\setlength{\parindent}{20pt}
$t_{i}$ sample randomly from $\psi(t)\sim t^{-\sigma}$

$x_{i}\leftarrow x_{i-1}+N(0,\sqrt{D})$

$\tau\leftarrow\tau +t_{i}$

$i\leftarrow i+1$

\setlength{\parindent}{10pt}
\textbf{end while}

\textbf{Return:} \textbf{$\vec x$}, \textbf{$\vec t$}

\subsection{Annealed transient time motion}

The annealed transient time motion (ATTM) implements the motion of a Brownian particle with time-dependent diffusivity \cite{massignan2014nonergodic}. The observable performs Brownian motion for a random time $t_{1}$ with a random diffusion coefficient $D_{1}$, then for $t_{2}$ with $D_{2}$, and so on. The diffusion coefficients follow a distribution such that $P(D)\propto D^{\sigma-1}$ with $\sigma>0$ as $D\to0$, and that decays rapidly for large $D$. If the random times $t$ are sampled from a distribution with expected value $E[t|D]=D^{-\gamma}$, with $\sigma<\gamma<\sigma+1$, the anomalous exponent is $\beta=\sigma/\gamma$. Here, we consider that the distribution is a delta function, $P_{t}(t|D)=\delta(t - D^{-\gamma})$. Hence, the time $t_{i}$ in which the observable performs Brownian motion with a random diffusion coefficient $D_{i}$ is $t_{i}=D_{i}^{-\gamma}$, with $D_{i}$ extracted from $P(D)$.

\setlength{\parindent}{10pt}The numerical implementation of ATTM is presented in \textit{Algorithm 3}. In contrast to CTRW and LW, the only output is $\vec x$ because the trajectory is produced at regular intervals.\\

\setlength{\parindent}{0pt}
\textbf{Algorithm 3:} Generate ATTM trajectory

\setlength{\parindent}{10pt}
\textbf{Input:}

\setlength{\parindent}{20pt}
length of the trajectory $T$

anomalous exponent $\beta$

sampling time $\Delta t$

\setlength{\parindent}{10pt}
\textbf{Define:}

\setlength{\parindent}{20pt}
\textbf{While} $\sigma>\gamma$ and $\gamma>\sigma+1$ \textbf{do}

\setlength{\parindent}{30pt}
$\sigma\leftarrow$ uniform random number $\in(0,3]$

$\gamma=\sigma/\beta$
\setlength{\parindent}{20pt}

\textbf{end while}

BM($D_{i}$, $t_{i}$, $\Delta t$) $\rightarrow$ generates a Brownian motion trajectory of length $t_{i}$ with diffusion coefficient $D_{i}$, sampled at time intervals $\Delta t$

\setlength{\parindent}{10pt}
\textbf{While} $\tau<T$ \textbf{do}

\setlength{\parindent}{20pt}
$D_{i}\leftarrow$ sample randomly from $P(D)\propto D^{\sigma-1}$

$t_{i}\leftarrow D_{i}^{-\gamma}$

number of steps $N_{i}=\mathrm{round}(t_{i}/\Delta t)$

$x_{i},\dots,x_{i+N_{i}}\leftarrow$ BM($D_{i}$,$t_{i}$,$\Delta t$)

$i\leftarrow i+N_{i}+1$

$\tau=\tau+N_{i}\Delta t$

\setlength{\parindent}{10pt}
\textbf{end while}

\textbf{Return:} $\vec x$

\subsection{Lévy walk}

The Lévy walk (LW) is a particular superdiffusive CTRW. Like subdiffusive CTRW, the flight time, that is, the time between steps, for LW is irregular \cite{klafter1994levy}, but, in contrast to subdiffusive CTRW, the distribution of displacements for LW is not Gaussian. We considered the case in which the flight times follow the distribution $\psi(t)=t^{-\sigma-1}$. At each step, the displacement is $\Delta x$, and the step length is $|\Delta x|$. The displacements are correlated with the flight times such that the probability of moving a step $\Delta x$ at time $t$ and stopping at the new position to wait for a new random event to happen is $\psi(\Delta x,t)=\frac{1}{2}\delta(|\Delta x|-vt)\psi(t)$, where $v$ is the velocity. The anomalous exponent is given by
\begin{equation*}
\beta =
  \begin{cases}
  2, & \text{if }0<\sigma<1 \\
  3-\sigma, & \text{if }1<\sigma<2.
  \end{cases}
  \tag{A1}\label{eq:A1}
\end{equation*}

\setlength{\parindent}{10pt}The numerical implementation of LW is presented in \textit{Algorithm 4}. Notice that we use a random number $r$, which can take values 0 or 1, to decide in which sense the step is performed. The output vectors $\vec x$ represent irregularly sampled positions and times.\\

\setlength{\parindent}{0pt}
\textbf{Algorithm 4:} Generate LW trajectory

\setlength{\parindent}{10pt}
\textbf{Input:}

\setlength{\parindent}{20pt}
length of the trajectory $T$

anomalous exponent $\beta$

\setlength{\parindent}{10pt}
\textbf{Define:}

\setlength{\parindent}{20pt}
$\vec x\rightarrow$ empty vector

$\vec t\rightarrow$ empty vector

$v\rightarrow$ random number $\in(0,10]$

$i=0$

\setlength{\parindent}{10pt}
\textbf{While} $\tau<T$ \textbf{do}

\setlength{\parindent}{20pt}
$t_{i}\leftarrow$ sample randomly from $\psi(t)\sim t^{-\sigma-1}$

$x_{i}\leftarrow(-1)^{r}vt_{i}$, where random $r$ is 0 or 1 with equal probability.

$\tau\leftarrow\tau+t_{i}$

$i\leftarrow i+1$

\setlength{\parindent}{10pt}
\textbf{end while}

\textbf{Return:} $\vec x,\vec t$

\section{Traditional features} \label{TraditionalFeatures}

This appendix briefly introduces the definitions of the traditional features from \cite{kowalek2019classification, kowalek2022boosting} listed in section \ref{Section: TraditionalFeatures}.

\subsection{Original features}

\subsubsection{Anomalous exponent}

Four estimates for the anomalous diffusion exponent $\beta$ constituted separate features:
\begin{enumerate}
    \itemsep0em
    \item the standard estimation, based on fitting the empirical TAMSD to Eq.~(\ref{eq: 1}), 
    \item 3 estimation methods proposed for trajectories with noise, which is normally distributed with zero mean \cite{lanoiselee2018optimal},
    \begin{enumerate}
        \item using the estimator
        \begin{widetext}
        \begin{equation}
            \hat{\beta}=\frac{n_\mathrm{max} \sum_{n=1}^{n_\mathrm{max}}\log(n)\log(
            \langle\mathbf{r}^2(n\Delta t)\rangle)-\sum_{n=1}^{n_\mathrm{max}}\log(n)\left(
            \sum_{n=1}^{n_\mathrm{max}}\log(\langle\mathbf{r}^2(n\cdot\Delta t)\rangle) \right)}
            {n_\mathrm{max}\sum_{n=1}^{n_\mathrm{max}}\log^2(n)-\left(\sum_{n=1}^{
            n_\mathrm{max}}\log(n)\right)^2},
        \end{equation}
        \end{widetext}
        where $n$ denotes time lag with $n_\mathrm{max}=N/10$---where $N$ is $T/\Delta t$---rounded to
        the nearest lower integer (but not less than $4$),
        \item simultaneous fitting of the parameters $\hat{D}$, $\hat{\beta}$, and $\hat{\sigma}$ in the relation
        \begin{equation}
            \langle\mathbf{r}^2(t)\rangle=2d\hat{D}t^{\hat{\beta}}+\hat{\sigma}^2, \tag{B2}\label{eq: B2}
        \end{equation}
        where $d$ denotes the embedding dimension, $D$ is the diffusion coefficient, and $\sigma^2$ is the variance of noise,
        \item simultaneous fitting of the parameters $\hat{D}$ and $\hat{\beta}$ in the equation
        \begin{equation}
            \langle\mathbf{r}^2(n\Delta t)\rangle=2d\hat{D}(n^{\hat{\beta}}-1)(\Delta t)^{\hat{\beta}} .
        \end{equation}
    \end{enumerate}
\end{enumerate}

\subsubsection{Diffusion coefficient}

An estimator of the diffusion coefficient was extracted from the fit of the empirical TAMSD to Eq.~(\ref{eq: B2}).

\subsubsection{Efficiency}

The efficiency $E$ relates the net squared displacement to the sum of squared step lengths,
\begin{equation}
    E(N,1)=\frac{|x_{N}-x_{1}|^{2}}{(N-1)\sum_{i=1}^{N-1}|x_{i+1}-x_{i}|^{2}}. \tag{B4}\label{eq: B4}
\end{equation} 
Efficiency ranges from $0$ to $1$ and should help detect directed motion, which takes values close to $1$.

\subsubsection{Empirical velocity autocorrelation function}

The empirical velocity autocorrelation function \cite{weber2010bacterial} for lag $1$ and point $n$ is in one dimension
\begin{equation}
    \chi_n=\frac{1}{N-1-n}\sum^{N-1-n}_{i=1}(x_{i+1+n}-x_{i+n})(x_{i+1}-x_{i}),
\end{equation}
it can be used to distinguish different subdiffusive processes. In Ref. \cite{kowalek2022boosting}, $\chi_n$ for points $n=1$ and $n=2$ was used, as well as in the present study.

\subsubsection{Fractal dimension}

The fractal dimension measures the space-filling capacity of a pattern (a trajectory in our case). For a planar trajectory, it may be calculated as 
\begin{equation}
    D_f=\frac{\log N}{\log(NdL^{-1})},
\end{equation}
where $L=\sum^{N}_{i}|\Delta x_{i}|$ is the total distance traveled, $N$ is the number of steps, and $d$ is the largest distance between any two positions \cite{katz1985fractals}. It usually takes values around $1$ for directed motion and around $2$ for normal diffusion. For subdiffusive CTRW, it is also around $2$, while for FBM, it is larger than $2$.

\subsubsection{Maximal excursion}

The maximal excursion of the particle is
\begin{equation}
    \mathrm{ME}=\frac{\max_{i}(|x_{i+1}-x_{i}|)}{x_{N}-x_{1}}.
\end{equation}
It detects relatively long jumps (in comparison to the overall displacement).

\subsubsection{Mean maximal excursion}

The mean maximal excursion can replace the MSD as the observable used to determine the anomalous exponent \cite{tejedor2010quantitative}. It is defined as the standardized value of the largest distance traveled by a particle,
\begin{equation}
    T_n=\frac{\max_{i}(|x_{i}-x_{1}|)}{\sqrt{\hat{\sigma}^{2}_{N}(t_{N}-t_{1})}}.
\end{equation}
The parameter $\hat{\sigma}_N$ is a consistent estimator of the standard deviation,
\begin{equation}
    \hat{\sigma}^{2}_{N}=\frac{1}{2(N-1)\Delta t}\sum^{N}_{j=2}|x_j-x_{j-1}|^{2}.
\end{equation}

\subsubsection{Mean Gaussianity}

The Gaussianity $g(n)$ checks the Gaussian statistics of increments of a trajectory \cite{ernst2014probing} as
\begin{equation}
    g(n)=\frac{\langle r_n^4\rangle}{3\langle r_n^2\rangle^2}-1,
\end{equation}
where $\langle r_n^k\rangle$ denotes the $k$th moment of the trajectory at time lag $n$. The Gaussianity for normal diffusion is equal to $0$. The same result should be obtained for FBM since its increments follow a Gaussian distribution. Other types of motion should show deviations from that value.

Instead of looking at Gaussianities at single-time lags, in Ref. \cite{kowalek2022boosting} and here, the mean Gaussianity across all lags was used as one of the features,
\begin{equation}
    \langle g\rangle=\frac{1}{N}\sum^{N}_{i=1}g(n).
\end{equation}

\subsubsection{Mean-squared displacement ratio}

The MSD ratio gives information about the shape of the corresponding MSD curve. We will define it as
\begin{equation}
    \mathrm{MSDR}(n_1,n_2)=\frac{\langle r_{n_1}^2\rangle}{\langle r_{n_2}^2\rangle}-\frac{n_1}{n_2}, \tag{B10}\label{eq: B10}
\end{equation}
where $n_1<n_2$. $\mathrm{MSDR}$ is zero for normal diffusion ($\beta=1$). We should get $\mathrm{MSDR}\leq0$ for sub- and $\mathrm{MSDR}\geq0$ for superdiffusion. in Ref. \cite{kowalek2022boosting} and the present study, $n_{2}=n_{1}+1$ was taken, and then the averaged ratio across all $n_{1}=1,2,\ldots, N-1$ was calculated for every trajectory.

\subsubsection{Kurtosis}

The kurtosis gives insight into the asymmetry and peakedness of the distribution of points within a trajectory \cite{helmuth2007novel}. It is defined as the fourth moment,
\begin{equation}
    K=\frac{1}{N}\sum_{i=1}^{N}\frac{(x_{i}-\bar{x})^{4}}{\sigma^{4}_{x}}, \tag{B11}\label{eq: B11}
\end{equation}
where $\bar{x}$ is the mean position and $\sigma_{x}$ the standard deviation.

\subsubsection{Statistics based on $p$-variation}

The empirical $p$-variation is given by the formula \cite{burnecki2010fractional}
\begin{equation}
    V_{m}^{(p)}=\sum_{k=1}^{\frac{N}{m}-1}|x_{(k+1)m}-x_{km}|^{p}\propto m^{\gamma_p}.
\end{equation}
This statistic can be used to detect fractional L\'{e}vy stable motion (including FBM). Ten features based on $V_{m}^{(p)}$ were used for the classification of trajectories:
\begin{enumerate}
    \itemsep0em
    \item the power $\gamma_{p}$ fitted to $p$-variation for lags $1$ to $5$ for nine values of $p$,
    \item the statistic $P$ used in Ref. \cite{loch2020impact}, based on the monotonicity changes of $V_{m}^{(p)}$ as a function of $m$ as indicated by the sign of $\gamma_{p}$:
    \begin{equation}
        P=\left\{\begin{array}{rl} 0 & \textrm{if $V_{m}^{(p)}$ does not change the monotonicity},\\ 
        1 & \textrm{if $\gamma_{p}$ changes from negative to positive},\\
        -1 & \textrm{if $\gamma_{p}$ changes from positive to negative}.
    \end{array} 
    \right.
    \end{equation}
\end{enumerate}

\subsubsection{Straightness}

The straightness $S$ measures the average direction change between subsequent steps. It relates the net displacement of a particle to the sum of all step lengths,
\begin{equation}
    S=\frac{|x_{N}-x_{1}|}{\sum_{i=1}^{N-1}|x_{i+1}-x_{i}|}. 
\end{equation}

\subsubsection{Trappedness}

The trappedness is the probability that a diffusing particle is trapped in a bounded region with radius $r_{0}$ up to some observation time $t$. \cite{saxton1993lateral} estimated this probability with
\begin{equation}
    P(D,t,r_0)\approx10^{0.2048-2.5117(Dt/r_{0}^{2})}.
\end{equation}
$r_{0}$ is approximated by half of the maximum distance between any two positions along a given trajectory, $D$ is estimated by fitting the first two points of the MSD curve (i.e., the so-called short-time diffusion coefficient), and $t$ is chosen as the total observation time $T$.

\subsection{Additional features}

\subsubsection{d'Agostino-Pearson test statistic}

The d'Agostino-Pearson $\kappa^{2}$ test statistic \cite{d1973tests} measures the departure of a given sample from normality,
\begin{equation}
    \kappa^{2}=Z_{1}(g_{1})+Z_{2}(K),
\end{equation}
where $K$ is the sample kurtosis given by Eq.~(\ref{eq: B11}) and $g_{1}=m_{3}/m_{2}^{3/2}$ is the sample skewness with $m_{j}$ being the $j$th sample central moment. The transformations $Z_{1}$ and $Z_{2}$ bring the distributions of the skewness and kurtosis as close to the standard normal as possible. This feature must help distinguish SBM and ATTM from other trajectories.

\subsubsection{Kolmogorov-Smirnov (KS) statistic against $\chi^{2}$ distribution}

The KS statistic quantifies the distance between the empirical distribution function of the sample $F_{T}(x)$ and the cumulative  function $G_{T}(x)$ of a reference distribution,
\begin{equation}
    D_{T}=\sup_x|F_{T}(x)-G_{T}(x)|.
\end{equation}
The next feature to consider is the statistic calculated by comparing the empirical distribution of squared increments from a trajectory to a sampled $\chi^{2}$ distribution. This choice is rooted in the concept that a Gaussian trajectory should theoretically yield a distribution of squared increments closely resembling the $\chi^{2}$ distribution.

\subsubsection{Joseph, Noah, and Moses exponents}

Processes featuring stationary increments can manifest anomalous scaling of MSD via two mechanisms that, in principle, defy the Gaussian central limit theorem. These mechanisms include long-time increment correlations, known as the Joseph effect, and a flat-tailed increment distribution, referred to as the Noah effect \cite{aghion2021moses, mandelbrot1968noah}. Notably, FBM typifies the first effect, while LW embodies the latter. Furthermore, nonstationary increment distributions can induce anomalous scaling, giving rise to the Moses effect \cite{aghion2021moses}. The Moses effect plays a pivotal role in identifying SBM and ATTM trajectories.

All three effects may be quantified by exponents, which can be used as features. Given a stochastic process $x_{t}$ and the corresponding increment process $\delta_{t}(\tau)=x_{t+\tau}-x_{t}$, the Joseph, Moses, and Noah exponents are defined as follows:
\begin{enumerate}
    \itemsep0em
    \item The Joseph exponent $J$ is estimated from the ensemble average of the rescaled range statistics,
    \begin{equation}
        \mathbb{E}\left[\frac{\max_{1\leq i\leq n}[x_{i}-\frac{i}{n}x_{n}]-\min_{1\leq i\leq n}[
        x_{i}-\frac{i}{n}x_{n}]}{\sigma_{n}}\right]\sim n^{J},
    \end{equation}
    where $\sigma_{j}$ is the standard deviation of the process $x_{j}$.
    \item The Moses exponent $M$ is determined from the scaling of the ensemble probability distribution of the sum of the absolute value of the increments, which can be estimated by the scaling of the median of the probability distribution of $Y_{n}=\sum^{n}_{i=1}|\delta_{i}|$,
    \begin{equation}
        \mathbb{E}[Y_{n}]\sim n^{M+\frac{1}{2}}.
    \end{equation}
    \item The Noah exponent $L$ is extracted from the scaling of the ensemble probability distribution of the sum of squared increments, which can be estimated by the scaling of the median of the probability distribution of $Z_{n}=\sum^{n}_{i=1}\delta_{i}^{2}$:
    \begin{equation}
        \mathbb{E}[Z_{n}]\sim n^{2L+2M-1}.
    \end{equation}
\end{enumerate}
The \{J,M,L\} exponents are related to the anomalous exponent $\beta$, \cite{chen2017anomalous, vilk2022unravelling}
\begin{equation}
    \beta/2 = J+M+L-1.
\end{equation}

\subsubsection{Detrending moving average}

The detrending moving average (DMA) statistic \cite{balcerek2021discriminating} is given by
\begin{equation}
    \mathrm{DMA}(\tau)=\frac{1}{N-\tau}\sum_{i=\tau+1}^N\left(x_{i}-\overline{x}^{\tau}_{i}\right)^{2},
\end{equation}
for $\tau=\{1,2,\ldots$\}, where $\overline{x}^{\tau}_{i}$ is a moving average of $\tau$ observations, that is, $\overline{x}^{\tau}_{i}=\frac{1}{\tau+1}\sum_{j=0}^{\tau} x_{i-j}$. According to Ref. \cite{balcerek2021discriminating}, a DMA-based statistical test can help detect SBM. In Ref. \cite{kowalek2022boosting} and in this work, $\mathrm{DMA}(1)$ and $\mathrm{DMA}(2)$ were used as features.

\subsubsection{Average moving window characteristics}

Let us define the following moving window characteristic
\begin{multline}
    \mathrm{MW}_m=\frac{1}{2(N-m-2)}\sum_{t=1}^{N-m-2}\Biggl|\mathrm{sgn}\left(\overline{x}_{t+2}^{(m)}-\overline{x}_{t+1}^{(m)}\right)\\
    -\mathrm{sgn}\left(\overline{x}_{t+1}^{(m)}
    -\overline{x}_{t}^{(m)}\right)\Biggl|,
\end{multline}
where $\overline{x}^{(m)}$ denotes a statistic of the process calculated within the window of length $m$ and $\mathrm{sgn}$ is the sign function. We here use four attributes calculating $\mathrm{MW}_m$ using the mean and standard deviation for $\overline{x}$ with windows of lengths $m=10$ and $m=20$.
    
\subsubsection{Maximum standard deviation}

The last two features from the extended feature set rely on the standard deviation $\sigma_{m}$ of the process calculated within windows of length $m$,
\begin{equation}
    \mathrm{MXM}_m=\frac{\min(\sigma_{m}(t))}{\max(\sigma_{m}(t))}
\end{equation}
and
\begin{equation}
    \mathrm{MXC}_m=\frac{\max|\sigma_{m}(t+1)-\sigma_{m}(t)|}{\sigma},
\end{equation}
where $\sigma$ denotes the sample standard deviation over the whole trajectory and $\sigma_m(t)$ the standard deviation within the window starting at $t$ and ending at $t+m\Delta t$. We used $m=3$. These features must improve the detection of ATTM-type movements.

\section*{Acknowledgments}

R.M. acknowledges funding from the German Ministry for Education and Research (NSF-BMBF project STAXS) and the German Science Foundation (DFG, grant no. ME 1535/12-1). M.M. was supported by the Center for Research in Human Movement Variability at the University of Nebraska at Omaha, funded by the National Institute of General Medical Sciences (NIGMS, grant no.  P20GM109090).

\bibliography{apssamp}

\providecommand{\noopsort}[1]{}\providecommand{\singleletter}[1]{#1}%
\begin{thebibliography}{179}%
\makeatletter
\providecommand \@ifxundefined [1]{%
 \@ifx{#1\undefined}
}%
\providecommand \@ifnum [1]{%
 \ifnum #1\expandafter \@firstoftwo
 \else \expandafter \@secondoftwo
 \fi
}%
\providecommand \@ifx [1]{%
 \ifx #1\expandafter \@firstoftwo
 \else \expandafter \@secondoftwo
 \fi
}%
\providecommand \natexlab [1]{#1}%
\providecommand \enquote  [1]{``#1''}%
\providecommand \bibnamefont  [1]{#1}%
\providecommand \bibfnamefont [1]{#1}%
\providecommand \citenamefont [1]{#1}%
\providecommand \href@noop [0]{\@secondoftwo}%
\providecommand \href [0]{\begingroup \@sanitize@url \@href}%
\providecommand \@href[1]{\@@startlink{#1}\@@href}%
\providecommand \@@href[1]{\endgroup#1\@@endlink}%
\providecommand \@sanitize@url [0]{\catcode `\\12\catcode `\$12\catcode
  `\&12\catcode `\#12\catcode `\^12\catcode `\_12\catcode `\%12\relax}%
\providecommand \@@startlink[1]{}%
\providecommand \@@endlink[0]{}%
\providecommand \url  [0]{\begingroup\@sanitize@url \@url }%
\providecommand \@url [1]{\endgroup\@href {#1}{\urlprefix }}%
\providecommand \urlprefix  [0]{URL }%
\providecommand \Eprint [0]{\href }%
\providecommand \doibase [0]{https://doi.org/}%
\providecommand \selectlanguage [0]{\@gobble}%
\providecommand \bibinfo  [0]{\@secondoftwo}%
\providecommand \bibfield  [0]{\@secondoftwo}%
\providecommand \translation [1]{[#1]}%
\providecommand \BibitemOpen [0]{}%
\providecommand \bibitemStop [0]{}%
\providecommand \bibitemNoStop [0]{.\EOS\space}%
\providecommand \EOS [0]{\spacefactor3000\relax}%
\providecommand \BibitemShut  [1]{\csname bibitem#1\endcsname}%
\let\auto@bib@innerbib\@empty
%</preamble>
\bibitem [{\citenamefont {Sagi}\ \emph {et~al.}(2012)\citenamefont {Sagi},
  \citenamefont {Brook}, \citenamefont {Almog},\ and\ \citenamefont
  {Davidson}}]{sagi2012observation}%
  \BibitemOpen
  \bibfield  {author} {\bibinfo {author} {\bibfnamefont {Y.}~\bibnamefont
  {Sagi}}, \bibinfo {author} {\bibfnamefont {M.}~\bibnamefont {Brook}},
  \bibinfo {author} {\bibfnamefont {I.}~\bibnamefont {Almog}},\ and\ \bibinfo
  {author} {\bibfnamefont {N.}~\bibnamefont {Davidson}},\ }\bibfield  {title}
  {\bibinfo {title} {Observation of anomalous diffusion and fractional
  self-similarity in one dimension},\ }\href@noop {} {\bibfield  {journal}
  {\bibinfo  {journal} {Physical Review Letters}\ }\textbf {\bibinfo {volume}
  {108}},\ \bibinfo {pages} {093002} (\bibinfo {year} {2012})}\BibitemShut
  {NoStop}%
\bibitem [{\citenamefont {Zhao}\ \emph {et~al.}(2014)\citenamefont {Zhao},
  \citenamefont {Deng}, \citenamefont {Avdoshenko}, \citenamefont {Fu},
  \citenamefont {Eckert},\ and\ \citenamefont {R{\"u}mmeli}}]{zhao2014direct}%
  \BibitemOpen
  \bibfield  {author} {\bibinfo {author} {\bibfnamefont {J.}~\bibnamefont
  {Zhao}}, \bibinfo {author} {\bibfnamefont {Q.}~\bibnamefont {Deng}}, \bibinfo
  {author} {\bibfnamefont {S.~M.}\ \bibnamefont {Avdoshenko}}, \bibinfo
  {author} {\bibfnamefont {L.}~\bibnamefont {Fu}}, \bibinfo {author}
  {\bibfnamefont {J.}~\bibnamefont {Eckert}},\ and\ \bibinfo {author}
  {\bibfnamefont {M.~H.}\ \bibnamefont {R{\"u}mmeli}},\ }\bibfield  {title}
  {\bibinfo {title} {Direct in situ observations of single {F}e atom catalytic
  processes and anomalous diffusion at graphene edges},\ }\href@noop {}
  {\bibfield  {journal} {\bibinfo  {journal} {Proceedings of the National
  Academy of Sciences}\ }\textbf {\bibinfo {volume} {111}},\ \bibinfo {pages}
  {15641} (\bibinfo {year} {2014})}\BibitemShut {NoStop}%
\bibitem [{\citenamefont {Banks}\ and\ \citenamefont
  {Fradin}(2005)}]{banks2005anomalous}%
  \BibitemOpen
  \bibfield  {author} {\bibinfo {author} {\bibfnamefont {D.~S.}\ \bibnamefont
  {Banks}}\ and\ \bibinfo {author} {\bibfnamefont {C.}~\bibnamefont {Fradin}},\
  }\bibfield  {title} {\bibinfo {title} {Anomalous diffusion of proteins due to
  molecular crowding},\ }\href@noop {} {\bibfield  {journal} {\bibinfo
  {journal} {Biophysical Journal}\ }\textbf {\bibinfo {volume} {89}},\ \bibinfo
  {pages} {2960} (\bibinfo {year} {2005})}\BibitemShut {NoStop}%
\bibitem [{\citenamefont {Barkai}\ \emph {et~al.}(2012)\citenamefont {Barkai},
  \citenamefont {Garini},\ and\ \citenamefont {Metzler}}]{barkai2012single}%
  \BibitemOpen
  \bibfield  {author} {\bibinfo {author} {\bibfnamefont {E.}~\bibnamefont
  {Barkai}}, \bibinfo {author} {\bibfnamefont {Y.}~\bibnamefont {Garini}},\
  and\ \bibinfo {author} {\bibfnamefont {R.}~\bibnamefont {Metzler}},\
  }\bibfield  {title} {\bibinfo {title} {Strange kinetics of single molecules
  in living cells},\ }\href@noop {} {\bibfield  {journal} {\bibinfo  {journal}
  {Physics Today}\ }\textbf {\bibinfo {volume} {65}},\ \bibinfo {pages} {29}
  (\bibinfo {year} {2012})}\BibitemShut {NoStop}%
\bibitem [{\citenamefont {Guigas}\ and\ \citenamefont
  {Weiss}(2008)}]{guigas2008sampling}%
  \BibitemOpen
  \bibfield  {author} {\bibinfo {author} {\bibfnamefont {G.}~\bibnamefont
  {Guigas}}\ and\ \bibinfo {author} {\bibfnamefont {M.}~\bibnamefont {Weiss}},\
  }\bibfield  {title} {\bibinfo {title} {Sampling the cell with anomalous
  diffusion—{T}he discovery of slowness},\ }\href@noop {} {\bibfield
  {journal} {\bibinfo  {journal} {Biophysical Journal}\ }\textbf {\bibinfo
  {volume} {94}},\ \bibinfo {pages} {90} (\bibinfo {year} {2008})}\BibitemShut
  {NoStop}%
\bibitem [{\citenamefont {H{\"o}fling}\ and\ \citenamefont
  {Franosch}(2013)}]{hofling2013anomalous}%
  \BibitemOpen
  \bibfield  {author} {\bibinfo {author} {\bibfnamefont {F.}~\bibnamefont
  {H{\"o}fling}}\ and\ \bibinfo {author} {\bibfnamefont {T.}~\bibnamefont
  {Franosch}},\ }\bibfield  {title} {\bibinfo {title} {Anomalous transport in
  the crowded world of biological cells},\ }\href@noop {} {\bibfield  {journal}
  {\bibinfo  {journal} {Reports on Progress in Physics}\ }\textbf {\bibinfo
  {volume} {76}},\ \bibinfo {pages} {046602} (\bibinfo {year}
  {2013})}\BibitemShut {NoStop}%
\bibitem [{\citenamefont {Jeon}\ \emph {et~al.}(2011)\citenamefont {Jeon},
  \citenamefont {Tejedor}, \citenamefont {Burov}, \citenamefont {Barkai},
  \citenamefont {Selhuber-Unkel}, \citenamefont {Berg-S{\o}rensen},
  \citenamefont {Oddershede},\ and\ \citenamefont {Metzler}}]{jeon2011vivo}%
  \BibitemOpen
  \bibfield  {author} {\bibinfo {author} {\bibfnamefont {J.-H.}\ \bibnamefont
  {Jeon}}, \bibinfo {author} {\bibfnamefont {V.}~\bibnamefont {Tejedor}},
  \bibinfo {author} {\bibfnamefont {S.}~\bibnamefont {Burov}}, \bibinfo
  {author} {\bibfnamefont {E.}~\bibnamefont {Barkai}}, \bibinfo {author}
  {\bibfnamefont {C.}~\bibnamefont {Selhuber-Unkel}}, \bibinfo {author}
  {\bibfnamefont {K.}~\bibnamefont {Berg-S{\o}rensen}}, \bibinfo {author}
  {\bibfnamefont {L.}~\bibnamefont {Oddershede}},\ and\ \bibinfo {author}
  {\bibfnamefont {R.}~\bibnamefont {Metzler}},\ }\bibfield  {title} {\bibinfo
  {title} {\textit{In vivo} anomalous diffusion and weak ergodicity breaking of
  lipid granules},\ }\href@noop {} {\bibfield  {journal} {\bibinfo  {journal}
  {Physical Review Letters}\ }\textbf {\bibinfo {volume} {106}},\ \bibinfo
  {pages} {048103} (\bibinfo {year} {2011})}\BibitemShut {NoStop}%
\bibitem [{\citenamefont {Jeon}\ \emph {et~al.}(2012)\citenamefont {Jeon},
  \citenamefont {Monne}, \citenamefont {Javanainen},\ and\ \citenamefont
  {Metzler}}]{jeon2012anomalous}%
  \BibitemOpen
  \bibfield  {author} {\bibinfo {author} {\bibfnamefont {J.-H.}\ \bibnamefont
  {Jeon}}, \bibinfo {author} {\bibfnamefont {H.~M.-S.}\ \bibnamefont {Monne}},
  \bibinfo {author} {\bibfnamefont {M.}~\bibnamefont {Javanainen}},\ and\
  \bibinfo {author} {\bibfnamefont {R.}~\bibnamefont {Metzler}},\ }\bibfield
  {title} {\bibinfo {title} {Anomalous diffusion of phospholipids and
  cholesterols in a lipid bilayer and its origins},\ }\href@noop {} {\bibfield
  {journal} {\bibinfo  {journal} {Physical Review Letters}\ }\textbf {\bibinfo
  {volume} {109}},\ \bibinfo {pages} {188103} (\bibinfo {year}
  {2012})}\BibitemShut {NoStop}%
\bibitem [{\citenamefont {Krapf}\ and\ \citenamefont
  {Metzler}(2019)}]{krapf2019strange}%
  \BibitemOpen
  \bibfield  {author} {\bibinfo {author} {\bibfnamefont {D.}~\bibnamefont
  {Krapf}}\ and\ \bibinfo {author} {\bibfnamefont {R.}~\bibnamefont
  {Metzler}},\ }\bibfield  {title} {\bibinfo {title} {Strange interfacial
  molecular dynamics},\ }\href@noop {} {\bibfield  {journal} {\bibinfo
  {journal} {Physics Today}\ }\textbf {\bibinfo {volume} {72}},\ \bibinfo
  {pages} {48} (\bibinfo {year} {2019})}\BibitemShut {NoStop}%
\bibitem [{\citenamefont {Metzler}\ and\ \citenamefont
  {Klafter}(2000)}]{metzler2000random}%
  \BibitemOpen
  \bibfield  {author} {\bibinfo {author} {\bibfnamefont {R.}~\bibnamefont
  {Metzler}}\ and\ \bibinfo {author} {\bibfnamefont {J.}~\bibnamefont
  {Klafter}},\ }\bibfield  {title} {\bibinfo {title} {The random walk's guide
  to anomalous diffusion: {A} fractional dynamics approach},\ }\href@noop {}
  {\bibfield  {journal} {\bibinfo  {journal} {Physics Reports}\ }\textbf
  {\bibinfo {volume} {339}},\ \bibinfo {pages} {1} (\bibinfo {year}
  {2000})}\BibitemShut {NoStop}%
\bibitem [{\citenamefont {Ritchie}\ \emph {et~al.}(2005)\citenamefont
  {Ritchie}, \citenamefont {Shan}, \citenamefont {Kondo}, \citenamefont
  {Iwasawa}, \citenamefont {Fujiwara},\ and\ \citenamefont
  {Kusumi}}]{ritchie2005detection}%
  \BibitemOpen
  \bibfield  {author} {\bibinfo {author} {\bibfnamefont {K.}~\bibnamefont
  {Ritchie}}, \bibinfo {author} {\bibfnamefont {X.-Y.}\ \bibnamefont {Shan}},
  \bibinfo {author} {\bibfnamefont {J.}~\bibnamefont {Kondo}}, \bibinfo
  {author} {\bibfnamefont {K.}~\bibnamefont {Iwasawa}}, \bibinfo {author}
  {\bibfnamefont {T.}~\bibnamefont {Fujiwara}},\ and\ \bibinfo {author}
  {\bibfnamefont {A.}~\bibnamefont {Kusumi}},\ }\bibfield  {title} {\bibinfo
  {title} {Detection of non-{B}rownian diffusion in the cell membrane in single
  molecule tracking},\ }\href@noop {} {\bibfield  {journal} {\bibinfo
  {journal} {Biophysical Journal}\ }\textbf {\bibinfo {volume} {88}},\ \bibinfo
  {pages} {2266} (\bibinfo {year} {2005})}\BibitemShut {NoStop}%
\bibitem [{\citenamefont {Toli{\'c}-N{\o}rrelykke}\ \emph
  {et~al.}(2004)\citenamefont {Toli{\'c}-N{\o}rrelykke}, \citenamefont
  {Munteanu}, \citenamefont {Thon}, \citenamefont {Oddershede},\ and\
  \citenamefont {Berg-S{\o}rensen}}]{tolic2004anomalous}%
  \BibitemOpen
  \bibfield  {author} {\bibinfo {author} {\bibfnamefont {I.~M.}\ \bibnamefont
  {Toli{\'c}-N{\o}rrelykke}}, \bibinfo {author} {\bibfnamefont {E.-L.}\
  \bibnamefont {Munteanu}}, \bibinfo {author} {\bibfnamefont {G.}~\bibnamefont
  {Thon}}, \bibinfo {author} {\bibfnamefont {L.}~\bibnamefont {Oddershede}},\
  and\ \bibinfo {author} {\bibfnamefont {K.}~\bibnamefont {Berg-S{\o}rensen}},\
  }\bibfield  {title} {\bibinfo {title} {Anomalous diffusion in living yeast
  cells},\ }\href@noop {} {\bibfield  {journal} {\bibinfo  {journal} {Physical
  Review Letters}\ }\textbf {\bibinfo {volume} {93}},\ \bibinfo {pages}
  {078102} (\bibinfo {year} {2004})}\BibitemShut {NoStop}%
\bibitem [{\citenamefont {Angelini}\ \emph {et~al.}(2011)\citenamefont
  {Angelini}, \citenamefont {Hannezo}, \citenamefont {Trepat}, \citenamefont
  {Marquez}, \citenamefont {Fredberg},\ and\ \citenamefont
  {Weitz}}]{angelini2011glass}%
  \BibitemOpen
  \bibfield  {author} {\bibinfo {author} {\bibfnamefont {T.~E.}\ \bibnamefont
  {Angelini}}, \bibinfo {author} {\bibfnamefont {E.}~\bibnamefont {Hannezo}},
  \bibinfo {author} {\bibfnamefont {X.}~\bibnamefont {Trepat}}, \bibinfo
  {author} {\bibfnamefont {M.}~\bibnamefont {Marquez}}, \bibinfo {author}
  {\bibfnamefont {J.~J.}\ \bibnamefont {Fredberg}},\ and\ \bibinfo {author}
  {\bibfnamefont {D.~A.}\ \bibnamefont {Weitz}},\ }\bibfield  {title} {\bibinfo
  {title} {Glass-like dynamics of collective cell migration},\ }\href@noop {}
  {\bibfield  {journal} {\bibinfo  {journal} {Proceedings of the National
  Academy of Sciences}\ }\textbf {\bibinfo {volume} {108}},\ \bibinfo {pages}
  {4714} (\bibinfo {year} {2011})}\BibitemShut {NoStop}%
\bibitem [{\citenamefont {Dieterich}\ \emph {et~al.}(2008)\citenamefont
  {Dieterich}, \citenamefont {Klages}, \citenamefont {Preuss},\ and\
  \citenamefont {Schwab}}]{dieterich2008anomalous}%
  \BibitemOpen
  \bibfield  {author} {\bibinfo {author} {\bibfnamefont {P.}~\bibnamefont
  {Dieterich}}, \bibinfo {author} {\bibfnamefont {R.}~\bibnamefont {Klages}},
  \bibinfo {author} {\bibfnamefont {R.}~\bibnamefont {Preuss}},\ and\ \bibinfo
  {author} {\bibfnamefont {A.}~\bibnamefont {Schwab}},\ }\bibfield  {title}
  {\bibinfo {title} {Anomalous dynamics of cell migration},\ }\href@noop {}
  {\bibfield  {journal} {\bibinfo  {journal} {Proceedings of the National
  Academy of Sciences}\ }\textbf {\bibinfo {volume} {105}},\ \bibinfo {pages}
  {459} (\bibinfo {year} {2008})}\BibitemShut {NoStop}%
\bibitem [{\citenamefont {Dieterich}\ \emph {et~al.}(2022)\citenamefont
  {Dieterich}, \citenamefont {Lindemann}, \citenamefont {Moskopp},
  \citenamefont {Tauzin}, \citenamefont {Huttenlocher}, \citenamefont {Klages},
  \citenamefont {Chechkin},\ and\ \citenamefont
  {Schwab}}]{dieterich2022anomalous}%
  \BibitemOpen
  \bibfield  {author} {\bibinfo {author} {\bibfnamefont {P.}~\bibnamefont
  {Dieterich}}, \bibinfo {author} {\bibfnamefont {O.}~\bibnamefont
  {Lindemann}}, \bibinfo {author} {\bibfnamefont {M.~L.}\ \bibnamefont
  {Moskopp}}, \bibinfo {author} {\bibfnamefont {S.}~\bibnamefont {Tauzin}},
  \bibinfo {author} {\bibfnamefont {A.}~\bibnamefont {Huttenlocher}}, \bibinfo
  {author} {\bibfnamefont {R.}~\bibnamefont {Klages}}, \bibinfo {author}
  {\bibfnamefont {A.}~\bibnamefont {Chechkin}},\ and\ \bibinfo {author}
  {\bibfnamefont {A.}~\bibnamefont {Schwab}},\ }\bibfield  {title} {\bibinfo
  {title} {Anomalous diffusion and asymmetric tempering memory in neutrophil
  chemotaxis},\ }\href@noop {} {\bibfield  {journal} {\bibinfo  {journal} {PLOS
  Computational Biology}\ }\textbf {\bibinfo {volume} {18}},\ \bibinfo {pages}
  {e1010089} (\bibinfo {year} {2022})}\BibitemShut {NoStop}%
\bibitem [{\citenamefont {Golding}\ and\ \citenamefont
  {Cox}(2006)}]{golding2006physical}%
  \BibitemOpen
  \bibfield  {author} {\bibinfo {author} {\bibfnamefont {I.}~\bibnamefont
  {Golding}}\ and\ \bibinfo {author} {\bibfnamefont {E.~C.}\ \bibnamefont
  {Cox}},\ }\bibfield  {title} {\bibinfo {title} {Physical nature of bacterial
  cytoplasm},\ }\href@noop {} {\bibfield  {journal} {\bibinfo  {journal}
  {Physical Review Letters}\ }\textbf {\bibinfo {volume} {96}},\ \bibinfo
  {pages} {098102} (\bibinfo {year} {2006})}\BibitemShut {NoStop}%
\bibitem [{\citenamefont {Hapca}\ \emph {et~al.}(2009)\citenamefont {Hapca},
  \citenamefont {Crawford},\ and\ \citenamefont {Young}}]{hapca2009anomalous}%
  \BibitemOpen
  \bibfield  {author} {\bibinfo {author} {\bibfnamefont {S.}~\bibnamefont
  {Hapca}}, \bibinfo {author} {\bibfnamefont {J.~W.}\ \bibnamefont
  {Crawford}},\ and\ \bibinfo {author} {\bibfnamefont {I.~M.}\ \bibnamefont
  {Young}},\ }\bibfield  {title} {\bibinfo {title} {Anomalous diffusion of
  heterogeneous populations characterized by normal diffusion at the individual
  level},\ }\href@noop {} {\bibfield  {journal} {\bibinfo  {journal} {Journal
  of the Royal Society Interface}\ }\textbf {\bibinfo {volume} {6}},\ \bibinfo
  {pages} {111} (\bibinfo {year} {2009})}\BibitemShut {NoStop}%
\bibitem [{\citenamefont {Lagarde}\ \emph {et~al.}(2020)\citenamefont
  {Lagarde}, \citenamefont {Dag{\`e}s}, \citenamefont {Nemoto}, \citenamefont
  {D{\'e}mery}, \citenamefont {Bartolo},\ and\ \citenamefont
  {Gibaud}}]{lagarde2020colloidal}%
  \BibitemOpen
  \bibfield  {author} {\bibinfo {author} {\bibfnamefont {A.}~\bibnamefont
  {Lagarde}}, \bibinfo {author} {\bibfnamefont {N.}~\bibnamefont {Dag{\`e}s}},
  \bibinfo {author} {\bibfnamefont {T.}~\bibnamefont {Nemoto}}, \bibinfo
  {author} {\bibfnamefont {V.}~\bibnamefont {D{\'e}mery}}, \bibinfo {author}
  {\bibfnamefont {D.}~\bibnamefont {Bartolo}},\ and\ \bibinfo {author}
  {\bibfnamefont {T.}~\bibnamefont {Gibaud}},\ }\bibfield  {title} {\bibinfo
  {title} {Colloidal transport in bacteria suspensions: From bacteria collision
  to anomalous and enhanced diffusion},\ }\href@noop {} {\bibfield  {journal}
  {\bibinfo  {journal} {Soft Matter}\ }\textbf {\bibinfo {volume} {16}},\
  \bibinfo {pages} {7503} (\bibinfo {year} {2020})}\BibitemShut {NoStop}%
\bibitem [{\citenamefont {Mukherjee}\ \emph {et~al.}(2021)\citenamefont
  {Mukherjee}, \citenamefont {Singh}, \citenamefont {James},\ and\
  \citenamefont {Ray}}]{mukherjee2021anomalous}%
  \BibitemOpen
  \bibfield  {author} {\bibinfo {author} {\bibfnamefont {S.}~\bibnamefont
  {Mukherjee}}, \bibinfo {author} {\bibfnamefont {R.~K.}\ \bibnamefont
  {Singh}}, \bibinfo {author} {\bibfnamefont {M.}~\bibnamefont {James}},\ and\
  \bibinfo {author} {\bibfnamefont {S.~S.}\ \bibnamefont {Ray}},\ }\bibfield
  {title} {\bibinfo {title} {Anomalous diffusion and {L}{\'e}vy walks
  distinguish active from inertial turbulence},\ }\href@noop {} {\bibfield
  {journal} {\bibinfo  {journal} {Physical Review Letters}\ }\textbf {\bibinfo
  {volume} {127}},\ \bibinfo {pages} {118001} (\bibinfo {year}
  {2021})}\BibitemShut {NoStop}%
\bibitem [{\citenamefont {Petrovskii}\ and\ \citenamefont
  {Morozov}(2009)}]{petrovskii2009dispersal}%
  \BibitemOpen
  \bibfield  {author} {\bibinfo {author} {\bibfnamefont {S.}~\bibnamefont
  {Petrovskii}}\ and\ \bibinfo {author} {\bibfnamefont {A.}~\bibnamefont
  {Morozov}},\ }\bibfield  {title} {\bibinfo {title} {Dispersal in a
  statistically structured population: {F}at tails revisited},\ }\href@noop {}
  {\bibfield  {journal} {\bibinfo  {journal} {The American Naturalist}\
  }\textbf {\bibinfo {volume} {173}},\ \bibinfo {pages} {278} (\bibinfo {year}
  {2009})}\BibitemShut {NoStop}%
\bibitem [{\citenamefont {Benhamou}(2007)}]{benhamou2007many}%
  \BibitemOpen
  \bibfield  {author} {\bibinfo {author} {\bibfnamefont {S.}~\bibnamefont
  {Benhamou}},\ }\bibfield  {title} {\bibinfo {title} {How many animals really
  do the {L}{\'e}vy walk?},\ }\href@noop {} {\bibfield  {journal} {\bibinfo
  {journal} {Ecology}\ }\textbf {\bibinfo {volume} {88}},\ \bibinfo {pages}
  {1962} (\bibinfo {year} {2007})}\BibitemShut {NoStop}%
\bibitem [{\citenamefont {James}\ \emph {et~al.}(2011)\citenamefont {James},
  \citenamefont {Plank},\ and\ \citenamefont {Edwards}}]{james2011assessing}%
  \BibitemOpen
  \bibfield  {author} {\bibinfo {author} {\bibfnamefont {A.}~\bibnamefont
  {James}}, \bibinfo {author} {\bibfnamefont {M.~J.}\ \bibnamefont {Plank}},\
  and\ \bibinfo {author} {\bibfnamefont {A.~M.}\ \bibnamefont {Edwards}},\
  }\bibfield  {title} {\bibinfo {title} {Assessing {L}{\'e}vy walks as models
  of animal foraging},\ }\href@noop {} {\bibfield  {journal} {\bibinfo
  {journal} {Journal of the Royal Society Interface}\ }\textbf {\bibinfo
  {volume} {8}},\ \bibinfo {pages} {1233} (\bibinfo {year} {2011})}\BibitemShut
  {NoStop}%
\bibitem [{\citenamefont {Reynolds}\ and\ \citenamefont
  {Rhodes}(2009)}]{reynolds2009levy}%
  \BibitemOpen
  \bibfield  {author} {\bibinfo {author} {\bibfnamefont {A.~M.}\ \bibnamefont
  {Reynolds}}\ and\ \bibinfo {author} {\bibfnamefont {C.~J.}\ \bibnamefont
  {Rhodes}},\ }\bibfield  {title} {\bibinfo {title} {The {L}{\'e}vy flight
  paradigm: {R}andom search patterns and mechanisms},\ }\href@noop {}
  {\bibfield  {journal} {\bibinfo  {journal} {Ecology}\ }\textbf {\bibinfo
  {volume} {90}},\ \bibinfo {pages} {877} (\bibinfo {year} {2009})}\BibitemShut
  {NoStop}%
\bibitem [{\citenamefont {Brown}\ \emph {et~al.}(2007)\citenamefont {Brown},
  \citenamefont {Liebovitch},\ and\ \citenamefont {Glendon}}]{brown2007levy}%
  \BibitemOpen
  \bibfield  {author} {\bibinfo {author} {\bibfnamefont {C.~T.}\ \bibnamefont
  {Brown}}, \bibinfo {author} {\bibfnamefont {L.~S.}\ \bibnamefont
  {Liebovitch}},\ and\ \bibinfo {author} {\bibfnamefont {R.}~\bibnamefont
  {Glendon}},\ }\bibfield  {title} {\bibinfo {title} {L{\'e}vy flights in
  {D}obe {J}u/’hoansi foraging patterns},\ }\href@noop {} {\bibfield
  {journal} {\bibinfo  {journal} {Human Ecology}\ }\textbf {\bibinfo {volume}
  {35}},\ \bibinfo {pages} {129} (\bibinfo {year} {2007})}\BibitemShut
  {NoStop}%
\bibitem [{\citenamefont {Raichlen}\ \emph {et~al.}(2014)\citenamefont
  {Raichlen}, \citenamefont {Wood}, \citenamefont {Gordon}, \citenamefont
  {Mabulla}, \citenamefont {Marlowe},\ and\ \citenamefont
  {Pontzer}}]{raichlen2014evidence}%
  \BibitemOpen
  \bibfield  {author} {\bibinfo {author} {\bibfnamefont {D.~A.}\ \bibnamefont
  {Raichlen}}, \bibinfo {author} {\bibfnamefont {B.~M.}\ \bibnamefont {Wood}},
  \bibinfo {author} {\bibfnamefont {A.~D.}\ \bibnamefont {Gordon}}, \bibinfo
  {author} {\bibfnamefont {A.~Z.}\ \bibnamefont {Mabulla}}, \bibinfo {author}
  {\bibfnamefont {F.~W.}\ \bibnamefont {Marlowe}},\ and\ \bibinfo {author}
  {\bibfnamefont {H.}~\bibnamefont {Pontzer}},\ }\bibfield  {title} {\bibinfo
  {title} {Evidence of l{\'e}vy walk foraging patterns in human
  hunter--gatherers},\ }\href@noop {} {\bibfield  {journal} {\bibinfo
  {journal} {Proceedings of the National Academy of Sciences}\ }\textbf
  {\bibinfo {volume} {111}},\ \bibinfo {pages} {728} (\bibinfo {year}
  {2014})}\BibitemShut {NoStop}%
\bibitem [{\citenamefont {Cherstvy}\ \emph
  {et~al.}(2021{\natexlab{a}})\citenamefont {Cherstvy}, \citenamefont {Vinod},
  \citenamefont {Aghion}, \citenamefont {Sokolov},\ and\ \citenamefont
  {Metzler}}]{cherstvy2021scaled}%
  \BibitemOpen
  \bibfield  {author} {\bibinfo {author} {\bibfnamefont {A.~G.}\ \bibnamefont
  {Cherstvy}}, \bibinfo {author} {\bibfnamefont {D.}~\bibnamefont {Vinod}},
  \bibinfo {author} {\bibfnamefont {E.}~\bibnamefont {Aghion}}, \bibinfo
  {author} {\bibfnamefont {I.~M.}\ \bibnamefont {Sokolov}},\ and\ \bibinfo
  {author} {\bibfnamefont {R.}~\bibnamefont {Metzler}},\ }\bibfield  {title}
  {\bibinfo {title} {Scaled geometric {B}rownian motion features sub-or
  superexponential ensemble-averaged, but linear time-averaged mean-squared
  displacements},\ }\href@noop {} {\bibfield  {journal} {\bibinfo  {journal}
  {Physical Review E}\ }\textbf {\bibinfo {volume} {103}},\ \bibinfo {pages}
  {062127} (\bibinfo {year} {2021}{\natexlab{a}})}\BibitemShut {NoStop}%
\bibitem [{\citenamefont {Plerou}\ \emph {et~al.}(2000)\citenamefont {Plerou},
  \citenamefont {Gopikrishnan}, \citenamefont {Amaral}, \citenamefont
  {Gabaix},\ and\ \citenamefont {Stanley}}]{plerou2000economic}%
  \BibitemOpen
  \bibfield  {author} {\bibinfo {author} {\bibfnamefont {V.}~\bibnamefont
  {Plerou}}, \bibinfo {author} {\bibfnamefont {P.}~\bibnamefont
  {Gopikrishnan}}, \bibinfo {author} {\bibfnamefont {L.~A.~N.}\ \bibnamefont
  {Amaral}}, \bibinfo {author} {\bibfnamefont {X.}~\bibnamefont {Gabaix}},\
  and\ \bibinfo {author} {\bibfnamefont {H.~E.}\ \bibnamefont {Stanley}},\
  }\bibfield  {title} {\bibinfo {title} {Economic fluctuations and anomalous
  diffusion},\ }\href@noop {} {\bibfield  {journal} {\bibinfo  {journal}
  {Physical Review E}\ }\textbf {\bibinfo {volume} {62}},\ \bibinfo {pages}
  {R3023} (\bibinfo {year} {2000})}\BibitemShut {NoStop}%
\bibitem [{\citenamefont {V{\'a}zquez}\ \emph {et~al.}(2006)\citenamefont
  {V{\'a}zquez}, \citenamefont {Oliveira}, \citenamefont {Dezs{\"o}},
  \citenamefont {Goh}, \citenamefont {Kondor},\ and\ \citenamefont
  {Barab{\'a}si}}]{vazquez2006modeling}%
  \BibitemOpen
  \bibfield  {author} {\bibinfo {author} {\bibfnamefont {A.}~\bibnamefont
  {V{\'a}zquez}}, \bibinfo {author} {\bibfnamefont {J.~G.}\ \bibnamefont
  {Oliveira}}, \bibinfo {author} {\bibfnamefont {Z.}~\bibnamefont {Dezs{\"o}}},
  \bibinfo {author} {\bibfnamefont {K.-I.}\ \bibnamefont {Goh}}, \bibinfo
  {author} {\bibfnamefont {I.}~\bibnamefont {Kondor}},\ and\ \bibinfo {author}
  {\bibfnamefont {A.-L.}\ \bibnamefont {Barab{\'a}si}},\ }\bibfield  {title}
  {\bibinfo {title} {Modeling bursts and heavy tails in human dynamics},\
  }\href@noop {} {\bibfield  {journal} {\bibinfo  {journal} {Physical Review
  E}\ }\textbf {\bibinfo {volume} {73}},\ \bibinfo {pages} {036127} (\bibinfo
  {year} {2006})}\BibitemShut {NoStop}%
\bibitem [{\citenamefont {Oliveira}\ \emph {et~al.}(2019)\citenamefont
  {Oliveira}, \citenamefont {Ferreira}, \citenamefont {Lapas},\ and\
  \citenamefont {Vainstein}}]{oliveira2019anomalous}%
  \BibitemOpen
  \bibfield  {author} {\bibinfo {author} {\bibfnamefont {F.~A.}\ \bibnamefont
  {Oliveira}}, \bibinfo {author} {\bibfnamefont {R.~M.}\ \bibnamefont
  {Ferreira}}, \bibinfo {author} {\bibfnamefont {L.~C.}\ \bibnamefont
  {Lapas}},\ and\ \bibinfo {author} {\bibfnamefont {M.~H.}\ \bibnamefont
  {Vainstein}},\ }\bibfield  {title} {\bibinfo {title} {Anomalous diffusion: A
  basic mechanism for the evolution of inhomogeneous systems},\ }\href@noop {}
  {\bibfield  {journal} {\bibinfo  {journal} {Frontiers in Physics}\ }\textbf
  {\bibinfo {volume} {7}},\ \bibinfo {pages} {18} (\bibinfo {year}
  {2019})}\BibitemShut {NoStop}%
\bibitem [{\citenamefont {Sokolov}\ and\ \citenamefont
  {Klafter}(2005)}]{sokolov2005diffusion}%
  \BibitemOpen
  \bibfield  {author} {\bibinfo {author} {\bibfnamefont {I.~M.}\ \bibnamefont
  {Sokolov}}\ and\ \bibinfo {author} {\bibfnamefont {J.}~\bibnamefont
  {Klafter}},\ }\bibfield  {title} {\bibinfo {title} {From diffusion to
  anomalous diffusion: {A} century after {E}instein’s {B}rownian motion},\
  }\href@noop {} {\bibfield  {journal} {\bibinfo  {journal} {Chaos}\ }\textbf
  {\bibinfo {volume} {15}},\ \bibinfo {pages} {026103} (\bibinfo {year}
  {2005})}\BibitemShut {NoStop}%
\bibitem [{\citenamefont {Timashev}\ \emph {et~al.}(2010)\citenamefont
  {Timashev}, \citenamefont {Polyakov}, \citenamefont {Misurkin},\ and\
  \citenamefont {Lakeev}}]{timashev2010anomalous}%
  \BibitemOpen
  \bibfield  {author} {\bibinfo {author} {\bibfnamefont {S.~F.}\ \bibnamefont
  {Timashev}}, \bibinfo {author} {\bibfnamefont {Y.~S.}\ \bibnamefont
  {Polyakov}}, \bibinfo {author} {\bibfnamefont {P.~I.}\ \bibnamefont
  {Misurkin}},\ and\ \bibinfo {author} {\bibfnamefont {S.~G.}\ \bibnamefont
  {Lakeev}},\ }\bibfield  {title} {\bibinfo {title} {Anomalous diffusion as a
  stochastic component in the dynamics of complex processes},\ }\href@noop {}
  {\bibfield  {journal} {\bibinfo  {journal} {Physical Review E}\ }\textbf
  {\bibinfo {volume} {81}},\ \bibinfo {pages} {041128} (\bibinfo {year}
  {2010})}\BibitemShut {NoStop}%
\bibitem [{\citenamefont {Vilk}\ \emph {et~al.}(2022)\citenamefont {Vilk},
  \citenamefont {Aghion}, \citenamefont {Avgar}, \citenamefont {Beta},
  \citenamefont {Nagel}, \citenamefont {Sabri}, \citenamefont {Sarfati},
  \citenamefont {Schwartz}, \citenamefont {Weiss}, \citenamefont {Krapf} \emph
  {et~al.}}]{vilk2022unravelling}%
  \BibitemOpen
  \bibfield  {author} {\bibinfo {author} {\bibfnamefont {O.}~\bibnamefont
  {Vilk}}, \bibinfo {author} {\bibfnamefont {E.}~\bibnamefont {Aghion}},
  \bibinfo {author} {\bibfnamefont {T.}~\bibnamefont {Avgar}}, \bibinfo
  {author} {\bibfnamefont {C.}~\bibnamefont {Beta}}, \bibinfo {author}
  {\bibfnamefont {O.}~\bibnamefont {Nagel}}, \bibinfo {author} {\bibfnamefont
  {A.}~\bibnamefont {Sabri}}, \bibinfo {author} {\bibfnamefont
  {R.}~\bibnamefont {Sarfati}}, \bibinfo {author} {\bibfnamefont {D.~K.}\
  \bibnamefont {Schwartz}}, \bibinfo {author} {\bibfnamefont {M.}~\bibnamefont
  {Weiss}}, \bibinfo {author} {\bibfnamefont {D.}~\bibnamefont {Krapf}}, \emph
  {et~al.},\ }\bibfield  {title} {\bibinfo {title} {Unravelling the origins of
  anomalous diffusion: {F}rom molecules to migrating storks},\ }\href@noop {}
  {\bibfield  {journal} {\bibinfo  {journal} {Physical Review Research}\
  }\textbf {\bibinfo {volume} {4}},\ \bibinfo {pages} {033055} (\bibinfo {year}
  {2022})}\BibitemShut {NoStop}%
\bibitem [{\citenamefont {Weigel}\ \emph {et~al.}(2011)\citenamefont {Weigel},
  \citenamefont {Simon}, \citenamefont {Tamkun},\ and\ \citenamefont
  {Krapf}}]{weigel2011ergodic}%
  \BibitemOpen
  \bibfield  {author} {\bibinfo {author} {\bibfnamefont {A.~V.}\ \bibnamefont
  {Weigel}}, \bibinfo {author} {\bibfnamefont {B.}~\bibnamefont {Simon}},
  \bibinfo {author} {\bibfnamefont {M.~M.}\ \bibnamefont {Tamkun}},\ and\
  \bibinfo {author} {\bibfnamefont {D.}~\bibnamefont {Krapf}},\ }\bibfield
  {title} {\bibinfo {title} {Ergodic and nonergodic processes coexist in the
  plasma membrane as observed by single-molecule tracking},\ }\href@noop {}
  {\bibfield  {journal} {\bibinfo  {journal} {Proceedings of the National
  Academy of Sciences}\ }\textbf {\bibinfo {volume} {108}},\ \bibinfo {pages}
  {6438} (\bibinfo {year} {2011})}\BibitemShut {NoStop}%
\bibitem [{\citenamefont {Bancaud}\ \emph {et~al.}(2009)\citenamefont
  {Bancaud}, \citenamefont {Huet}, \citenamefont {Daigle}, \citenamefont
  {Mozziconacci}, \citenamefont {Beaudouin},\ and\ \citenamefont
  {Ellenberg}}]{bancaud2009molecular}%
  \BibitemOpen
  \bibfield  {author} {\bibinfo {author} {\bibfnamefont {A.}~\bibnamefont
  {Bancaud}}, \bibinfo {author} {\bibfnamefont {S.}~\bibnamefont {Huet}},
  \bibinfo {author} {\bibfnamefont {N.}~\bibnamefont {Daigle}}, \bibinfo
  {author} {\bibfnamefont {J.}~\bibnamefont {Mozziconacci}}, \bibinfo {author}
  {\bibfnamefont {J.}~\bibnamefont {Beaudouin}},\ and\ \bibinfo {author}
  {\bibfnamefont {J.}~\bibnamefont {Ellenberg}},\ }\bibfield  {title} {\bibinfo
  {title} {Molecular crowding affects diffusion and binding of nuclear proteins
  in heterochromatin and reveals the fractal organization of chromatin},\
  }\href@noop {} {\bibfield  {journal} {\bibinfo  {journal} {EMBO Journal}\
  }\textbf {\bibinfo {volume} {28}},\ \bibinfo {pages} {3785} (\bibinfo {year}
  {2009})}\BibitemShut {NoStop}%
\bibitem [{\citenamefont {Bouchaud}\ and\ \citenamefont
  {Georges}(1990)}]{bouchaud1990anomalous}%
  \BibitemOpen
  \bibfield  {author} {\bibinfo {author} {\bibfnamefont {J.-P.}\ \bibnamefont
  {Bouchaud}}\ and\ \bibinfo {author} {\bibfnamefont {A.}~\bibnamefont
  {Georges}},\ }\bibfield  {title} {\bibinfo {title} {Anomalous diffusion in
  disordered media: {S}tatistical mechanisms, models and physical
  applications},\ }\href@noop {} {\bibfield  {journal} {\bibinfo  {journal}
  {Physics Reports}\ }\textbf {\bibinfo {volume} {195}},\ \bibinfo {pages}
  {127} (\bibinfo {year} {1990})}\BibitemShut {NoStop}%
\bibitem [{\citenamefont {Bronstein}\ \emph {et~al.}(2009)\citenamefont
  {Bronstein}, \citenamefont {Israel}, \citenamefont {Kepten}, \citenamefont
  {Mai}, \citenamefont {Shav-Tal}, \citenamefont {Barkai},\ and\ \citenamefont
  {Garini}}]{bronstein2009transient}%
  \BibitemOpen
  \bibfield  {author} {\bibinfo {author} {\bibfnamefont {I.}~\bibnamefont
  {Bronstein}}, \bibinfo {author} {\bibfnamefont {Y.}~\bibnamefont {Israel}},
  \bibinfo {author} {\bibfnamefont {E.}~\bibnamefont {Kepten}}, \bibinfo
  {author} {\bibfnamefont {S.}~\bibnamefont {Mai}}, \bibinfo {author}
  {\bibfnamefont {Y.}~\bibnamefont {Shav-Tal}}, \bibinfo {author}
  {\bibfnamefont {E.}~\bibnamefont {Barkai}},\ and\ \bibinfo {author}
  {\bibfnamefont {Y.}~\bibnamefont {Garini}},\ }\bibfield  {title} {\bibinfo
  {title} {Transient anomalous diffusion of telomeres in the nucleus of
  mammalian cells},\ }\href@noop {} {\bibfield  {journal} {\bibinfo  {journal}
  {Physical Review Letters}\ }\textbf {\bibinfo {volume} {103}},\ \bibinfo
  {pages} {018102} (\bibinfo {year} {2009})}\BibitemShut {NoStop}%
\bibitem [{\citenamefont {Caspi}\ \emph {et~al.}(2000)\citenamefont {Caspi},
  \citenamefont {Granek},\ and\ \citenamefont {Elbaum}}]{caspi2000enhanced}%
  \BibitemOpen
  \bibfield  {author} {\bibinfo {author} {\bibfnamefont {A.}~\bibnamefont
  {Caspi}}, \bibinfo {author} {\bibfnamefont {R.}~\bibnamefont {Granek}},\ and\
  \bibinfo {author} {\bibfnamefont {M.}~\bibnamefont {Elbaum}},\ }\bibfield
  {title} {\bibinfo {title} {Enhanced diffusion in active intracellular
  transport},\ }\href@noop {} {\bibfield  {journal} {\bibinfo  {journal}
  {Physical Review Letters}\ }\textbf {\bibinfo {volume} {85}},\ \bibinfo
  {pages} {5655} (\bibinfo {year} {2000})}\BibitemShut {NoStop}%
\bibitem [{\citenamefont {Fern{\'a}ndez}\ \emph {et~al.}(2020)\citenamefont
  {Fern{\'a}ndez}, \citenamefont {Charchar}, \citenamefont {Cherstvy},
  \citenamefont {Metzler},\ and\ \citenamefont
  {Finnis}}]{fernandez2020diffusion}%
  \BibitemOpen
  \bibfield  {author} {\bibinfo {author} {\bibfnamefont {A.~D.}\ \bibnamefont
  {Fern{\'a}ndez}}, \bibinfo {author} {\bibfnamefont {P.}~\bibnamefont
  {Charchar}}, \bibinfo {author} {\bibfnamefont {A.~G.}\ \bibnamefont
  {Cherstvy}}, \bibinfo {author} {\bibfnamefont {R.}~\bibnamefont {Metzler}},\
  and\ \bibinfo {author} {\bibfnamefont {M.~W.}\ \bibnamefont {Finnis}},\
  }\bibfield  {title} {\bibinfo {title} {The diffusion of doxorubicin drug
  molecules in silica nanoslits is non-{G}aussian, intermittent and
  anticorrelated},\ }\href@noop {} {\bibfield  {journal} {\bibinfo  {journal}
  {Physical Chemistry Chemical Physics}\ }\textbf {\bibinfo {volume} {22}},\
  \bibinfo {pages} {27955} (\bibinfo {year} {2020})}\BibitemShut {NoStop}%
\bibitem [{\citenamefont {H{\"o}fling}\ \emph {et~al.}(2011)\citenamefont
  {H{\"o}fling}, \citenamefont {Bamberg},\ and\ \citenamefont
  {Franosch}}]{hofling2011anomalous}%
  \BibitemOpen
  \bibfield  {author} {\bibinfo {author} {\bibfnamefont {F.}~\bibnamefont
  {H{\"o}fling}}, \bibinfo {author} {\bibfnamefont {K.-U.}\ \bibnamefont
  {Bamberg}},\ and\ \bibinfo {author} {\bibfnamefont {T.}~\bibnamefont
  {Franosch}},\ }\bibfield  {title} {\bibinfo {title} {Anomalous transport
  resolved in space and time by fluorescence correlation spectroscopy},\
  }\href@noop {} {\bibfield  {journal} {\bibinfo  {journal} {Soft Matter}\
  }\textbf {\bibinfo {volume} {7}},\ \bibinfo {pages} {1358} (\bibinfo {year}
  {2011})}\BibitemShut {NoStop}%
\bibitem [{\citenamefont {Jeon}\ \emph {et~al.}(2016)\citenamefont {Jeon},
  \citenamefont {Javanainen}, \citenamefont {Martinez-Seara}, \citenamefont
  {Metzler},\ and\ \citenamefont {Vattulainen}}]{jeon2016protein}%
  \BibitemOpen
  \bibfield  {author} {\bibinfo {author} {\bibfnamefont {J.-H.}\ \bibnamefont
  {Jeon}}, \bibinfo {author} {\bibfnamefont {M.}~\bibnamefont {Javanainen}},
  \bibinfo {author} {\bibfnamefont {H.}~\bibnamefont {Martinez-Seara}},
  \bibinfo {author} {\bibfnamefont {R.}~\bibnamefont {Metzler}},\ and\ \bibinfo
  {author} {\bibfnamefont {I.}~\bibnamefont {Vattulainen}},\ }\bibfield
  {title} {\bibinfo {title} {Protein crowding in lipid bilayers gives rise to
  non-{G}aussian anomalous lateral diffusion of phospholipids and proteins},\
  }\href@noop {} {\bibfield  {journal} {\bibinfo  {journal} {Physical Review
  X}\ }\textbf {\bibinfo {volume} {6}},\ \bibinfo {pages} {021006} (\bibinfo
  {year} {2016})}\BibitemShut {NoStop}%
\bibitem [{\citenamefont {Saxton}(2001)}]{saxton2001anomalous}%
  \BibitemOpen
  \bibfield  {author} {\bibinfo {author} {\bibfnamefont {M.~J.}\ \bibnamefont
  {Saxton}},\ }\bibfield  {title} {\bibinfo {title} {Anomalous subdiffusion in
  fluorescence photobleaching recovery: {A} monte carlo study},\ }\href@noop {}
  {\bibfield  {journal} {\bibinfo  {journal} {Biophysical Journal}\ }\textbf
  {\bibinfo {volume} {81}},\ \bibinfo {pages} {2226} (\bibinfo {year}
  {2001})}\BibitemShut {NoStop}%
\bibitem [{\citenamefont {Seisenberger}\ \emph {et~al.}(2001)\citenamefont
  {Seisenberger}, \citenamefont {Ried}, \citenamefont {Endress}, \citenamefont
  {Buning}, \citenamefont {Hallek},\ and\ \citenamefont
  {Brauchle}}]{seisenberger2001real}%
  \BibitemOpen
  \bibfield  {author} {\bibinfo {author} {\bibfnamefont {G.}~\bibnamefont
  {Seisenberger}}, \bibinfo {author} {\bibfnamefont {M.~U.}\ \bibnamefont
  {Ried}}, \bibinfo {author} {\bibfnamefont {T.}~\bibnamefont {Endress}},
  \bibinfo {author} {\bibfnamefont {H.}~\bibnamefont {Buning}}, \bibinfo
  {author} {\bibfnamefont {M.}~\bibnamefont {Hallek}},\ and\ \bibinfo {author}
  {\bibfnamefont {C.}~\bibnamefont {Brauchle}},\ }\bibfield  {title} {\bibinfo
  {title} {Real-time single-molecule imaging of the infection pathway of an
  adeno-associated virus},\ }\href@noop {} {\bibfield  {journal} {\bibinfo
  {journal} {Science}\ }\textbf {\bibinfo {volume} {294}},\ \bibinfo {pages}
  {1929} (\bibinfo {year} {2001})}\BibitemShut {NoStop}%
\bibitem [{\citenamefont {Smith}\ \emph {et~al.}(1999)\citenamefont {Smith},
  \citenamefont {Morrison}, \citenamefont {Wilson}, \citenamefont {Fernandez},\
  and\ \citenamefont {Cherry}}]{smith1999anomalous}%
  \BibitemOpen
  \bibfield  {author} {\bibinfo {author} {\bibfnamefont {P.~R.}\ \bibnamefont
  {Smith}}, \bibinfo {author} {\bibfnamefont {I.~E.}\ \bibnamefont {Morrison}},
  \bibinfo {author} {\bibfnamefont {K.~M.}\ \bibnamefont {Wilson}}, \bibinfo
  {author} {\bibfnamefont {N.}~\bibnamefont {Fernandez}},\ and\ \bibinfo
  {author} {\bibfnamefont {R.~J.}\ \bibnamefont {Cherry}},\ }\bibfield  {title}
  {\bibinfo {title} {Anomalous diffusion of major histocompatibility complex
  class {I} molecules on {H}e{L}a cells determined by single particle
  tracking},\ }\href@noop {} {\bibfield  {journal} {\bibinfo  {journal}
  {Biophysical Journal}\ }\textbf {\bibinfo {volume} {76}},\ \bibinfo {pages}
  {3331} (\bibinfo {year} {1999})}\BibitemShut {NoStop}%
\bibitem [{\citenamefont {Weber}\ \emph {et~al.}(2010)\citenamefont {Weber},
  \citenamefont {Spakowitz},\ and\ \citenamefont
  {Theriot}}]{weber2010bacterial}%
  \BibitemOpen
  \bibfield  {author} {\bibinfo {author} {\bibfnamefont {S.~C.}\ \bibnamefont
  {Weber}}, \bibinfo {author} {\bibfnamefont {A.~J.}\ \bibnamefont
  {Spakowitz}},\ and\ \bibinfo {author} {\bibfnamefont {J.~A.}\ \bibnamefont
  {Theriot}},\ }\bibfield  {title} {\bibinfo {title} {Bacterial chromosomal
  loci move subdiffusively through a viscoelastic cytoplasm},\ }\href@noop {}
  {\bibfield  {journal} {\bibinfo  {journal} {Physical Review Letters}\
  }\textbf {\bibinfo {volume} {104}},\ \bibinfo {pages} {238102} (\bibinfo
  {year} {2010})}\BibitemShut {NoStop}%
\bibitem [{\citenamefont {Weiss}(2013)}]{weiss2013single}%
  \BibitemOpen
  \bibfield  {author} {\bibinfo {author} {\bibfnamefont {M.}~\bibnamefont
  {Weiss}},\ }\bibfield  {title} {\bibinfo {title} {Single-particle tracking
  data reveal anticorrelated fractional {B}rownian motion in crowded fluids},\
  }\href@noop {} {\bibfield  {journal} {\bibinfo  {journal} {Physical Review
  E}\ }\textbf {\bibinfo {volume} {88}},\ \bibinfo {pages} {010101} (\bibinfo
  {year} {2013})}\BibitemShut {NoStop}%
\bibitem [{\citenamefont {Arcizet}\ \emph {et~al.}(2008)\citenamefont
  {Arcizet}, \citenamefont {Meier}, \citenamefont {Sackmann}, \citenamefont
  {R{\"a}dler},\ and\ \citenamefont {Heinrich}}]{arcizet2008temporal}%
  \BibitemOpen
  \bibfield  {author} {\bibinfo {author} {\bibfnamefont {D.}~\bibnamefont
  {Arcizet}}, \bibinfo {author} {\bibfnamefont {B.}~\bibnamefont {Meier}},
  \bibinfo {author} {\bibfnamefont {E.}~\bibnamefont {Sackmann}}, \bibinfo
  {author} {\bibfnamefont {J.~O.}\ \bibnamefont {R{\"a}dler}},\ and\ \bibinfo
  {author} {\bibfnamefont {D.}~\bibnamefont {Heinrich}},\ }\bibfield  {title}
  {\bibinfo {title} {Temporal analysis of active and passive transport in
  living cells},\ }\href@noop {} {\bibfield  {journal} {\bibinfo  {journal}
  {Physical Review Letters}\ }\textbf {\bibinfo {volume} {101}},\ \bibinfo
  {pages} {248103} (\bibinfo {year} {2008})}\BibitemShut {NoStop}%
\bibitem [{\citenamefont {Caspi}\ \emph {et~al.}(2002)\citenamefont {Caspi},
  \citenamefont {Granek},\ and\ \citenamefont {Elbaum}}]{caspi2002diffusion}%
  \BibitemOpen
  \bibfield  {author} {\bibinfo {author} {\bibfnamefont {A.}~\bibnamefont
  {Caspi}}, \bibinfo {author} {\bibfnamefont {R.}~\bibnamefont {Granek}},\ and\
  \bibinfo {author} {\bibfnamefont {M.}~\bibnamefont {Elbaum}},\ }\bibfield
  {title} {\bibinfo {title} {Diffusion and directed motion in cellular
  transport},\ }\href@noop {} {\bibfield  {journal} {\bibinfo  {journal}
  {Physical Review E}\ }\textbf {\bibinfo {volume} {66}},\ \bibinfo {pages}
  {011916} (\bibinfo {year} {2002})}\BibitemShut {NoStop}%
\bibitem [{\citenamefont {de~Jager}\ \emph {et~al.}(2011)\citenamefont
  {de~Jager}, \citenamefont {Weissing}, \citenamefont {Herman}, \citenamefont
  {Nolet},\ and\ \citenamefont {van~de Koppel}}]{de2011levy}%
  \BibitemOpen
  \bibfield  {author} {\bibinfo {author} {\bibfnamefont {M.}~\bibnamefont
  {de~Jager}}, \bibinfo {author} {\bibfnamefont {F.~J.}\ \bibnamefont
  {Weissing}}, \bibinfo {author} {\bibfnamefont {P.~M.}\ \bibnamefont
  {Herman}}, \bibinfo {author} {\bibfnamefont {B.~A.}\ \bibnamefont {Nolet}},\
  and\ \bibinfo {author} {\bibfnamefont {J.}~\bibnamefont {van~de Koppel}},\
  }\bibfield  {title} {\bibinfo {title} {L{\'e}vy walks evolve through
  interaction between movement and environmental complexity},\ }\href@noop {}
  {\bibfield  {journal} {\bibinfo  {journal} {Science}\ }\textbf {\bibinfo
  {volume} {332}},\ \bibinfo {pages} {1551} (\bibinfo {year}
  {2011})}\BibitemShut {NoStop}%
\bibitem [{\citenamefont {Duits}\ \emph {et~al.}(2009)\citenamefont {Duits},
  \citenamefont {Li}, \citenamefont {Vanapalli},\ and\ \citenamefont
  {Mugele}}]{duits2009mapping}%
  \BibitemOpen
  \bibfield  {author} {\bibinfo {author} {\bibfnamefont {M.~H.}\ \bibnamefont
  {Duits}}, \bibinfo {author} {\bibfnamefont {Y.}~\bibnamefont {Li}}, \bibinfo
  {author} {\bibfnamefont {S.~A.}\ \bibnamefont {Vanapalli}},\ and\ \bibinfo
  {author} {\bibfnamefont {F.}~\bibnamefont {Mugele}},\ }\bibfield  {title}
  {\bibinfo {title} {Mapping of spatiotemporal heterogeneous particle dynamics
  in living cells},\ }\href@noop {} {\bibfield  {journal} {\bibinfo  {journal}
  {Physical Review E}\ }\textbf {\bibinfo {volume} {79}},\ \bibinfo {pages}
  {051910} (\bibinfo {year} {2009})}\BibitemShut {NoStop}%
\bibitem [{\citenamefont {Gonzalez}\ \emph {et~al.}(2008)\citenamefont
  {Gonzalez}, \citenamefont {Hidalgo},\ and\ \citenamefont
  {Barabasi}}]{gonzalez2008understanding}%
  \BibitemOpen
  \bibfield  {author} {\bibinfo {author} {\bibfnamefont {M.~C.}\ \bibnamefont
  {Gonzalez}}, \bibinfo {author} {\bibfnamefont {C.~A.}\ \bibnamefont
  {Hidalgo}},\ and\ \bibinfo {author} {\bibfnamefont {A.-L.}\ \bibnamefont
  {Barabasi}},\ }\bibfield  {title} {\bibinfo {title} {Understanding individual
  human mobility patterns},\ }\href@noop {} {\bibfield  {journal} {\bibinfo
  {journal} {Nature}\ }\textbf {\bibinfo {volume} {453}},\ \bibinfo {pages}
  {779} (\bibinfo {year} {2008})}\BibitemShut {NoStop}%
\bibitem [{\citenamefont {Leoni}\ and\ \citenamefont
  {Franzese}(2014)}]{leoni2014structural}%
  \BibitemOpen
  \bibfield  {author} {\bibinfo {author} {\bibfnamefont {F.}~\bibnamefont
  {Leoni}}\ and\ \bibinfo {author} {\bibfnamefont {G.}~\bibnamefont
  {Franzese}},\ }\bibfield  {title} {\bibinfo {title} {Structural behavior and
  dynamics of an anomalous fluid between attractive and repulsive walls:
  {T}emplating, molding, and superdiffusion},\ }\href@noop {} {\bibfield
  {journal} {\bibinfo  {journal} {Journal of Chemical Physics}\ }\textbf
  {\bibinfo {volume} {141}},\ \bibinfo {pages} {174501} (\bibinfo {year}
  {2014})}\BibitemShut {NoStop}%
\bibitem [{\citenamefont {Mashanova}\ \emph {et~al.}(2010)\citenamefont
  {Mashanova}, \citenamefont {Oliver},\ and\ \citenamefont
  {Jansen}}]{mashanova2010evidence}%
  \BibitemOpen
  \bibfield  {author} {\bibinfo {author} {\bibfnamefont {A.}~\bibnamefont
  {Mashanova}}, \bibinfo {author} {\bibfnamefont {T.~H.}\ \bibnamefont
  {Oliver}},\ and\ \bibinfo {author} {\bibfnamefont {V.~A.}\ \bibnamefont
  {Jansen}},\ }\bibfield  {title} {\bibinfo {title} {Evidence for intermittency
  and a truncated power law from highly resolved aphid movement data},\
  }\href@noop {} {\bibfield  {journal} {\bibinfo  {journal} {Journal of the
  Royal Society Interface}\ }\textbf {\bibinfo {volume} {7}},\ \bibinfo {pages}
  {199} (\bibinfo {year} {2010})}\BibitemShut {NoStop}%
\bibitem [{\citenamefont {Meyer}\ \emph {et~al.}(2023)\citenamefont {Meyer},
  \citenamefont {Cherstvy}, \citenamefont {Seckler}, \citenamefont {Hering},
  \citenamefont {Blaum}, \citenamefont {Jeltsch},\ and\ \citenamefont
  {Metzler}}]{meyer2023directedeness}%
  \BibitemOpen
  \bibfield  {author} {\bibinfo {author} {\bibfnamefont {P.~G.}\ \bibnamefont
  {Meyer}}, \bibinfo {author} {\bibfnamefont {A.~G.}\ \bibnamefont {Cherstvy}},
  \bibinfo {author} {\bibfnamefont {H.}~\bibnamefont {Seckler}}, \bibinfo
  {author} {\bibfnamefont {R.}~\bibnamefont {Hering}}, \bibinfo {author}
  {\bibfnamefont {N.}~\bibnamefont {Blaum}}, \bibinfo {author} {\bibfnamefont
  {F.}~\bibnamefont {Jeltsch}},\ and\ \bibinfo {author} {\bibfnamefont
  {R.}~\bibnamefont {Metzler}},\ }\bibfield  {title} {\bibinfo {title}
  {Directedeness, correlations, and daily cycles in springbok motion: {F}rom
  data via stochastic models to movement prediction},\ }\href@noop {}
  {\bibfield  {journal} {\bibinfo  {journal} {Physical Review Research}\
  }\textbf {\bibinfo {volume} {5}},\ \bibinfo {pages} {043129} (\bibinfo {year}
  {2023})}\BibitemShut {NoStop}%
\bibitem [{\citenamefont {Nathan}\ \emph {et~al.}(2008)\citenamefont {Nathan},
  \citenamefont {Getz}, \citenamefont {Revilla}, \citenamefont {Holyoak},
  \citenamefont {Kadmon}, \citenamefont {Saltz},\ and\ \citenamefont
  {Smouse}}]{nathan2008movement}%
  \BibitemOpen
  \bibfield  {author} {\bibinfo {author} {\bibfnamefont {R.}~\bibnamefont
  {Nathan}}, \bibinfo {author} {\bibfnamefont {W.~M.}\ \bibnamefont {Getz}},
  \bibinfo {author} {\bibfnamefont {E.}~\bibnamefont {Revilla}}, \bibinfo
  {author} {\bibfnamefont {M.}~\bibnamefont {Holyoak}}, \bibinfo {author}
  {\bibfnamefont {R.}~\bibnamefont {Kadmon}}, \bibinfo {author} {\bibfnamefont
  {D.}~\bibnamefont {Saltz}},\ and\ \bibinfo {author} {\bibfnamefont {P.~E.}\
  \bibnamefont {Smouse}},\ }\bibfield  {title} {\bibinfo {title} {A movement
  ecology paradigm for unifying organismal movement research},\ }\href@noop {}
  {\bibfield  {journal} {\bibinfo  {journal} {Proceedings of the National
  Academy of Sciences}\ }\textbf {\bibinfo {volume} {105}},\ \bibinfo {pages}
  {19052} (\bibinfo {year} {2008})}\BibitemShut {NoStop}%
\bibitem [{\citenamefont {Ott}\ \emph {et~al.}(1990)\citenamefont {Ott},
  \citenamefont {Bouchaud}, \citenamefont {Langevin},\ and\ \citenamefont
  {Urbach}}]{ott1990anomalous}%
  \BibitemOpen
  \bibfield  {author} {\bibinfo {author} {\bibfnamefont {A.}~\bibnamefont
  {Ott}}, \bibinfo {author} {\bibfnamefont {J.-P.}\ \bibnamefont {Bouchaud}},
  \bibinfo {author} {\bibfnamefont {D.}~\bibnamefont {Langevin}},\ and\
  \bibinfo {author} {\bibfnamefont {W.}~\bibnamefont {Urbach}},\ }\bibfield
  {title} {\bibinfo {title} {Anomalous diffusion in ``living polymers:'' {A}
  genuine {L}evy flight?},\ }\href@noop {} {\bibfield  {journal} {\bibinfo
  {journal} {Physical Review Letters}\ }\textbf {\bibinfo {volume} {65}},\
  \bibinfo {pages} {2201} (\bibinfo {year} {1990})}\BibitemShut {NoStop}%
\bibitem [{\citenamefont {Krog}\ and\ \citenamefont
  {Lomholt}(2017)}]{krog2017bayesian}%
  \BibitemOpen
  \bibfield  {author} {\bibinfo {author} {\bibfnamefont {J.}~\bibnamefont
  {Krog}}\ and\ \bibinfo {author} {\bibfnamefont {M.~A.}\ \bibnamefont
  {Lomholt}},\ }\bibfield  {title} {\bibinfo {title} {Bayesian inference with
  information content model check for {L}angevin equations},\ }\href@noop {}
  {\bibfield  {journal} {\bibinfo  {journal} {Physical Review E}\ }\textbf
  {\bibinfo {volume} {96}},\ \bibinfo {pages} {062106} (\bibinfo {year}
  {2017})}\BibitemShut {NoStop}%
\bibitem [{\citenamefont {Krog}\ \emph {et~al.}(2018)\citenamefont {Krog},
  \citenamefont {Jacobsen}, \citenamefont {Lund}, \citenamefont {W{\"u}stner},\
  and\ \citenamefont {Lomholt}}]{krog2018bayesian}%
  \BibitemOpen
  \bibfield  {author} {\bibinfo {author} {\bibfnamefont {J.}~\bibnamefont
  {Krog}}, \bibinfo {author} {\bibfnamefont {L.~H.}\ \bibnamefont {Jacobsen}},
  \bibinfo {author} {\bibfnamefont {F.~W.}\ \bibnamefont {Lund}}, \bibinfo
  {author} {\bibfnamefont {D.}~\bibnamefont {W{\"u}stner}},\ and\ \bibinfo
  {author} {\bibfnamefont {M.~A.}\ \bibnamefont {Lomholt}},\ }\bibfield
  {title} {\bibinfo {title} {Bayesian model selection with fractional
  {B}rownian motion},\ }\href@noop {} {\bibfield  {journal} {\bibinfo
  {journal} {Journal of Statistical Mechanics: Theory and Experiment}\ }\textbf
  {\bibinfo {volume} {2018}},\ \bibinfo {pages} {093501} (\bibinfo {year}
  {2018})}\BibitemShut {NoStop}%
\bibitem [{\citenamefont {Park}\ \emph {et~al.}(2021)\citenamefont {Park},
  \citenamefont {Thapa}, \citenamefont {Kim}, \citenamefont {Lomholt},\ and\
  \citenamefont {Jeon}}]{park2021bayesian}%
  \BibitemOpen
  \bibfield  {author} {\bibinfo {author} {\bibfnamefont {S.}~\bibnamefont
  {Park}}, \bibinfo {author} {\bibfnamefont {S.}~\bibnamefont {Thapa}},
  \bibinfo {author} {\bibfnamefont {Y.}~\bibnamefont {Kim}}, \bibinfo {author}
  {\bibfnamefont {M.~A.}\ \bibnamefont {Lomholt}},\ and\ \bibinfo {author}
  {\bibfnamefont {J.-H.}\ \bibnamefont {Jeon}},\ }\bibfield  {title} {\bibinfo
  {title} {Bayesian inference of {L}{\'e}vy walks via hidden markov models},\
  }\href@noop {} {\bibfield  {journal} {\bibinfo  {journal} {Journal of Physics
  A: Mathematical and Theoretical}\ }\textbf {\bibinfo {volume} {54}},\
  \bibinfo {pages} {484001} (\bibinfo {year} {2021})}\BibitemShut {NoStop}%
\bibitem [{\citenamefont {Thapa}\ \emph {et~al.}(2018)\citenamefont {Thapa},
  \citenamefont {Lomholt}, \citenamefont {Krog}, \citenamefont {Cherstvy},\
  and\ \citenamefont {Metzler}}]{thapa2018bayesian}%
  \BibitemOpen
  \bibfield  {author} {\bibinfo {author} {\bibfnamefont {S.}~\bibnamefont
  {Thapa}}, \bibinfo {author} {\bibfnamefont {M.~A.}\ \bibnamefont {Lomholt}},
  \bibinfo {author} {\bibfnamefont {J.}~\bibnamefont {Krog}}, \bibinfo {author}
  {\bibfnamefont {A.~G.}\ \bibnamefont {Cherstvy}},\ and\ \bibinfo {author}
  {\bibfnamefont {R.}~\bibnamefont {Metzler}},\ }\bibfield  {title} {\bibinfo
  {title} {Bayesian analysis of single-particle tracking data using the
  nested-sampling algorithm: {M}aximum-likelihood model selection applied to
  stochastic-diffusivity data},\ }\href@noop {} {\bibfield  {journal} {\bibinfo
   {journal} {Physical Chemistry Chemical Physics}\ }\textbf {\bibinfo {volume}
  {20}},\ \bibinfo {pages} {29018} (\bibinfo {year} {2018})}\BibitemShut
  {NoStop}%
\bibitem [{\citenamefont {Thapa}\ \emph {et~al.}(2022)\citenamefont {Thapa},
  \citenamefont {Park}, \citenamefont {Kim}, \citenamefont {Jeon},
  \citenamefont {Metzler},\ and\ \citenamefont {Lomholt}}]{thapa2022bayesian}%
  \BibitemOpen
  \bibfield  {author} {\bibinfo {author} {\bibfnamefont {S.}~\bibnamefont
  {Thapa}}, \bibinfo {author} {\bibfnamefont {S.}~\bibnamefont {Park}},
  \bibinfo {author} {\bibfnamefont {Y.}~\bibnamefont {Kim}}, \bibinfo {author}
  {\bibfnamefont {J.-H.}\ \bibnamefont {Jeon}}, \bibinfo {author}
  {\bibfnamefont {R.}~\bibnamefont {Metzler}},\ and\ \bibinfo {author}
  {\bibfnamefont {M.~A.}\ \bibnamefont {Lomholt}},\ }\bibfield  {title}
  {\bibinfo {title} {Bayesian inference of scaled versus fractional {B}rownian
  motion},\ }\href@noop {} {\bibfield  {journal} {\bibinfo  {journal} {Journal
  of Physics A: Mathematical and Theoretical}\ }\textbf {\bibinfo {volume}
  {55}},\ \bibinfo {pages} {194003} (\bibinfo {year} {2022})}\BibitemShut
  {NoStop}%
\bibitem [{\citenamefont {Bo}\ \emph {et~al.}(2019)\citenamefont {Bo},
  \citenamefont {Schmidt}, \citenamefont {Eichhorn},\ and\ \citenamefont
  {Volpe}}]{bo2019measurement}%
  \BibitemOpen
  \bibfield  {author} {\bibinfo {author} {\bibfnamefont {S.}~\bibnamefont
  {Bo}}, \bibinfo {author} {\bibfnamefont {F.}~\bibnamefont {Schmidt}},
  \bibinfo {author} {\bibfnamefont {R.}~\bibnamefont {Eichhorn}},\ and\
  \bibinfo {author} {\bibfnamefont {G.}~\bibnamefont {Volpe}},\ }\bibfield
  {title} {\bibinfo {title} {Measurement of anomalous diffusion using recurrent
  neural networks},\ }\href@noop {} {\bibfield  {journal} {\bibinfo  {journal}
  {Physical Review E}\ }\textbf {\bibinfo {volume} {100}},\ \bibinfo {pages}
  {010102} (\bibinfo {year} {2019})}\BibitemShut {NoStop}%
\bibitem [{\citenamefont {Cichos}\ \emph {et~al.}(2020)\citenamefont {Cichos},
  \citenamefont {Gustavsson}, \citenamefont {Mehlig},\ and\ \citenamefont
  {Volpe}}]{cichos2020machine}%
  \BibitemOpen
  \bibfield  {author} {\bibinfo {author} {\bibfnamefont {F.}~\bibnamefont
  {Cichos}}, \bibinfo {author} {\bibfnamefont {K.}~\bibnamefont {Gustavsson}},
  \bibinfo {author} {\bibfnamefont {B.}~\bibnamefont {Mehlig}},\ and\ \bibinfo
  {author} {\bibfnamefont {G.}~\bibnamefont {Volpe}},\ }\bibfield  {title}
  {\bibinfo {title} {Machine learning for active matter},\ }\href@noop {}
  {\bibfield  {journal} {\bibinfo  {journal} {Nature Machine Intelligence}\
  }\textbf {\bibinfo {volume} {2}},\ \bibinfo {pages} {94} (\bibinfo {year}
  {2020})}\BibitemShut {NoStop}%
\bibitem [{\citenamefont {Gajowczyk}\ and\ \citenamefont
  {Szwabi{\'n}ski}(2021)}]{gajowczyk2021detection}%
  \BibitemOpen
  \bibfield  {author} {\bibinfo {author} {\bibfnamefont {M.}~\bibnamefont
  {Gajowczyk}}\ and\ \bibinfo {author} {\bibfnamefont {J.}~\bibnamefont
  {Szwabi{\'n}ski}},\ }\bibfield  {title} {\bibinfo {title} {Detection of
  anomalous diffusion with deep residual networks},\ }\href@noop {} {\bibfield
  {journal} {\bibinfo  {journal} {Entropy}\ }\textbf {\bibinfo {volume} {23}},\
  \bibinfo {pages} {649} (\bibinfo {year} {2021})}\BibitemShut {NoStop}%
\bibitem [{\citenamefont {Gentili}\ and\ \citenamefont
  {Volpe}(2021)}]{gentili2021characterization}%
  \BibitemOpen
  \bibfield  {author} {\bibinfo {author} {\bibfnamefont {A.}~\bibnamefont
  {Gentili}}\ and\ \bibinfo {author} {\bibfnamefont {G.}~\bibnamefont
  {Volpe}},\ }\bibfield  {title} {\bibinfo {title} {Characterization of
  anomalous diffusion classical statistics powered by deep learning
  ({CONDOR})},\ }\href@noop {} {\bibfield  {journal} {\bibinfo  {journal}
  {Journal of Physics A: Mathematical and Theoretical}\ }\textbf {\bibinfo
  {volume} {54}},\ \bibinfo {pages} {314003} (\bibinfo {year}
  {2021})}\BibitemShut {NoStop}%
\bibitem [{\citenamefont {Granik}\ \emph {et~al.}(2019)\citenamefont {Granik},
  \citenamefont {Weiss}, \citenamefont {Nehme}, \citenamefont {Levin},
  \citenamefont {Chein}, \citenamefont {Perlson}, \citenamefont {Roichman},\
  and\ \citenamefont {Shechtman}}]{granik2019single}%
  \BibitemOpen
  \bibfield  {author} {\bibinfo {author} {\bibfnamefont {N.}~\bibnamefont
  {Granik}}, \bibinfo {author} {\bibfnamefont {L.~E.}\ \bibnamefont {Weiss}},
  \bibinfo {author} {\bibfnamefont {E.}~\bibnamefont {Nehme}}, \bibinfo
  {author} {\bibfnamefont {M.}~\bibnamefont {Levin}}, \bibinfo {author}
  {\bibfnamefont {M.}~\bibnamefont {Chein}}, \bibinfo {author} {\bibfnamefont
  {E.}~\bibnamefont {Perlson}}, \bibinfo {author} {\bibfnamefont
  {Y.}~\bibnamefont {Roichman}},\ and\ \bibinfo {author} {\bibfnamefont
  {Y.}~\bibnamefont {Shechtman}},\ }\bibfield  {title} {\bibinfo {title}
  {Single-particle diffusion characterization by deep learning},\ }\href@noop
  {} {\bibfield  {journal} {\bibinfo  {journal} {Biophysical Journal}\ }\textbf
  {\bibinfo {volume} {117}},\ \bibinfo {pages} {185} (\bibinfo {year}
  {2019})}\BibitemShut {NoStop}%
\bibitem [{\citenamefont {Janczura}\ \emph {et~al.}(2020)\citenamefont
  {Janczura}, \citenamefont {Kowalek}, \citenamefont {Loch-Olszewska},
  \citenamefont {Szwabi{\'n}ski},\ and\ \citenamefont
  {Weron}}]{janczura2020classification}%
  \BibitemOpen
  \bibfield  {author} {\bibinfo {author} {\bibfnamefont {J.}~\bibnamefont
  {Janczura}}, \bibinfo {author} {\bibfnamefont {P.}~\bibnamefont {Kowalek}},
  \bibinfo {author} {\bibfnamefont {H.}~\bibnamefont {Loch-Olszewska}},
  \bibinfo {author} {\bibfnamefont {J.}~\bibnamefont {Szwabi{\'n}ski}},\ and\
  \bibinfo {author} {\bibfnamefont {A.}~\bibnamefont {Weron}},\ }\bibfield
  {title} {\bibinfo {title} {Classification of particle trajectories in living
  cells: {M}achine learning versus statistical testing hypothesis for
  fractional anomalous diffusion},\ }\href@noop {} {\bibfield  {journal}
  {\bibinfo  {journal} {Physical Review E}\ }\textbf {\bibinfo {volume}
  {102}},\ \bibinfo {pages} {032402} (\bibinfo {year} {2020})}\BibitemShut
  {NoStop}%
\bibitem [{\citenamefont {Kowalek}\ \emph {et~al.}(2022)\citenamefont
  {Kowalek}, \citenamefont {Loch-Olszewska}, \citenamefont {{\L}aszczuk},
  \citenamefont {Opa{\l}a},\ and\ \citenamefont
  {Szwabi{\'n}ski}}]{kowalek2022boosting}%
  \BibitemOpen
  \bibfield  {author} {\bibinfo {author} {\bibfnamefont {P.}~\bibnamefont
  {Kowalek}}, \bibinfo {author} {\bibfnamefont {H.}~\bibnamefont
  {Loch-Olszewska}}, \bibinfo {author} {\bibfnamefont {{\L}.}~\bibnamefont
  {{\L}aszczuk}}, \bibinfo {author} {\bibfnamefont {J.}~\bibnamefont
  {Opa{\l}a}},\ and\ \bibinfo {author} {\bibfnamefont {J.}~\bibnamefont
  {Szwabi{\'n}ski}},\ }\bibfield  {title} {\bibinfo {title} {Boosting the
  performance of anomalous diffusion classifiers with the proper choice of
  features},\ }\href@noop {} {\bibfield  {journal} {\bibinfo  {journal}
  {Journal of Physics A: Mathematical and Theoretical}\ }\textbf {\bibinfo
  {volume} {55}},\ \bibinfo {pages} {244005} (\bibinfo {year}
  {2022})}\BibitemShut {NoStop}%
\bibitem [{\citenamefont {Mu{\~n}oz-Gil}\ \emph {et~al.}(2020)\citenamefont
  {Mu{\~n}oz-Gil}, \citenamefont {Garcia-March}, \citenamefont {Manzo},
  \citenamefont {Mart{\'\i}n-Guerrero},\ and\ \citenamefont
  {Lewenstein}}]{munoz2020single}%
  \BibitemOpen
  \bibfield  {author} {\bibinfo {author} {\bibfnamefont {G.}~\bibnamefont
  {Mu{\~n}oz-Gil}}, \bibinfo {author} {\bibfnamefont {M.~A.}\ \bibnamefont
  {Garcia-March}}, \bibinfo {author} {\bibfnamefont {C.}~\bibnamefont {Manzo}},
  \bibinfo {author} {\bibfnamefont {J.~D.}\ \bibnamefont
  {Mart{\'\i}n-Guerrero}},\ and\ \bibinfo {author} {\bibfnamefont
  {M.}~\bibnamefont {Lewenstein}},\ }\bibfield  {title} {\bibinfo {title}
  {Single trajectory characterization via machine learning},\ }\href@noop {}
  {\bibfield  {journal} {\bibinfo  {journal} {New Journal of Physics}\ }\textbf
  {\bibinfo {volume} {22}},\ \bibinfo {pages} {013010} (\bibinfo {year}
  {2020})}\BibitemShut {NoStop}%
\bibitem [{\citenamefont {Mu{\~n}oz-Gil}\ \emph
  {et~al.}(2021{\natexlab{a}})\citenamefont {Mu{\~n}oz-Gil}, \citenamefont
  {Volpe}, \citenamefont {Garcia-March}, \citenamefont {Aghion}, \citenamefont
  {Argun}, \citenamefont {Hong}, \citenamefont {Bland}, \citenamefont {Bo},
  \citenamefont {Conejero}, \citenamefont {Firbas} \emph
  {et~al.}}]{munoz2021objective}%
  \BibitemOpen
  \bibfield  {author} {\bibinfo {author} {\bibfnamefont {G.}~\bibnamefont
  {Mu{\~n}oz-Gil}}, \bibinfo {author} {\bibfnamefont {G.}~\bibnamefont
  {Volpe}}, \bibinfo {author} {\bibfnamefont {M.~A.}\ \bibnamefont
  {Garcia-March}}, \bibinfo {author} {\bibfnamefont {E.}~\bibnamefont
  {Aghion}}, \bibinfo {author} {\bibfnamefont {A.}~\bibnamefont {Argun}},
  \bibinfo {author} {\bibfnamefont {C.~B.}\ \bibnamefont {Hong}}, \bibinfo
  {author} {\bibfnamefont {T.}~\bibnamefont {Bland}}, \bibinfo {author}
  {\bibfnamefont {S.}~\bibnamefont {Bo}}, \bibinfo {author} {\bibfnamefont
  {J.~A.}\ \bibnamefont {Conejero}}, \bibinfo {author} {\bibfnamefont
  {N.}~\bibnamefont {Firbas}}, \emph {et~al.},\ }\bibfield  {title} {\bibinfo
  {title} {Objective comparison of methods to decode anomalous diffusion},\
  }\href@noop {} {\bibfield  {journal} {\bibinfo  {journal} {Nature
  Communications}\ }\textbf {\bibinfo {volume} {12}},\ \bibinfo {pages} {1}
  (\bibinfo {year} {2021}{\natexlab{a}})}\BibitemShut {NoStop}%
\bibitem [{\citenamefont {Seckler}\ and\ \citenamefont
  {Metzler}(2022)}]{seckler2022bayesian}%
  \BibitemOpen
  \bibfield  {author} {\bibinfo {author} {\bibfnamefont {H.}~\bibnamefont
  {Seckler}}\ and\ \bibinfo {author} {\bibfnamefont {R.}~\bibnamefont
  {Metzler}},\ }\bibfield  {title} {\bibinfo {title} {Bayesian deep learning
  for error estimation in the analysis of anomalous diffusion},\ }\href@noop {}
  {\bibfield  {journal} {\bibinfo  {journal} {Nature Communications}\ }\textbf
  {\bibinfo {volume} {13}},\ \bibinfo {pages} {6717} (\bibinfo {year}
  {2022})}\BibitemShut {NoStop}%
\bibitem [{\citenamefont {Seckler}\ \emph {et~al.}(2023)\citenamefont
  {Seckler}, \citenamefont {Szwabi{\'n}ski},\ and\ \citenamefont
  {Metzler}}]{seckler2023machine}%
  \BibitemOpen
  \bibfield  {author} {\bibinfo {author} {\bibfnamefont {H.}~\bibnamefont
  {Seckler}}, \bibinfo {author} {\bibfnamefont {J.}~\bibnamefont
  {Szwabi{\'n}ski}},\ and\ \bibinfo {author} {\bibfnamefont {R.}~\bibnamefont
  {Metzler}},\ }\bibfield  {title} {\bibinfo {title} {Machine-learning
  solutions for the analysis of single-particle diffusion trajectories},\
  }\href@noop {} {\bibfield  {journal} {\bibinfo  {journal} {Journal of
  Physical Chemistry Letters}\ }\textbf {\bibinfo {volume} {14}},\ \bibinfo
  {pages} {7910} (\bibinfo {year} {2023})}\BibitemShut {NoStop}%
\bibitem [{\citenamefont {Pineda}\ \emph {et~al.}(2022)\citenamefont {Pineda},
  \citenamefont {Midtvedt}, \citenamefont {Bachimanchi}, \citenamefont
  {No{\'e}}, \citenamefont {Midtvedt}, \citenamefont {Volpe},\ and\
  \citenamefont {Manzo}}]{pineda2022geometric}%
  \BibitemOpen
  \bibfield  {author} {\bibinfo {author} {\bibfnamefont {J.}~\bibnamefont
  {Pineda}}, \bibinfo {author} {\bibfnamefont {B.}~\bibnamefont {Midtvedt}},
  \bibinfo {author} {\bibfnamefont {H.}~\bibnamefont {Bachimanchi}}, \bibinfo
  {author} {\bibfnamefont {S.}~\bibnamefont {No{\'e}}}, \bibinfo {author}
  {\bibfnamefont {D.}~\bibnamefont {Midtvedt}}, \bibinfo {author}
  {\bibfnamefont {G.}~\bibnamefont {Volpe}},\ and\ \bibinfo {author}
  {\bibfnamefont {C.}~\bibnamefont {Manzo}},\ }\bibfield  {title} {\bibinfo
  {title} {Geometric deep learning reveals the spatiotemporal fingerprint of
  microscopic motion},\ }\href@noop {} {\bibfield  {journal} {\bibinfo
  {journal} {arXiv preprint arXiv:2202.06355}\ } (\bibinfo {year}
  {2022})}\BibitemShut {NoStop}%
\bibitem [{\citenamefont {Mu{\~n}oz-Gil}\ \emph
  {et~al.}(2021{\natexlab{b}})\citenamefont {Mu{\~n}oz-Gil}, \citenamefont
  {i~Corominas},\ and\ \citenamefont {Lewenstein}}]{munoz2021unsupervised}%
  \BibitemOpen
  \bibfield  {author} {\bibinfo {author} {\bibfnamefont {G.}~\bibnamefont
  {Mu{\~n}oz-Gil}}, \bibinfo {author} {\bibfnamefont {G.~G.}\ \bibnamefont
  {i~Corominas}},\ and\ \bibinfo {author} {\bibfnamefont {M.}~\bibnamefont
  {Lewenstein}},\ }\bibfield  {title} {\bibinfo {title} {Unsupervised learning
  of anomalous diffusion data an anomaly detection approach},\ }\href@noop {}
  {\bibfield  {journal} {\bibinfo  {journal} {Journal of Physics A:
  Mathematical and Theoretical}\ }\textbf {\bibinfo {volume} {54}},\ \bibinfo
  {pages} {504001} (\bibinfo {year} {2021}{\natexlab{b}})}\BibitemShut
  {NoStop}%
\bibitem [{\citenamefont {Kowalek}\ \emph {et~al.}(2019)\citenamefont
  {Kowalek}, \citenamefont {Loch-Olszewska},\ and\ \citenamefont
  {Szwabi{\'n}ski}}]{kowalek2019classification}%
  \BibitemOpen
  \bibfield  {author} {\bibinfo {author} {\bibfnamefont {P.}~\bibnamefont
  {Kowalek}}, \bibinfo {author} {\bibfnamefont {H.}~\bibnamefont
  {Loch-Olszewska}},\ and\ \bibinfo {author} {\bibfnamefont {J.}~\bibnamefont
  {Szwabi{\'n}ski}},\ }\bibfield  {title} {\bibinfo {title} {Classification of
  diffusion modes in single-particle tracking data: {F}eature-based versus
  deep-learning approach},\ }\href@noop {} {\bibfield  {journal} {\bibinfo
  {journal} {Physical Review E}\ }\textbf {\bibinfo {volume} {100}},\ \bibinfo
  {pages} {032410} (\bibinfo {year} {2019})}\BibitemShut {NoStop}%
\bibitem [{\citenamefont {Loch-Olszewska}\ and\ \citenamefont
  {Szwabi{\'n}ski}(2020)}]{loch2020impact}%
  \BibitemOpen
  \bibfield  {author} {\bibinfo {author} {\bibfnamefont {H.}~\bibnamefont
  {Loch-Olszewska}}\ and\ \bibinfo {author} {\bibfnamefont {J.}~\bibnamefont
  {Szwabi{\'n}ski}},\ }\bibfield  {title} {\bibinfo {title} {Impact of feature
  choice on machine learning classification of fractional anomalous
  diffusion},\ }\href@noop {} {\bibfield  {journal} {\bibinfo  {journal}
  {Entropy}\ }\textbf {\bibinfo {volume} {22}},\ \bibinfo {pages} {1436}
  (\bibinfo {year} {2020})}\BibitemShut {NoStop}%
\bibitem [{\citenamefont {Mangalam}\ \emph
  {et~al.}(2023{\natexlab{a}})\citenamefont {Mangalam}, \citenamefont
  {Metzler},\ and\ \citenamefont {Kelty-Stephen}}]{mangalam2023ergodic}%
  \BibitemOpen
  \bibfield  {author} {\bibinfo {author} {\bibfnamefont {M.}~\bibnamefont
  {Mangalam}}, \bibinfo {author} {\bibfnamefont {R.}~\bibnamefont {Metzler}},\
  and\ \bibinfo {author} {\bibfnamefont {D.~G.}\ \bibnamefont
  {Kelty-Stephen}},\ }\bibfield  {title} {\bibinfo {title} {Ergodic
  characterization of nonergodic anomalous diffusion processes},\ }\href@noop
  {} {\bibfield  {journal} {\bibinfo  {journal} {Physical Review Research}\
  }\textbf {\bibinfo {volume} {5}},\ \bibinfo {pages} {023144} (\bibinfo {year}
  {2023}{\natexlab{a}})}\BibitemShut {NoStop}%
\bibitem [{\citenamefont {Ihlen}(2012)}]{ihlen2012introduction}%
  \BibitemOpen
  \bibfield  {author} {\bibinfo {author} {\bibfnamefont {E.~A. F.~E.}\
  \bibnamefont {Ihlen}},\ }\bibfield  {title} {\bibinfo {title} {Introduction
  to multifractal detrended fluctuation analysis in {M}atlab},\ }\href@noop {}
  {\bibfield  {journal} {\bibinfo  {journal} {Frontiers in Physiology}\
  }\textbf {\bibinfo {volume} {3}},\ \bibinfo {pages} {141} (\bibinfo {year}
  {2012})}\BibitemShut {NoStop}%
\bibitem [{\citenamefont {Kelty-Stephen}\ \emph {et~al.}(2013)\citenamefont
  {Kelty-Stephen}, \citenamefont {Palatinus}, \citenamefont {Saltzman},\ and\
  \citenamefont {Dixon}}]{kelty2013tutorial}%
  \BibitemOpen
  \bibfield  {author} {\bibinfo {author} {\bibfnamefont {D.~G.}\ \bibnamefont
  {Kelty-Stephen}}, \bibinfo {author} {\bibfnamefont {K.}~\bibnamefont
  {Palatinus}}, \bibinfo {author} {\bibfnamefont {E.}~\bibnamefont
  {Saltzman}},\ and\ \bibinfo {author} {\bibfnamefont {J.~A.}\ \bibnamefont
  {Dixon}},\ }\bibfield  {title} {\bibinfo {title} {A tutorial on
  multifractality, cascades, and interactivity for empirical time series in
  ecological science},\ }\href@noop {} {\bibfield  {journal} {\bibinfo
  {journal} {Ecological Psychology}\ }\textbf {\bibinfo {volume} {25}},\
  \bibinfo {pages} {1} (\bibinfo {year} {2013})}\BibitemShut {NoStop}%
\bibitem [{\citenamefont {Kelty-Stephen}\ \emph {et~al.}(2023)\citenamefont
  {Kelty-Stephen}, \citenamefont {Lane}, \citenamefont {Bloomfield},\ and\
  \citenamefont {Mangalam}}]{kelty2023multifractaltest}%
  \BibitemOpen
  \bibfield  {author} {\bibinfo {author} {\bibfnamefont {D.~G.}\ \bibnamefont
  {Kelty-Stephen}}, \bibinfo {author} {\bibfnamefont {E.}~\bibnamefont {Lane}},
  \bibinfo {author} {\bibfnamefont {L.}~\bibnamefont {Bloomfield}},\ and\
  \bibinfo {author} {\bibfnamefont {M.}~\bibnamefont {Mangalam}},\ }\bibfield
  {title} {\bibinfo {title} {Multifractal test for nonlinearity of interactions
  across scales in time series},\ }\href@noop {} {\bibfield  {journal}
  {\bibinfo  {journal} {Behavior Research Methods}\ }\textbf {\bibinfo {volume}
  {55}},\ \bibinfo {pages} {2249} (\bibinfo {year} {2023})}\BibitemShut
  {NoStop}%
\bibitem [{\citenamefont {Ritschel}\ \emph {et~al.}(2021)\citenamefont
  {Ritschel}, \citenamefont {Cherstvy},\ and\ \citenamefont
  {Metzler}}]{ritschel2021universality}%
  \BibitemOpen
  \bibfield  {author} {\bibinfo {author} {\bibfnamefont {S.}~\bibnamefont
  {Ritschel}}, \bibinfo {author} {\bibfnamefont {A.~G.}\ \bibnamefont
  {Cherstvy}},\ and\ \bibinfo {author} {\bibfnamefont {R.}~\bibnamefont
  {Metzler}},\ }\bibfield  {title} {\bibinfo {title} {Universality of
  delay-time averages for financial time series: {A}nalytical results, computer
  simulations, and analysis of historical stock-market prices},\ }\href@noop {}
  {\bibfield  {journal} {\bibinfo  {journal} {Journal of Physics: Complexity}\
  }\textbf {\bibinfo {volume} {2}},\ \bibinfo {pages} {045003} (\bibinfo {year}
  {2021})}\BibitemShut {NoStop}%
\bibitem [{\citenamefont {Vinod}\ \emph {et~al.}(2022)\citenamefont {Vinod},
  \citenamefont {Cherstvy}, \citenamefont {Wang}, \citenamefont {Metzler},\
  and\ \citenamefont {Sokolov}}]{vinod2022nonergodicity}%
  \BibitemOpen
  \bibfield  {author} {\bibinfo {author} {\bibfnamefont {D.}~\bibnamefont
  {Vinod}}, \bibinfo {author} {\bibfnamefont {A.~G.}\ \bibnamefont {Cherstvy}},
  \bibinfo {author} {\bibfnamefont {W.}~\bibnamefont {Wang}}, \bibinfo {author}
  {\bibfnamefont {R.}~\bibnamefont {Metzler}},\ and\ \bibinfo {author}
  {\bibfnamefont {I.~M.}\ \bibnamefont {Sokolov}},\ }\bibfield  {title}
  {\bibinfo {title} {Nonergodicity of reset geometric {B}rownian motion},\
  }\href@noop {} {\bibfield  {journal} {\bibinfo  {journal} {Physical Review
  E}\ }\textbf {\bibinfo {volume} {105}},\ \bibinfo {pages} {L012106} (\bibinfo
  {year} {2022})}\BibitemShut {NoStop}%
\bibitem [{\citenamefont {Wang}\ \emph
  {et~al.}(2022{\natexlab{a}})\citenamefont {Wang}, \citenamefont {Cherstvy},
  \citenamefont {Metzler},\ and\ \citenamefont {Sokolov}}]{wang2022restoring}%
  \BibitemOpen
  \bibfield  {author} {\bibinfo {author} {\bibfnamefont {W.}~\bibnamefont
  {Wang}}, \bibinfo {author} {\bibfnamefont {A.~G.}\ \bibnamefont {Cherstvy}},
  \bibinfo {author} {\bibfnamefont {R.}~\bibnamefont {Metzler}},\ and\ \bibinfo
  {author} {\bibfnamefont {I.~M.}\ \bibnamefont {Sokolov}},\ }\bibfield
  {title} {\bibinfo {title} {Restoring ergodicity of stochastically reset
  anomalous-diffusion processes},\ }\href@noop {} {\bibfield  {journal}
  {\bibinfo  {journal} {Physical Review Research}\ }\textbf {\bibinfo {volume}
  {4}},\ \bibinfo {pages} {013161} (\bibinfo {year}
  {2022}{\natexlab{a}})}\BibitemShut {NoStop}%
\bibitem [{\citenamefont {Mandelbrot}(1974)}]{mandelbrot1974intermittent}%
  \BibitemOpen
  \bibfield  {author} {\bibinfo {author} {\bibfnamefont {B.~B.}\ \bibnamefont
  {Mandelbrot}},\ }\bibfield  {title} {\bibinfo {title} {Intermittent
  turbulence in self-similar cascades: {D}ivergence of high moments and
  dimension of the carrier},\ }\href@noop {} {\bibfield  {journal} {\bibinfo
  {journal} {Journal of Fluid Mechanics}\ }\textbf {\bibinfo {volume} {62}},\
  \bibinfo {pages} {331} (\bibinfo {year} {1974})}\BibitemShut {NoStop}%
\bibitem [{\citenamefont {Schertzer}\ \emph {et~al.}(1997)\citenamefont
  {Schertzer}, \citenamefont {Lovejoy}, \citenamefont {Schmitt}, \citenamefont
  {Chigirinskaya},\ and\ \citenamefont {Marsan}}]{schertzer1997multifractal}%
  \BibitemOpen
  \bibfield  {author} {\bibinfo {author} {\bibfnamefont {D.}~\bibnamefont
  {Schertzer}}, \bibinfo {author} {\bibfnamefont {S.}~\bibnamefont {Lovejoy}},
  \bibinfo {author} {\bibfnamefont {F.}~\bibnamefont {Schmitt}}, \bibinfo
  {author} {\bibfnamefont {Y.}~\bibnamefont {Chigirinskaya}},\ and\ \bibinfo
  {author} {\bibfnamefont {D.}~\bibnamefont {Marsan}},\ }\bibfield  {title}
  {\bibinfo {title} {Multifractal cascade dynamics and turbulent
  intermittency},\ }\href@noop {} {\bibfield  {journal} {\bibinfo  {journal}
  {Fractals}\ }\textbf {\bibinfo {volume} {5}},\ \bibinfo {pages} {427}
  (\bibinfo {year} {1997})}\BibitemShut {NoStop}%
\bibitem [{\citenamefont {Shlesinger}\ \emph {et~al.}(1987)\citenamefont
  {Shlesinger}, \citenamefont {West},\ and\ \citenamefont
  {Klafter}}]{shlesinger1987levy}%
  \BibitemOpen
  \bibfield  {author} {\bibinfo {author} {\bibfnamefont {M.~F.}\ \bibnamefont
  {Shlesinger}}, \bibinfo {author} {\bibfnamefont {B.}~\bibnamefont {West}},\
  and\ \bibinfo {author} {\bibfnamefont {J.}~\bibnamefont {Klafter}},\
  }\bibfield  {title} {\bibinfo {title} {L{\'e}vy dynamics of enhanced
  diffusion: Application to turbulence},\ }\href@noop {} {\bibfield  {journal}
  {\bibinfo  {journal} {Physical Review Letters}\ }\textbf {\bibinfo {volume}
  {58}},\ \bibinfo {pages} {1100} (\bibinfo {year} {1987})}\BibitemShut
  {NoStop}%
\bibitem [{\citenamefont {Mu{\~n}oz}\ \emph {et~al.}(2023)\citenamefont
  {Mu{\~n}oz}, \citenamefont {Bachimanchi}, \citenamefont {Pineda},
  \citenamefont {Midtvedt}, \citenamefont {Lewenstein}, \citenamefont
  {Metzler}, \citenamefont {Krapf}, \citenamefont {Volpe},\ and\ \citenamefont
  {Manzo}}]{munoz2023quantitative}%
  \BibitemOpen
  \bibfield  {author} {\bibinfo {author} {\bibfnamefont {G.}~\bibnamefont
  {Mu{\~n}oz}}, \bibinfo {author} {\bibfnamefont {H.}~\bibnamefont
  {Bachimanchi}}, \bibinfo {author} {\bibfnamefont {J.}~\bibnamefont {Pineda}},
  \bibinfo {author} {\bibfnamefont {B.}~\bibnamefont {Midtvedt}}, \bibinfo
  {author} {\bibfnamefont {M.}~\bibnamefont {Lewenstein}}, \bibinfo {author}
  {\bibfnamefont {R.}~\bibnamefont {Metzler}}, \bibinfo {author} {\bibfnamefont
  {D.}~\bibnamefont {Krapf}}, \bibinfo {author} {\bibfnamefont
  {G.}~\bibnamefont {Volpe}},\ and\ \bibinfo {author} {\bibfnamefont
  {C.}~\bibnamefont {Manzo}},\ }\bibfield  {title} {\bibinfo {title}
  {Quantitative evaluation of methods to analyze motion changes in
  single-particle experiments},\ }\href@noop {} {\bibfield  {journal} {\bibinfo
   {journal} {arXiv preprint arXiv:2311.18100}\ } (\bibinfo {year}
  {2023})}\BibitemShut {NoStop}%
\bibitem [{\citenamefont {Chhabra}\ and\ \citenamefont
  {Jensen}(1989)}]{chhabra1989direct}%
  \BibitemOpen
  \bibfield  {author} {\bibinfo {author} {\bibfnamefont {A.}~\bibnamefont
  {Chhabra}}\ and\ \bibinfo {author} {\bibfnamefont {R.~V.}\ \bibnamefont
  {Jensen}},\ }\bibfield  {title} {\bibinfo {title} {Direct determination of
  the f($\alpha$) singularity spectrum},\ }\href@noop {} {\bibfield  {journal}
  {\bibinfo  {journal} {Physical Review Letters}\ }\textbf {\bibinfo {volume}
  {62}},\ \bibinfo {pages} {1327} (\bibinfo {year} {1989})}\BibitemShut
  {NoStop}%
\bibitem [{\citenamefont {Mandelbrot}\ and\ \citenamefont
  {Mandelbrot}(1982)}]{mandelbrot1982fractal}%
  \BibitemOpen
  \bibfield  {author} {\bibinfo {author} {\bibfnamefont {B.~B.}\ \bibnamefont
  {Mandelbrot}}\ and\ \bibinfo {author} {\bibfnamefont {B.~B.}\ \bibnamefont
  {Mandelbrot}},\ }\href@noop {} {\emph {\bibinfo {title} {The {F}ractal
  {G}eometry of {N}ature}}},\ Vol.~\bibinfo {volume} {1}\ (\bibinfo
  {publisher} {{WH} {F}reeman, New York, NY},\ \bibinfo {year}
  {1982})\BibitemShut {NoStop}%
\bibitem [{\citenamefont {Halsey}\ \emph {et~al.}(1986)\citenamefont {Halsey},
  \citenamefont {Jensen}, \citenamefont {Kadanoff}, \citenamefont {Procaccia},\
  and\ \citenamefont {Shraiman}}]{halsey1986fractal}%
  \BibitemOpen
  \bibfield  {author} {\bibinfo {author} {\bibfnamefont {T.~C.}\ \bibnamefont
  {Halsey}}, \bibinfo {author} {\bibfnamefont {M.~H.}\ \bibnamefont {Jensen}},
  \bibinfo {author} {\bibfnamefont {L.~P.}\ \bibnamefont {Kadanoff}}, \bibinfo
  {author} {\bibfnamefont {I.}~\bibnamefont {Procaccia}},\ and\ \bibinfo
  {author} {\bibfnamefont {B.~I.}\ \bibnamefont {Shraiman}},\ }\bibfield
  {title} {\bibinfo {title} {Fractal measures and their singularities: {T}he
  characterization of strange sets},\ }\href@noop {} {\bibfield  {journal}
  {\bibinfo  {journal} {Physical Review A}\ }\textbf {\bibinfo {volume} {33}},\
  \bibinfo {pages} {1141} (\bibinfo {year} {1986})}\BibitemShut {NoStop}%
\bibitem [{\citenamefont {Mandelbrot}(2013)}]{mandelbrot2013fractals}%
  \BibitemOpen
  \bibfield  {author} {\bibinfo {author} {\bibfnamefont {B.~B.}\ \bibnamefont
  {Mandelbrot}},\ }\href@noop {} {\emph {\bibinfo {title} {Fractals and
  {S}caling in {F}inance: {D}iscontinuity, {C}oncentration, {R}isk}}}\
  (\bibinfo  {publisher} {Springer, New York, NY},\ \bibinfo {year}
  {2013})\BibitemShut {NoStop}%
\bibitem [{\citenamefont {Zamir}(2003)}]{zamir2003critique}%
  \BibitemOpen
  \bibfield  {author} {\bibinfo {author} {\bibfnamefont {M.}~\bibnamefont
  {Zamir}},\ }\bibfield  {title} {\bibinfo {title} {Critique of the test of
  multifractality as applied to biological data},\ }\href@noop {} {\bibfield
  {journal} {\bibinfo  {journal} {Journal of Theoretical Biology}\ }\textbf
  {\bibinfo {volume} {225}},\ \bibinfo {pages} {407} (\bibinfo {year}
  {2003})}\BibitemShut {NoStop}%
\bibitem [{\citenamefont {Schreiber}\ and\ \citenamefont
  {Schmitz}(1996)}]{schreiber1996improved}%
  \BibitemOpen
  \bibfield  {author} {\bibinfo {author} {\bibfnamefont {T.}~\bibnamefont
  {Schreiber}}\ and\ \bibinfo {author} {\bibfnamefont {A.}~\bibnamefont
  {Schmitz}},\ }\bibfield  {title} {\bibinfo {title} {Improved surrogate data
  for nonlinearity tests},\ }\href@noop {} {\bibfield  {journal} {\bibinfo
  {journal} {Physical Review Letters}\ }\textbf {\bibinfo {volume} {77}},\
  \bibinfo {pages} {635} (\bibinfo {year} {1996})}\BibitemShut {NoStop}%
\bibitem [{\citenamefont {Fausett}(1994)}]{fausett1994fundamentals}%
  \BibitemOpen
  \bibfield  {author} {\bibinfo {author} {\bibfnamefont {L.~V.}\ \bibnamefont
  {Fausett}},\ }\href@noop {} {\emph {\bibinfo {title} {Fundamentals of
  {N}eural {N}etworks: {A}rchitectures, {A}lgorithms and {A}pplications}}}\
  (\bibinfo  {publisher} {Prentice Hall},\ \bibinfo {address} {Hoboken, NJ},\
  \bibinfo {year} {1994})\BibitemShut {NoStop}%
\bibitem [{\citenamefont {Alpaydin}(2020)}]{alpaydin2020introduction}%
  \BibitemOpen
  \bibfield  {author} {\bibinfo {author} {\bibfnamefont {E.}~\bibnamefont
  {Alpaydin}},\ }\href@noop {} {\emph {\bibinfo {title} {Introduction to
  {M}achine {L}earning}}}\ (\bibinfo  {publisher} {MIT Press},\ \bibinfo
  {address} {Cambridge, MA},\ \bibinfo {year} {2020})\BibitemShut {NoStop}%
\bibitem [{\citenamefont {Zhang}\ and\ \citenamefont
  {Sabuncu}(2018)}]{zhang2018generalized}%
  \BibitemOpen
  \bibfield  {author} {\bibinfo {author} {\bibfnamefont {Z.}~\bibnamefont
  {Zhang}}\ and\ \bibinfo {author} {\bibfnamefont {M.}~\bibnamefont
  {Sabuncu}},\ }\bibfield  {title} {\bibinfo {title} {Generalized cross entropy
  loss for training deep neural networks with noisy labels},\ }\href@noop {}
  {\bibfield  {journal} {\bibinfo  {journal} {Advances in Neural Information
  Processing Systems}\ }\textbf {\bibinfo {volume} {31}} (\bibinfo {year}
  {2018})}\BibitemShut {NoStop}%
\bibitem [{\citenamefont {Bottou}(2010)}]{bottou2010large}%
  \BibitemOpen
  \bibfield  {author} {\bibinfo {author} {\bibfnamefont {L.}~\bibnamefont
  {Bottou}},\ }\bibfield  {title} {\bibinfo {title} {Large-scale machine
  learning with stochastic gradient descent},\ }in\ \href@noop {} {\emph
  {\bibinfo {booktitle} {Proceedings of COMPSTAT'2010: 19th International
  Conference on Computational StatisticsParis France, August 22--27, 2010
  Keynote, Invited and Contributed Papers}}}\ (\bibinfo {organization}
  {Springer},\ \bibinfo {year} {2010})\ pp.\ \bibinfo {pages}
  {177--186}\BibitemShut {NoStop}%
\bibitem [{\citenamefont {Kingma}\ and\ \citenamefont
  {Ba}(2014)}]{kingma2014adam}%
  \BibitemOpen
  \bibfield  {author} {\bibinfo {author} {\bibfnamefont {D.~P.}\ \bibnamefont
  {Kingma}}\ and\ \bibinfo {author} {\bibfnamefont {J.}~\bibnamefont {Ba}},\
  }\bibfield  {title} {\bibinfo {title} {Adam: {A} method for stochastic
  optimization},\ }\href@noop {} {\bibfield  {journal} {\bibinfo  {journal}
  {arXiv preprint arXiv:1412.6980}\ } (\bibinfo {year} {2014})}\BibitemShut
  {NoStop}%
\bibitem [{\citenamefont {Maddox}\ \emph {et~al.}(2019)\citenamefont {Maddox},
  \citenamefont {Izmailov}, \citenamefont {Garipov}, \citenamefont {Vetrov},\
  and\ \citenamefont {Wilson}}]{maddox2019simple}%
  \BibitemOpen
  \bibfield  {author} {\bibinfo {author} {\bibfnamefont {W.~J.}\ \bibnamefont
  {Maddox}}, \bibinfo {author} {\bibfnamefont {P.}~\bibnamefont {Izmailov}},
  \bibinfo {author} {\bibfnamefont {T.}~\bibnamefont {Garipov}}, \bibinfo
  {author} {\bibfnamefont {D.~P.}\ \bibnamefont {Vetrov}},\ and\ \bibinfo
  {author} {\bibfnamefont {A.~G.}\ \bibnamefont {Wilson}},\ }\bibfield  {title}
  {\bibinfo {title} {A simple baseline for bayesian uncertainty in deep
  learning},\ }\href@noop {} {\bibfield  {journal} {\bibinfo  {journal}
  {Advances in Neural Information Processing Systems}\ }\textbf {\bibinfo
  {volume} {32}} (\bibinfo {year} {2019})}\BibitemShut {NoStop}%
\bibitem [{\citenamefont {Hahnloser}\ \emph {et~al.}(2000)\citenamefont
  {Hahnloser}, \citenamefont {Sarpeshkar}, \citenamefont {Mahowald},
  \citenamefont {Douglas},\ and\ \citenamefont {Seung}}]{hahnloser2000digital}%
  \BibitemOpen
  \bibfield  {author} {\bibinfo {author} {\bibfnamefont {R.~H.}\ \bibnamefont
  {Hahnloser}}, \bibinfo {author} {\bibfnamefont {R.}~\bibnamefont
  {Sarpeshkar}}, \bibinfo {author} {\bibfnamefont {M.~A.}\ \bibnamefont
  {Mahowald}}, \bibinfo {author} {\bibfnamefont {R.~J.}\ \bibnamefont
  {Douglas}},\ and\ \bibinfo {author} {\bibfnamefont {H.~S.}\ \bibnamefont
  {Seung}},\ }\bibfield  {title} {\bibinfo {title} {Digital selection and
  analogue amplification coexist in a cortex-inspired silicon circuit},\
  }\href@noop {} {\bibfield  {journal} {\bibinfo  {journal} {Nature}\ }\textbf
  {\bibinfo {volume} {405}},\ \bibinfo {pages} {947} (\bibinfo {year}
  {2000})}\BibitemShut {NoStop}%
\bibitem [{\citenamefont {Gao}\ and\ \citenamefont
  {Pavel}(2017)}]{gao2017properties}%
  \BibitemOpen
  \bibfield  {author} {\bibinfo {author} {\bibfnamefont {B.}~\bibnamefont
  {Gao}}\ and\ \bibinfo {author} {\bibfnamefont {L.}~\bibnamefont {Pavel}},\
  }\bibfield  {title} {\bibinfo {title} {On the properties of the softmax
  function with application in game theory and reinforcement learning},\
  }\href@noop {} {\bibfield  {journal} {\bibinfo  {journal} {arXiv preprint
  arXiv:1704.00805}\ } (\bibinfo {year} {2017})}\BibitemShut {NoStop}%
\bibitem [{\citenamefont {Kian-Bostanabad}\ and\ \citenamefont
  {Azghani}(2017)}]{kian2017relationship}%
  \BibitemOpen
  \bibfield  {author} {\bibinfo {author} {\bibfnamefont {S.}~\bibnamefont
  {Kian-Bostanabad}}\ and\ \bibinfo {author} {\bibfnamefont {M.-R.}\
  \bibnamefont {Azghani}},\ }\bibfield  {title} {\bibinfo {title} {The
  relationship between {RMS} electromyography and thickness change in the
  skeletal muscles},\ }\href@noop {} {\bibfield  {journal} {\bibinfo  {journal}
  {Medical Engineering \& Physics}\ }\textbf {\bibinfo {volume} {43}},\
  \bibinfo {pages} {92} (\bibinfo {year} {2017})}\BibitemShut {NoStop}%
\bibitem [{\citenamefont {Vink}\ \emph {et~al.}(2017)\citenamefont {Vink},
  \citenamefont {Westover}, \citenamefont {Pascual-Leone},\ and\ \citenamefont
  {Shafi}}]{vink2017eeg}%
  \BibitemOpen
  \bibfield  {author} {\bibinfo {author} {\bibfnamefont {J.}~\bibnamefont
  {Vink}}, \bibinfo {author} {\bibfnamefont {M.}~\bibnamefont {Westover}},
  \bibinfo {author} {\bibfnamefont {A.}~\bibnamefont {Pascual-Leone}},\ and\
  \bibinfo {author} {\bibfnamefont {M.}~\bibnamefont {Shafi}},\ }\bibfield
  {title} {\bibinfo {title} {{EEG} functional connectivity predicts propagation
  of {TMS}-evoked potentials},\ }\href@noop {} {\bibfield  {journal} {\bibinfo
  {journal} {Brain Stimulation}\ }\textbf {\bibinfo {volume} {10}},\ \bibinfo
  {pages} {516} (\bibinfo {year} {2017})}\BibitemShut {NoStop}%
\bibitem [{\citenamefont {Vink}\ \emph {et~al.}(2020)\citenamefont {Vink},
  \citenamefont {Klooster}, \citenamefont {Ozdemir}, \citenamefont {Westover},
  \citenamefont {Pascual-Leone},\ and\ \citenamefont {Shafi}}]{vink2020eeg}%
  \BibitemOpen
  \bibfield  {author} {\bibinfo {author} {\bibfnamefont {J.~J.}\ \bibnamefont
  {Vink}}, \bibinfo {author} {\bibfnamefont {D.~C.}\ \bibnamefont {Klooster}},
  \bibinfo {author} {\bibfnamefont {R.~A.}\ \bibnamefont {Ozdemir}}, \bibinfo
  {author} {\bibfnamefont {M.~B.}\ \bibnamefont {Westover}}, \bibinfo {author}
  {\bibfnamefont {A.}~\bibnamefont {Pascual-Leone}},\ and\ \bibinfo {author}
  {\bibfnamefont {M.~M.}\ \bibnamefont {Shafi}},\ }\bibfield  {title} {\bibinfo
  {title} {{EEG} functional connectivity is a weak predictor of causal brain
  interactions},\ }\href@noop {} {\bibfield  {journal} {\bibinfo  {journal}
  {Brain Topography}\ }\textbf {\bibinfo {volume} {33}},\ \bibinfo {pages}
  {221} (\bibinfo {year} {2020})}\BibitemShut {NoStop}%
\bibitem [{\citenamefont {Kelty-Stephen}\ and\ \citenamefont
  {Mangalam}(2022)}]{kelty2022fractal}%
  \BibitemOpen
  \bibfield  {author} {\bibinfo {author} {\bibfnamefont {D.~G.}\ \bibnamefont
  {Kelty-Stephen}}\ and\ \bibinfo {author} {\bibfnamefont {M.}~\bibnamefont
  {Mangalam}},\ }\bibfield  {title} {\bibinfo {title} {Fractal and multifractal
  descriptors restore ergodicity broken by non-{G}aussianity in time series},\
  }\href {https://doi.org/https://doi.org/10.1016/j.chaos.2022.112568}
  {\bibfield  {journal} {\bibinfo  {journal} {Chaos, Solitons \& Fractals}\
  }\textbf {\bibinfo {volume} {163}},\ \bibinfo {pages} {112568} (\bibinfo
  {year} {2022})}\BibitemShut {NoStop}%
\bibitem [{\citenamefont {Kelty-Stephen}\ and\ \citenamefont
  {Mangalam}(2023)}]{kelty2023multifractal}%
  \BibitemOpen
  \bibfield  {author} {\bibinfo {author} {\bibfnamefont {D.~G.}\ \bibnamefont
  {Kelty-Stephen}}\ and\ \bibinfo {author} {\bibfnamefont {M.}~\bibnamefont
  {Mangalam}},\ }\bibfield  {title} {\bibinfo {title} {Multifractal descriptors
  ergodically characterize non-ergodic multiplicative cascade processes},\
  }\href@noop {} {\bibfield  {journal} {\bibinfo  {journal} {Physica A:
  Statistical Mechanics and its Applications}\ }\textbf {\bibinfo {volume}
  {617}},\ \bibinfo {pages} {128651} (\bibinfo {year} {2023})}\BibitemShut
  {NoStop}%
\bibitem [{\citenamefont {Mangalam}\ and\ \citenamefont
  {Kelty-Stephen}(2022)}]{mangalam2022ergodic}%
  \BibitemOpen
  \bibfield  {author} {\bibinfo {author} {\bibfnamefont {M.}~\bibnamefont
  {Mangalam}}\ and\ \bibinfo {author} {\bibfnamefont {D.~G.}\ \bibnamefont
  {Kelty-Stephen}},\ }\bibfield  {title} {\bibinfo {title} {Ergodic descriptors
  of non-ergodic stochastic processes},\ }\href@noop {} {\bibfield  {journal}
  {\bibinfo  {journal} {Journal of the Royal Society Interface}\ }\textbf
  {\bibinfo {volume} {19}},\ \bibinfo {pages} {20220095} (\bibinfo {year}
  {2022})}\BibitemShut {NoStop}%
\bibitem [{\citenamefont {Mangalam}\ \emph
  {et~al.}(2023{\natexlab{b}})\citenamefont {Mangalam}, \citenamefont {Sadri},
  \citenamefont {Hayano}, \citenamefont {Watanabe}, \citenamefont {Kiyono},\
  and\ \citenamefont {Kelty-Stephen}}]{mangalam2023multifractal}%
  \BibitemOpen
  \bibfield  {author} {\bibinfo {author} {\bibfnamefont {M.}~\bibnamefont
  {Mangalam}}, \bibinfo {author} {\bibfnamefont {A.}~\bibnamefont {Sadri}},
  \bibinfo {author} {\bibfnamefont {J.}~\bibnamefont {Hayano}}, \bibinfo
  {author} {\bibfnamefont {E.}~\bibnamefont {Watanabe}}, \bibinfo {author}
  {\bibfnamefont {K.}~\bibnamefont {Kiyono}},\ and\ \bibinfo {author}
  {\bibfnamefont {D.~G.}\ \bibnamefont {Kelty-Stephen}},\ }\bibfield  {title}
  {\bibinfo {title} {Multifractal foundations of biomarker discovery for heart
  disease and stroke},\ }\href@noop {} {\bibfield  {journal} {\bibinfo
  {journal} {Scientific Reports}\ }\textbf {\bibinfo {volume} {13}},\ \bibinfo
  {pages} {18316} (\bibinfo {year} {2023}{\natexlab{b}})},\ \bibinfo {note}
  {\url{https://doi.org/10.1038/s41598-023-45184-2}}\BibitemShut {NoStop}%
\bibitem [{\citenamefont {Bloomfield}\ \emph {et~al.}(2021)\citenamefont
  {Bloomfield}, \citenamefont {Lane}, \citenamefont {Mangalam},\ and\
  \citenamefont {Kelty-Stephen}}]{bloomfield2021perceiving}%
  \BibitemOpen
  \bibfield  {author} {\bibinfo {author} {\bibfnamefont {L.}~\bibnamefont
  {Bloomfield}}, \bibinfo {author} {\bibfnamefont {E.}~\bibnamefont {Lane}},
  \bibinfo {author} {\bibfnamefont {M.}~\bibnamefont {Mangalam}},\ and\
  \bibinfo {author} {\bibfnamefont {D.~G.}\ \bibnamefont {Kelty-Stephen}},\
  }\bibfield  {title} {\bibinfo {title} {Perceiving and remembering speech
  depend on multifractal nonlinearity in movements producing and exploring
  speech},\ }\href@noop {} {\bibfield  {journal} {\bibinfo  {journal} {Journal
  of the Royal Society Interface}\ }\textbf {\bibinfo {volume} {18}},\ \bibinfo
  {pages} {20210272} (\bibinfo {year} {2021})}\BibitemShut {NoStop}%
\bibitem [{\citenamefont {Booth}\ \emph {et~al.}(2018)\citenamefont {Booth},
  \citenamefont {Brown}, \citenamefont {Eason}, \citenamefont {Wallot},\ and\
  \citenamefont {Kelty-Stephen}}]{booth2018expectations}%
  \BibitemOpen
  \bibfield  {author} {\bibinfo {author} {\bibfnamefont {C.~R.}\ \bibnamefont
  {Booth}}, \bibinfo {author} {\bibfnamefont {H.~L.}\ \bibnamefont {Brown}},
  \bibinfo {author} {\bibfnamefont {E.~G.}\ \bibnamefont {Eason}}, \bibinfo
  {author} {\bibfnamefont {S.}~\bibnamefont {Wallot}},\ and\ \bibinfo {author}
  {\bibfnamefont {D.~G.}\ \bibnamefont {Kelty-Stephen}},\ }\bibfield  {title}
  {\bibinfo {title} {Expectations on hierarchical scales of discourse:
  {M}ultifractality predicts both short-and long-range effects of violating
  gender expectations in text reading},\ }\href@noop {} {\bibfield  {journal}
  {\bibinfo  {journal} {Discourse Processes}\ }\textbf {\bibinfo {volume}
  {55}},\ \bibinfo {pages} {12} (\bibinfo {year} {2018})}\BibitemShut {NoStop}%
\bibitem [{\citenamefont {Carver}\ \emph {et~al.}(2017)\citenamefont {Carver},
  \citenamefont {Bojovic},\ and\ \citenamefont
  {Kelty-Stephen}}]{carver2017multifractal}%
  \BibitemOpen
  \bibfield  {author} {\bibinfo {author} {\bibfnamefont {N.~S.}\ \bibnamefont
  {Carver}}, \bibinfo {author} {\bibfnamefont {D.}~\bibnamefont {Bojovic}},\
  and\ \bibinfo {author} {\bibfnamefont {D.~G.}\ \bibnamefont
  {Kelty-Stephen}},\ }\bibfield  {title} {\bibinfo {title} {Multifractal
  foundations of visually-guided aiming and adaptation to prismatic
  perturbation},\ }\href@noop {} {\bibfield  {journal} {\bibinfo  {journal}
  {Human Movement Science}\ }\textbf {\bibinfo {volume} {55}},\ \bibinfo
  {pages} {61} (\bibinfo {year} {2017})}\BibitemShut {NoStop}%
\bibitem [{\citenamefont {Dixon}\ \emph {et~al.}(2012)\citenamefont {Dixon},
  \citenamefont {Holden}, \citenamefont {Mirman},\ and\ \citenamefont
  {Stephen}}]{dixon2012multifractal}%
  \BibitemOpen
  \bibfield  {author} {\bibinfo {author} {\bibfnamefont {J.~A.}\ \bibnamefont
  {Dixon}}, \bibinfo {author} {\bibfnamefont {J.~G.}\ \bibnamefont {Holden}},
  \bibinfo {author} {\bibfnamefont {D.}~\bibnamefont {Mirman}},\ and\ \bibinfo
  {author} {\bibfnamefont {D.~G.}\ \bibnamefont {Stephen}},\ }\bibfield
  {title} {\bibinfo {title} {Multifractal dynamics in the emergence of
  cognitive structure},\ }\href@noop {} {\bibfield  {journal} {\bibinfo
  {journal} {Topics in Cognitive Science}\ }\textbf {\bibinfo {volume} {4}},\
  \bibinfo {pages} {51} (\bibinfo {year} {2012})}\BibitemShut {NoStop}%
\bibitem [{\citenamefont {Jacobson}\ \emph {et~al.}(2021)\citenamefont
  {Jacobson}, \citenamefont {Berleman-Paul}, \citenamefont {Mangalam},
  \citenamefont {Kelty-Stephen},\ and\ \citenamefont
  {Ralston}}]{jacobson2021multifractality}%
  \BibitemOpen
  \bibfield  {author} {\bibinfo {author} {\bibfnamefont {N.}~\bibnamefont
  {Jacobson}}, \bibinfo {author} {\bibfnamefont {Q.}~\bibnamefont
  {Berleman-Paul}}, \bibinfo {author} {\bibfnamefont {M.}~\bibnamefont
  {Mangalam}}, \bibinfo {author} {\bibfnamefont {D.~G.}\ \bibnamefont
  {Kelty-Stephen}},\ and\ \bibinfo {author} {\bibfnamefont {C.}~\bibnamefont
  {Ralston}},\ }\bibfield  {title} {\bibinfo {title} {Multifractality in
  postural sway supports quiet eye training in aiming tasks: {A} study of golf
  putting},\ }\href@noop {} {\bibfield  {journal} {\bibinfo  {journal} {Human
  Movement Science}\ }\textbf {\bibinfo {volume} {76}},\ \bibinfo {pages}
  {102752} (\bibinfo {year} {2021})}\BibitemShut {NoStop}%
\bibitem [{\citenamefont {Kelty-Stephen}\ and\ \citenamefont
  {Dixon}(2014)}]{kelty2014interwoven}%
  \BibitemOpen
  \bibfield  {author} {\bibinfo {author} {\bibfnamefont {D.~G.}\ \bibnamefont
  {Kelty-Stephen}}\ and\ \bibinfo {author} {\bibfnamefont {J.~A.}\ \bibnamefont
  {Dixon}},\ }\bibfield  {title} {\bibinfo {title} {Interwoven fluctuations
  during intermodal perception: {F}ractality in head sway supports the use of
  visual feedback in haptic perceptual judgments by manual wielding},\
  }\href@noop {} {\bibfield  {journal} {\bibinfo  {journal} {Journal of
  Experimental Psychology: Human Perception and Performance}\ }\textbf
  {\bibinfo {volume} {40}},\ \bibinfo {pages} {2289} (\bibinfo {year}
  {2014})}\BibitemShut {NoStop}%
\bibitem [{\citenamefont {Kelty-Stephen}\ \emph {et~al.}(2021)\citenamefont
  {Kelty-Stephen}, \citenamefont {Lee}, \citenamefont {Carver}, \citenamefont
  {Newell},\ and\ \citenamefont {Mangalam}}]{kelty2021multifractal}%
  \BibitemOpen
  \bibfield  {author} {\bibinfo {author} {\bibfnamefont {D.~G.}\ \bibnamefont
  {Kelty-Stephen}}, \bibinfo {author} {\bibfnamefont {I.~C.}\ \bibnamefont
  {Lee}}, \bibinfo {author} {\bibfnamefont {N.~S.}\ \bibnamefont {Carver}},
  \bibinfo {author} {\bibfnamefont {K.~M.}\ \bibnamefont {Newell}},\ and\
  \bibinfo {author} {\bibfnamefont {M.}~\bibnamefont {Mangalam}},\ }\bibfield
  {title} {\bibinfo {title} {Multifractal roots of suprapostural dexterity},\
  }\href@noop {} {\bibfield  {journal} {\bibinfo  {journal} {Human Movement
  Science}\ }\textbf {\bibinfo {volume} {76}},\ \bibinfo {pages} {102771}
  (\bibinfo {year} {2021})}\BibitemShut {NoStop}%
\bibitem [{\citenamefont {Mangalam}\ \emph
  {et~al.}(2020{\natexlab{a}})\citenamefont {Mangalam}, \citenamefont {Chen},
  \citenamefont {McHugh}, \citenamefont {Singh},\ and\ \citenamefont
  {Kelty-Stephen}}]{mangalam2020bodywide}%
  \BibitemOpen
  \bibfield  {author} {\bibinfo {author} {\bibfnamefont {M.}~\bibnamefont
  {Mangalam}}, \bibinfo {author} {\bibfnamefont {R.}~\bibnamefont {Chen}},
  \bibinfo {author} {\bibfnamefont {T.~R.}\ \bibnamefont {McHugh}}, \bibinfo
  {author} {\bibfnamefont {T.}~\bibnamefont {Singh}},\ and\ \bibinfo {author}
  {\bibfnamefont {D.~G.}\ \bibnamefont {Kelty-Stephen}},\ }\bibfield  {title}
  {\bibinfo {title} {Bodywide fluctuations support manual exploration: Fractal
  fluctuations in posture predict perception of heaviness and length via
  effortful touch by the hand},\ }\href@noop {} {\bibfield  {journal} {\bibinfo
   {journal} {Human Movement Science}\ }\textbf {\bibinfo {volume} {69}},\
  \bibinfo {pages} {102543} (\bibinfo {year} {2020}{\natexlab{a}})}\BibitemShut
  {NoStop}%
\bibitem [{\citenamefont {Mangalam}\ \emph
  {et~al.}(2020{\natexlab{b}})\citenamefont {Mangalam}, \citenamefont
  {Carver},\ and\ \citenamefont {Kelty-Stephen}}]{mangalam2020global}%
  \BibitemOpen
  \bibfield  {author} {\bibinfo {author} {\bibfnamefont {M.}~\bibnamefont
  {Mangalam}}, \bibinfo {author} {\bibfnamefont {N.~S.}\ \bibnamefont
  {Carver}},\ and\ \bibinfo {author} {\bibfnamefont {D.~G.}\ \bibnamefont
  {Kelty-Stephen}},\ }\bibfield  {title} {\bibinfo {title} {Global broadcasting
  of local fractal fluctuations in a bodywide distributed system supports
  perception via effortful touch},\ }\href@noop {} {\bibfield  {journal}
  {\bibinfo  {journal} {Chaos, Solitons \& Fractals}\ }\textbf {\bibinfo
  {volume} {135}},\ \bibinfo {pages} {109740} (\bibinfo {year}
  {2020}{\natexlab{b}})}\BibitemShut {NoStop}%
\bibitem [{\citenamefont {Mangalam}\ \emph
  {et~al.}(2020{\natexlab{c}})\citenamefont {Mangalam}, \citenamefont
  {Carver},\ and\ \citenamefont {Kelty-Stephen}}]{mangalam2020multifractal}%
  \BibitemOpen
  \bibfield  {author} {\bibinfo {author} {\bibfnamefont {M.}~\bibnamefont
  {Mangalam}}, \bibinfo {author} {\bibfnamefont {N.~S.}\ \bibnamefont
  {Carver}},\ and\ \bibinfo {author} {\bibfnamefont {D.~G.}\ \bibnamefont
  {Kelty-Stephen}},\ }\bibfield  {title} {\bibinfo {title} {Multifractal
  signatures of perceptual processing on anatomical sleeves of the human
  body},\ }\href@noop {} {\bibfield  {journal} {\bibinfo  {journal} {Journal of
  The Royal Society Interface}\ }\textbf {\bibinfo {volume} {17}},\ \bibinfo
  {pages} {20200328} (\bibinfo {year} {2020}{\natexlab{c}})}\BibitemShut
  {NoStop}%
\bibitem [{\citenamefont {Mangalam}\ and\ \citenamefont
  {Kelty-Stephen}(2020)}]{mangalam2020multiplicative}%
  \BibitemOpen
  \bibfield  {author} {\bibinfo {author} {\bibfnamefont {M.}~\bibnamefont
  {Mangalam}}\ and\ \bibinfo {author} {\bibfnamefont {D.~G.}\ \bibnamefont
  {Kelty-Stephen}},\ }\bibfield  {title} {\bibinfo {title}
  {Multiplicative-cascade dynamics supports whole-body coordination for
  perception via effortful touch},\ }\href@noop {} {\bibfield  {journal}
  {\bibinfo  {journal} {Human Movement Science}\ }\textbf {\bibinfo {volume}
  {70}},\ \bibinfo {pages} {102595} (\bibinfo {year} {2020})}\BibitemShut
  {NoStop}%
\bibitem [{\citenamefont {Wallot}\ and\ \citenamefont
  {Kelty-Stephen}(2018)}]{wallot2018interaction}%
  \BibitemOpen
  \bibfield  {author} {\bibinfo {author} {\bibfnamefont {S.}~\bibnamefont
  {Wallot}}\ and\ \bibinfo {author} {\bibfnamefont {D.~G.}\ \bibnamefont
  {Kelty-Stephen}},\ }\bibfield  {title} {\bibinfo {title}
  {Interaction-dominant causation in mind and brain, and its implication for
  questions of generalization and replication},\ }\href@noop {} {\bibfield
  {journal} {\bibinfo  {journal} {Minds and Machines}\ }\textbf {\bibinfo
  {volume} {28}},\ \bibinfo {pages} {353} (\bibinfo {year} {2018})}\BibitemShut
  {NoStop}%
\bibitem [{\citenamefont {Cardenas}\ \emph {et~al.}(2012)\citenamefont
  {Cardenas}, \citenamefont {Kumar},\ and\ \citenamefont
  {Mohanty}}]{cardenas2012dynamics}%
  \BibitemOpen
  \bibfield  {author} {\bibinfo {author} {\bibfnamefont {N.}~\bibnamefont
  {Cardenas}}, \bibinfo {author} {\bibfnamefont {S.}~\bibnamefont {Kumar}},\
  and\ \bibinfo {author} {\bibfnamefont {S.}~\bibnamefont {Mohanty}},\
  }\bibfield  {title} {\bibinfo {title} {Dynamics of cellular response to
  hypotonic stimulation revealed by quantitative phase microscopy and
  multi-fractal detrended fluctuation analysis},\ }\href@noop {} {\bibfield
  {journal} {\bibinfo  {journal} {Applied Physics Letters}\ }\textbf {\bibinfo
  {volume} {101}},\ \bibinfo {pages} {203702} (\bibinfo {year}
  {2012})}\BibitemShut {NoStop}%
\bibitem [{\citenamefont {Chaieb}\ \emph {et~al.}(2008)\citenamefont {Chaieb},
  \citenamefont {M{\'a}lkov{\'a}},\ and\ \citenamefont
  {Lal}}]{chaieb2008wrinkling}%
  \BibitemOpen
  \bibfield  {author} {\bibinfo {author} {\bibfnamefont {S.}~\bibnamefont
  {Chaieb}}, \bibinfo {author} {\bibfnamefont {{\v{S}}.}~\bibnamefont
  {M{\'a}lkov{\'a}}},\ and\ \bibinfo {author} {\bibfnamefont {J.}~\bibnamefont
  {Lal}},\ }\bibfield  {title} {\bibinfo {title} {Why the wrinkling transition
  in partially polymerized membranes is not universal? {F}ractal-multifractal
  hierarchy},\ }\href@noop {} {\bibfield  {journal} {\bibinfo  {journal}
  {Journal of Theoretical Biology}\ }\textbf {\bibinfo {volume} {251}},\
  \bibinfo {pages} {60} (\bibinfo {year} {2008})}\BibitemShut {NoStop}%
\bibitem [{\citenamefont {Rezania}\ \emph {et~al.}(2021)\citenamefont
  {Rezania}, \citenamefont {Sudirga},\ and\ \citenamefont
  {Tuszynski}}]{rezania2021multifractality}%
  \BibitemOpen
  \bibfield  {author} {\bibinfo {author} {\bibfnamefont {V.}~\bibnamefont
  {Rezania}}, \bibinfo {author} {\bibfnamefont {F.~C.}\ \bibnamefont
  {Sudirga}},\ and\ \bibinfo {author} {\bibfnamefont {J.~A.}\ \bibnamefont
  {Tuszynski}},\ }\bibfield  {title} {\bibinfo {title} {Multifractality nature
  of microtubule dynamic instability process},\ }\href@noop {} {\bibfield
  {journal} {\bibinfo  {journal} {Physica A: Statistical Mechanics and its
  Applications}\ }\textbf {\bibinfo {volume} {573}},\ \bibinfo {pages} {125929}
  (\bibinfo {year} {2021})}\BibitemShut {NoStop}%
\bibitem [{\citenamefont {Wawrzkiewicz-Ja{\l}owiecka}\ \emph
  {et~al.}(2020)\citenamefont {Wawrzkiewicz-Ja{\l}owiecka}, \citenamefont
  {Trybek}, \citenamefont {Dworakowska},\ and\ \citenamefont
  {Machura}}]{wawrzkiewicz2020multifractal}%
  \BibitemOpen
  \bibfield  {author} {\bibinfo {author} {\bibfnamefont {A.}~\bibnamefont
  {Wawrzkiewicz-Ja{\l}owiecka}}, \bibinfo {author} {\bibfnamefont
  {P.}~\bibnamefont {Trybek}}, \bibinfo {author} {\bibfnamefont
  {B.}~\bibnamefont {Dworakowska}},\ and\ \bibinfo {author} {\bibfnamefont
  {{\L}.}~\bibnamefont {Machura}},\ }\bibfield  {title} {\bibinfo {title}
  {Multifractal properties of {BK} channel currents in human glioblastoma
  cells},\ }\href@noop {} {\bibfield  {journal} {\bibinfo  {journal} {Journal
  of Physical Chemistry B}\ }\textbf {\bibinfo {volume} {124}},\ \bibinfo
  {pages} {2382} (\bibinfo {year} {2020})}\BibitemShut {NoStop}%
\bibitem [{\citenamefont {Guti{\'e}rrez}\ and\ \citenamefont
  {Cabrera}(2015)}]{gutierrez2015neural}%
  \BibitemOpen
  \bibfield  {author} {\bibinfo {author} {\bibfnamefont {E.~D.}\ \bibnamefont
  {Guti{\'e}rrez}}\ and\ \bibinfo {author} {\bibfnamefont {J.~L.}\ \bibnamefont
  {Cabrera}},\ }\bibfield  {title} {\bibinfo {title} {A neural coding scheme
  reproducing foraging trajectories},\ }\href@noop {} {\bibfield  {journal}
  {\bibinfo  {journal} {Scientific Reports}\ }\textbf {\bibinfo {volume} {5}},\
  \bibinfo {pages} {18009} (\bibinfo {year} {2015})}\BibitemShut {NoStop}%
\bibitem [{\citenamefont {Ikeda}\ \emph {et~al.}(2020)\citenamefont {Ikeda},
  \citenamefont {Jurica}, \citenamefont {Kimura}, \citenamefont {Takagi},
  \citenamefont {Struzik}, \citenamefont {Kiyono}, \citenamefont {Arata},\ and\
  \citenamefont {Sako}}]{ikeda2020c}%
  \BibitemOpen
  \bibfield  {author} {\bibinfo {author} {\bibfnamefont {Y.}~\bibnamefont
  {Ikeda}}, \bibinfo {author} {\bibfnamefont {P.}~\bibnamefont {Jurica}},
  \bibinfo {author} {\bibfnamefont {H.}~\bibnamefont {Kimura}}, \bibinfo
  {author} {\bibfnamefont {H.}~\bibnamefont {Takagi}}, \bibinfo {author}
  {\bibfnamefont {Z.~R.}\ \bibnamefont {Struzik}}, \bibinfo {author}
  {\bibfnamefont {K.}~\bibnamefont {Kiyono}}, \bibinfo {author} {\bibfnamefont
  {Y.}~\bibnamefont {Arata}},\ and\ \bibinfo {author} {\bibfnamefont
  {Y.}~\bibnamefont {Sako}},\ }\bibfield  {title} {\bibinfo {title} {\textit{C.
  elegans} episodic swimming is driven by multifractal kinetics},\ }\href@noop
  {} {\bibfield  {journal} {\bibinfo  {journal} {Scientific Reports}\ }\textbf
  {\bibinfo {volume} {10}},\ \bibinfo {pages} {14775} (\bibinfo {year}
  {2020})}\BibitemShut {NoStop}%
\bibitem [{\citenamefont {Schmitt}\ and\ \citenamefont
  {Seuront}(2001)}]{schmitt2001multifractal}%
  \BibitemOpen
  \bibfield  {author} {\bibinfo {author} {\bibfnamefont {F.~G.}\ \bibnamefont
  {Schmitt}}\ and\ \bibinfo {author} {\bibfnamefont {L.}~\bibnamefont
  {Seuront}},\ }\bibfield  {title} {\bibinfo {title} {Multifractal random walk
  in copepod behavior},\ }\href@noop {} {\bibfield  {journal} {\bibinfo
  {journal} {Physica A: Statistical Mechanics and its Applications}\ }\textbf
  {\bibinfo {volume} {301}},\ \bibinfo {pages} {375} (\bibinfo {year}
  {2001})}\BibitemShut {NoStop}%
\bibitem [{\citenamefont {Seuront}\ and\ \citenamefont
  {Stanley}(2014)}]{seuront2014anomalous}%
  \BibitemOpen
  \bibfield  {author} {\bibinfo {author} {\bibfnamefont {L.}~\bibnamefont
  {Seuront}}\ and\ \bibinfo {author} {\bibfnamefont {H.~E.}\ \bibnamefont
  {Stanley}},\ }\bibfield  {title} {\bibinfo {title} {Anomalous diffusion and
  multifractality enhance mating encounters in the ocean},\ }\href@noop {}
  {\bibfield  {journal} {\bibinfo  {journal} {Proceedings of the National
  Academy of Sciences}\ }\textbf {\bibinfo {volume} {111}},\ \bibinfo {pages}
  {2206} (\bibinfo {year} {2014})}\BibitemShut {NoStop}%
\bibitem [{\citenamefont {Balaban}\ \emph {et~al.}(2018)\citenamefont
  {Balaban}, \citenamefont {Lim}, \citenamefont {Gupta}, \citenamefont
  {Boedicker},\ and\ \citenamefont {Bogdan}}]{balaban2018quantifying}%
  \BibitemOpen
  \bibfield  {author} {\bibinfo {author} {\bibfnamefont {V.}~\bibnamefont
  {Balaban}}, \bibinfo {author} {\bibfnamefont {S.}~\bibnamefont {Lim}},
  \bibinfo {author} {\bibfnamefont {G.}~\bibnamefont {Gupta}}, \bibinfo
  {author} {\bibfnamefont {J.}~\bibnamefont {Boedicker}},\ and\ \bibinfo
  {author} {\bibfnamefont {P.}~\bibnamefont {Bogdan}},\ }\bibfield  {title}
  {\bibinfo {title} {Quantifying emergence and self-organisation of
  \textit{Enterobacter cloacae} microbial communities},\ }\href@noop {}
  {\bibfield  {journal} {\bibinfo  {journal} {Scientific Reports}\ }\textbf
  {\bibinfo {volume} {8}},\ \bibinfo {pages} {12416} (\bibinfo {year}
  {2018})}\BibitemShut {NoStop}%
\bibitem [{\citenamefont {Koorehdavoudi}\ \emph {et~al.}(2017)\citenamefont
  {Koorehdavoudi}, \citenamefont {Bogdan}, \citenamefont {Wei}, \citenamefont
  {Marculescu}, \citenamefont {Zhuang}, \citenamefont {Carlsen},\ and\
  \citenamefont {Sitti}}]{koorehdavoudi2017multi}%
  \BibitemOpen
  \bibfield  {author} {\bibinfo {author} {\bibfnamefont {H.}~\bibnamefont
  {Koorehdavoudi}}, \bibinfo {author} {\bibfnamefont {P.}~\bibnamefont
  {Bogdan}}, \bibinfo {author} {\bibfnamefont {G.}~\bibnamefont {Wei}},
  \bibinfo {author} {\bibfnamefont {R.}~\bibnamefont {Marculescu}}, \bibinfo
  {author} {\bibfnamefont {J.}~\bibnamefont {Zhuang}}, \bibinfo {author}
  {\bibfnamefont {R.~W.}\ \bibnamefont {Carlsen}},\ and\ \bibinfo {author}
  {\bibfnamefont {M.}~\bibnamefont {Sitti}},\ }\bibfield  {title} {\bibinfo
  {title} {Multi-fractal characterization of bacterial swimming dynamics: {A}
  case study on real and simulated \textit{Serratia marcescens}},\ }\href@noop
  {} {\bibfield  {journal} {\bibinfo  {journal} {Proceedings of the Royal
  Society A: Mathematical, Physical and Engineering Sciences}\ }\textbf
  {\bibinfo {volume} {473}},\ \bibinfo {pages} {20170154} (\bibinfo {year}
  {2017})}\BibitemShut {NoStop}%
\bibitem [{\citenamefont {Jeon}\ and\ \citenamefont
  {Metzler}(2010{\natexlab{a}})}]{jeon2010analysis}%
  \BibitemOpen
  \bibfield  {author} {\bibinfo {author} {\bibfnamefont {J.-H.}\ \bibnamefont
  {Jeon}}\ and\ \bibinfo {author} {\bibfnamefont {R.}~\bibnamefont {Metzler}},\
  }\bibfield  {title} {\bibinfo {title} {Analysis of short subdiffusive time
  series: {S}catter of the time-averaged mean-squared displacement},\
  }\href@noop {} {\bibfield  {journal} {\bibinfo  {journal} {Journal of Physics
  A: Mathematical and Theoretical}\ }\textbf {\bibinfo {volume} {43}},\
  \bibinfo {pages} {252001} (\bibinfo {year} {2010}{\natexlab{a}})}\BibitemShut
  {NoStop}%
\bibitem [{\citenamefont {Jeon}\ \emph {et~al.}(2014)\citenamefont {Jeon},
  \citenamefont {Chechkin},\ and\ \citenamefont {Metzler}}]{jeon2014scaled}%
  \BibitemOpen
  \bibfield  {author} {\bibinfo {author} {\bibfnamefont {J.-H.}\ \bibnamefont
  {Jeon}}, \bibinfo {author} {\bibfnamefont {A.~V.}\ \bibnamefont {Chechkin}},\
  and\ \bibinfo {author} {\bibfnamefont {R.}~\bibnamefont {Metzler}},\
  }\bibfield  {title} {\bibinfo {title} {Scaled {B}rownian motion: {A}
  paradoxical process with a time dependent diffusivity for the description of
  anomalous diffusion},\ }\href@noop {} {\bibfield  {journal} {\bibinfo
  {journal} {Physical Chemistry Chemical Physics}\ }\textbf {\bibinfo {volume}
  {16}},\ \bibinfo {pages} {15811} (\bibinfo {year} {2014})}\BibitemShut
  {NoStop}%
\bibitem [{\citenamefont {Cherstvy}\ and\ \citenamefont
  {Metzler}(2015)}]{cherstvy2015ergodicity}%
  \BibitemOpen
  \bibfield  {author} {\bibinfo {author} {\bibfnamefont {A.~G.}\ \bibnamefont
  {Cherstvy}}\ and\ \bibinfo {author} {\bibfnamefont {R.}~\bibnamefont
  {Metzler}},\ }\bibfield  {title} {\bibinfo {title} {Ergodicity breaking,
  ageing, and confinement in generalized diffusion processes with position and
  time dependent diffusivity},\ }\href@noop {} {\bibfield  {journal} {\bibinfo
  {journal} {Journal of Statistical Mechanics: Theory and Experiment}\ }\textbf
  {\bibinfo {volume} {2015}},\ \bibinfo {pages} {P05010} (\bibinfo {year}
  {2015})}\BibitemShut {NoStop}%
\bibitem [{\citenamefont {Wang}\ \emph {et~al.}(2020)\citenamefont {Wang},
  \citenamefont {Cherstvy}, \citenamefont {Liu},\ and\ \citenamefont
  {Metzler}}]{wang2020anomalous}%
  \BibitemOpen
  \bibfield  {author} {\bibinfo {author} {\bibfnamefont {W.}~\bibnamefont
  {Wang}}, \bibinfo {author} {\bibfnamefont {A.~G.}\ \bibnamefont {Cherstvy}},
  \bibinfo {author} {\bibfnamefont {X.}~\bibnamefont {Liu}},\ and\ \bibinfo
  {author} {\bibfnamefont {R.}~\bibnamefont {Metzler}},\ }\bibfield  {title}
  {\bibinfo {title} {Anomalous diffusion and nonergodicity for heterogeneous
  diffusion processes with fractional {G}aussian noise},\ }\href@noop {}
  {\bibfield  {journal} {\bibinfo  {journal} {Physical Review E}\ }\textbf
  {\bibinfo {volume} {102}},\ \bibinfo {pages} {012146} (\bibinfo {year}
  {2020})}\BibitemShut {NoStop}%
\bibitem [{\citenamefont {Wang}\ \emph
  {et~al.}(2022{\natexlab{b}})\citenamefont {Wang}, \citenamefont {Metzler},\
  and\ \citenamefont {Cherstvy}}]{wang2022anomalous}%
  \BibitemOpen
  \bibfield  {author} {\bibinfo {author} {\bibfnamefont {W.}~\bibnamefont
  {Wang}}, \bibinfo {author} {\bibfnamefont {R.}~\bibnamefont {Metzler}},\ and\
  \bibinfo {author} {\bibfnamefont {A.~G.}\ \bibnamefont {Cherstvy}},\
  }\bibfield  {title} {\bibinfo {title} {Anomalous diffusion, aging, and
  nonergodicity of scaled {B}rownian motion with fractional {G}aussian noise:
  {O}verview of related experimental observations and models},\ }\href@noop {}
  {\bibfield  {journal} {\bibinfo  {journal} {Physical Chemistry Chemical
  Physics}\ }\textbf {\bibinfo {volume} {24}},\ \bibinfo {pages} {18482}
  (\bibinfo {year} {2022}{\natexlab{b}})}\BibitemShut {NoStop}%
\bibitem [{\citenamefont {Lovejoy}\ \emph {et~al.}(1998)\citenamefont
  {Lovejoy}, \citenamefont {Schertzer},\ and\ \citenamefont
  {Silas}}]{lovejoy1998diffusion}%
  \BibitemOpen
  \bibfield  {author} {\bibinfo {author} {\bibfnamefont {S.}~\bibnamefont
  {Lovejoy}}, \bibinfo {author} {\bibfnamefont {D.}~\bibnamefont {Schertzer}},\
  and\ \bibinfo {author} {\bibfnamefont {P.}~\bibnamefont {Silas}},\ }\bibfield
   {title} {\bibinfo {title} {Diffusion in one-dimensional multifractal porous
  media},\ }\href@noop {} {\bibfield  {journal} {\bibinfo  {journal} {Water
  Resources Research}\ }\textbf {\bibinfo {volume} {34}},\ \bibinfo {pages}
  {3283} (\bibinfo {year} {1998})}\BibitemShut {NoStop}%
\bibitem [{\citenamefont {Mangalam}\ \emph
  {et~al.}(2023{\natexlab{c}})\citenamefont {Mangalam}, \citenamefont
  {Likens},\ and\ \citenamefont
  {Kelty-Stephen}}]{mangalam2023multifractalnonlinearity}%
  \BibitemOpen
  \bibfield  {author} {\bibinfo {author} {\bibfnamefont {M.}~\bibnamefont
  {Mangalam}}, \bibinfo {author} {\bibfnamefont {A.~D.}\ \bibnamefont
  {Likens}},\ and\ \bibinfo {author} {\bibfnamefont {D.~G.}\ \bibnamefont
  {Kelty-Stephen}},\ }\bibfield  {title} {\bibinfo {title} {Multifractal
  nonlinearity as a robust estimator of multiplicative cascade dynamics},\
  }\href@noop {} {\bibfield  {journal} {\bibinfo  {journal} {arXiv preprint
  arXiv:2312.05653}\ } (\bibinfo {year} {2023}{\natexlab{c}})}\BibitemShut
  {NoStop}%
\bibitem [{\citenamefont {Mangalam}\ and\ \citenamefont
  {Kelty-Stephen}(2024)}]{mangalam2024multifractal}%
  \BibitemOpen
  \bibfield  {author} {\bibinfo {author} {\bibfnamefont {M.}~\bibnamefont
  {Mangalam}}\ and\ \bibinfo {author} {\bibfnamefont {D.~G.}\ \bibnamefont
  {Kelty-Stephen}},\ }\bibfield  {title} {\bibinfo {title} {Multifractal
  emergent processes: Multiplicative interactions override nonlinear component
  properties},\ }\href@noop {} {\bibfield  {journal} {\bibinfo  {journal}
  {arXiv preprint arXiv:2401.05105}\ } (\bibinfo {year} {2024})}\BibitemShut
  {NoStop}%
\bibitem [{\citenamefont {Cherstvy}\ \emph
  {et~al.}(2021{\natexlab{b}})\citenamefont {Cherstvy}, \citenamefont {Wang},
  \citenamefont {Metzler},\ and\ \citenamefont
  {Sokolov}}]{cherstvy2021inertia}%
  \BibitemOpen
  \bibfield  {author} {\bibinfo {author} {\bibfnamefont {A.~G.}\ \bibnamefont
  {Cherstvy}}, \bibinfo {author} {\bibfnamefont {W.}~\bibnamefont {Wang}},
  \bibinfo {author} {\bibfnamefont {R.}~\bibnamefont {Metzler}},\ and\ \bibinfo
  {author} {\bibfnamefont {I.~M.}\ \bibnamefont {Sokolov}},\ }\bibfield
  {title} {\bibinfo {title} {Inertia triggers nonergodicity of fractional
  {B}rownian motion},\ }\href@noop {} {\bibfield  {journal} {\bibinfo
  {journal} {Physical Review E}\ }\textbf {\bibinfo {volume} {104}},\ \bibinfo
  {pages} {024115} (\bibinfo {year} {2021}{\natexlab{b}})}\BibitemShut
  {NoStop}%
\bibitem [{\citenamefont {Afanasiev}\ \emph {et~al.}(1991)\citenamefont
  {Afanasiev}, \citenamefont {Sagdeev},\ and\ \citenamefont
  {Zaslavsky}}]{afanasiev1991chaotic}%
  \BibitemOpen
  \bibfield  {author} {\bibinfo {author} {\bibfnamefont {V.~V.}\ \bibnamefont
  {Afanasiev}}, \bibinfo {author} {\bibfnamefont {R.~Z.}\ \bibnamefont
  {Sagdeev}},\ and\ \bibinfo {author} {\bibfnamefont {G.~M.}\ \bibnamefont
  {Zaslavsky}},\ }\bibfield  {title} {\bibinfo {title} {Chaotic jets with
  multifractal space-time random walk},\ }\href@noop {} {\bibfield  {journal}
  {\bibinfo  {journal} {Chaos}\ }\textbf {\bibinfo {volume} {1}},\ \bibinfo
  {pages} {143} (\bibinfo {year} {1991})}\BibitemShut {NoStop}%
\bibitem [{\citenamefont {Chen}\ \emph {et~al.}(2010)\citenamefont {Chen},
  \citenamefont {Sun}, \citenamefont {Zhang},\ and\ \citenamefont
  {Koro{\v{s}}ak}}]{chen2010anomalous}%
  \BibitemOpen
  \bibfield  {author} {\bibinfo {author} {\bibfnamefont {W.}~\bibnamefont
  {Chen}}, \bibinfo {author} {\bibfnamefont {H.}~\bibnamefont {Sun}}, \bibinfo
  {author} {\bibfnamefont {X.}~\bibnamefont {Zhang}},\ and\ \bibinfo {author}
  {\bibfnamefont {D.}~\bibnamefont {Koro{\v{s}}ak}},\ }\bibfield  {title}
  {\bibinfo {title} {Anomalous diffusion modeling by fractal and fractional
  derivatives},\ }\href@noop {} {\bibfield  {journal} {\bibinfo  {journal}
  {Computers \& Mathematics with Applications}\ }\textbf {\bibinfo {volume}
  {59}},\ \bibinfo {pages} {1754} (\bibinfo {year} {2010})}\BibitemShut
  {NoStop}%
\bibitem [{\citenamefont {de~Dreuzy}\ \emph {et~al.}(2004)\citenamefont
  {de~Dreuzy}, \citenamefont {Davy}, \citenamefont {Erhel},\ and\ \citenamefont
  {de~Br{\'e}mond~d’Ars}}]{de2004anomalous}%
  \BibitemOpen
  \bibfield  {author} {\bibinfo {author} {\bibfnamefont {J.-R.}\ \bibnamefont
  {de~Dreuzy}}, \bibinfo {author} {\bibfnamefont {P.}~\bibnamefont {Davy}},
  \bibinfo {author} {\bibfnamefont {J.}~\bibnamefont {Erhel}},\ and\ \bibinfo
  {author} {\bibfnamefont {J.}~\bibnamefont {de~Br{\'e}mond~d’Ars}},\
  }\bibfield  {title} {\bibinfo {title} {Anomalous diffusion exponents in
  continuous two-dimensional multifractal media},\ }\href@noop {} {\bibfield
  {journal} {\bibinfo  {journal} {Physical Review E}\ }\textbf {\bibinfo
  {volume} {70}},\ \bibinfo {pages} {016306} (\bibinfo {year}
  {2004})}\BibitemShut {NoStop}%
\bibitem [{\citenamefont {Gmachowski}(2015)}]{gmachowski2015fractal}%
  \BibitemOpen
  \bibfield  {author} {\bibinfo {author} {\bibfnamefont {L.}~\bibnamefont
  {Gmachowski}},\ }\bibfield  {title} {\bibinfo {title} {Fractal model of
  anomalous diffusion},\ }\href@noop {} {\bibfield  {journal} {\bibinfo
  {journal} {European Biophysics Journal}\ }\textbf {\bibinfo {volume} {44}},\
  \bibinfo {pages} {613} (\bibinfo {year} {2015})}\BibitemShut {NoStop}%
\bibitem [{\citenamefont {Lim}\ and\ \citenamefont
  {Muniandy}(2002)}]{lim2002self}%
  \BibitemOpen
  \bibfield  {author} {\bibinfo {author} {\bibfnamefont {S.}~\bibnamefont
  {Lim}}\ and\ \bibinfo {author} {\bibfnamefont {S.}~\bibnamefont {Muniandy}},\
  }\bibfield  {title} {\bibinfo {title} {Self-similar gaussian processes for
  modeling anomalous diffusion},\ }\href@noop {} {\bibfield  {journal}
  {\bibinfo  {journal} {Physical Review E}\ }\textbf {\bibinfo {volume} {66}},\
  \bibinfo {pages} {021114} (\bibinfo {year} {2002})}\BibitemShut {NoStop}%
\bibitem [{\citenamefont {Bickel}(1999)}]{bickel1999simple}%
  \BibitemOpen
  \bibfield  {author} {\bibinfo {author} {\bibfnamefont {D.~R.}\ \bibnamefont
  {Bickel}},\ }\bibfield  {title} {\bibinfo {title} {Simple estimation of
  intermittency in multifractal stochastic processes: {B}iomedical
  applications},\ }\href@noop {} {\bibfield  {journal} {\bibinfo  {journal}
  {Physics Letters A}\ }\textbf {\bibinfo {volume} {262}},\ \bibinfo {pages}
  {251} (\bibinfo {year} {1999})}\BibitemShut {NoStop}%
\bibitem [{\citenamefont {Lovejoy}\ and\ \citenamefont
  {Schertzer}(2018)}]{lovejoy2018weather}%
  \BibitemOpen
  \bibfield  {author} {\bibinfo {author} {\bibfnamefont {S.}~\bibnamefont
  {Lovejoy}}\ and\ \bibinfo {author} {\bibfnamefont {D.}~\bibnamefont
  {Schertzer}},\ }\href@noop {} {\emph {\bibinfo {title} {The {W}eather and
  {C}limate: {E}mergent {L}aws and {M}ultifractal {C}ascades}}}\ (\bibinfo
  {publisher} {Cambridge University Press, Cambridge, MA},\ \bibinfo {year}
  {2018})\BibitemShut {NoStop}%
\bibitem [{\citenamefont {Menu}\ and\ \citenamefont
  {Roscilde}(2020)}]{menu2020anomalous}%
  \BibitemOpen
  \bibfield  {author} {\bibinfo {author} {\bibfnamefont {R.}~\bibnamefont
  {Menu}}\ and\ \bibinfo {author} {\bibfnamefont {T.}~\bibnamefont
  {Roscilde}},\ }\bibfield  {title} {\bibinfo {title} {Anomalous diffusion and
  localization in a positionally disordered quantum spin array},\ }\href@noop
  {} {\bibfield  {journal} {\bibinfo  {journal} {Physical Review Letters}\
  }\textbf {\bibinfo {volume} {124}},\ \bibinfo {pages} {130604} (\bibinfo
  {year} {2020})}\BibitemShut {NoStop}%
\bibitem [{\citenamefont {Seuront}\ \emph {et~al.}(2004)\citenamefont
  {Seuront}, \citenamefont {Schmitt}, \citenamefont {Brewer}, \citenamefont
  {Strickler},\ and\ \citenamefont {Souissi}}]{seuront2004random}%
  \BibitemOpen
  \bibfield  {author} {\bibinfo {author} {\bibfnamefont {L.}~\bibnamefont
  {Seuront}}, \bibinfo {author} {\bibfnamefont {F.~G.}\ \bibnamefont
  {Schmitt}}, \bibinfo {author} {\bibfnamefont {M.~C.}\ \bibnamefont {Brewer}},
  \bibinfo {author} {\bibfnamefont {J.~R.}\ \bibnamefont {Strickler}},\ and\
  \bibinfo {author} {\bibfnamefont {S.}~\bibnamefont {Souissi}},\ }\bibfield
  {title} {\bibinfo {title} {From random walk to multifractal random walk in
  zooplankton swimming behavior},\ }\href@noop {} {\bibfield  {journal}
  {\bibinfo  {journal} {Zoological Studies}\ }\textbf {\bibinfo {volume}
  {43}},\ \bibinfo {pages} {498} (\bibinfo {year} {2004})}\BibitemShut
  {NoStop}%
\bibitem [{\citenamefont {Sharifi-Viand}\ \emph {et~al.}(2012)\citenamefont
  {Sharifi-Viand}, \citenamefont {Mahjani},\ and\ \citenamefont
  {Jafarian}}]{sharifi2012investigation}%
  \BibitemOpen
  \bibfield  {author} {\bibinfo {author} {\bibfnamefont {A.}~\bibnamefont
  {Sharifi-Viand}}, \bibinfo {author} {\bibfnamefont {M.}~\bibnamefont
  {Mahjani}},\ and\ \bibinfo {author} {\bibfnamefont {M.}~\bibnamefont
  {Jafarian}},\ }\bibfield  {title} {\bibinfo {title} {Investigation of
  anomalous diffusion and multifractal dimensions in polypyrrole film},\
  }\href@noop {} {\bibfield  {journal} {\bibinfo  {journal} {Journal of
  Electroanalytical Chemistry}\ }\textbf {\bibinfo {volume} {671}},\ \bibinfo
  {pages} {51} (\bibinfo {year} {2012})}\BibitemShut {NoStop}%
\bibitem [{\citenamefont {Shaebani}\ \emph {et~al.}(2020)\citenamefont
  {Shaebani}, \citenamefont {Wysocki}, \citenamefont {Winkler}, \citenamefont
  {Gompper},\ and\ \citenamefont {Rieger}}]{shaebani2020computational}%
  \BibitemOpen
  \bibfield  {author} {\bibinfo {author} {\bibfnamefont {M.~R.}\ \bibnamefont
  {Shaebani}}, \bibinfo {author} {\bibfnamefont {A.}~\bibnamefont {Wysocki}},
  \bibinfo {author} {\bibfnamefont {R.~G.}\ \bibnamefont {Winkler}}, \bibinfo
  {author} {\bibfnamefont {G.}~\bibnamefont {Gompper}},\ and\ \bibinfo {author}
  {\bibfnamefont {H.}~\bibnamefont {Rieger}},\ }\bibfield  {title} {\bibinfo
  {title} {Computational models for active matter},\ }\href@noop {} {\bibfield
  {journal} {\bibinfo  {journal} {Nature Reviews Physics}\ }\textbf {\bibinfo
  {volume} {2}},\ \bibinfo {pages} {181} (\bibinfo {year} {2020})}\BibitemShut
  {NoStop}%
\bibitem [{\citenamefont {Metzler}\ and\ \citenamefont
  {Klafter}(2004)}]{metzler2004restaurant}%
  \BibitemOpen
  \bibfield  {author} {\bibinfo {author} {\bibfnamefont {R.}~\bibnamefont
  {Metzler}}\ and\ \bibinfo {author} {\bibfnamefont {J.}~\bibnamefont
  {Klafter}},\ }\bibfield  {title} {\bibinfo {title} {The restaurant at the end
  of the random walk: {R}ecent developments in the description of anomalous
  transport by fractional dynamics},\ }\href@noop {} {\bibfield  {journal}
  {\bibinfo  {journal} {Journal of Physics A: Mathematical and General}\
  }\textbf {\bibinfo {volume} {37}},\ \bibinfo {pages} {R161} (\bibinfo {year}
  {2004})}\BibitemShut {NoStop}%
\bibitem [{\citenamefont {Metzler}\ \emph {et~al.}(2014)\citenamefont
  {Metzler}, \citenamefont {Jeon}, \citenamefont {Cherstvy},\ and\
  \citenamefont {Barkai}}]{metzler2014anomalous}%
  \BibitemOpen
  \bibfield  {author} {\bibinfo {author} {\bibfnamefont {R.}~\bibnamefont
  {Metzler}}, \bibinfo {author} {\bibfnamefont {J.-H.}\ \bibnamefont {Jeon}},
  \bibinfo {author} {\bibfnamefont {A.~G.}\ \bibnamefont {Cherstvy}},\ and\
  \bibinfo {author} {\bibfnamefont {E.}~\bibnamefont {Barkai}},\ }\bibfield
  {title} {\bibinfo {title} {Anomalous diffusion models and their properties:
  {N}on-stationarity, non-ergodicity, and ageing at the centenary of single
  particle tracking},\ }\href@noop {} {\bibfield  {journal} {\bibinfo
  {journal} {Physical Chemistry Chemical Physics}\ }\textbf {\bibinfo {volume}
  {16}},\ \bibinfo {pages} {24128} (\bibinfo {year} {2014})}\BibitemShut
  {NoStop}%
\bibitem [{\citenamefont {Stephen}\ \emph {et~al.}(2012)\citenamefont
  {Stephen}, \citenamefont {Hsu}, \citenamefont {Young}, \citenamefont
  {Saltzman}, \citenamefont {Holt}, \citenamefont {Newman}, \citenamefont
  {Weinberg}, \citenamefont {Wood}, \citenamefont {Nagpal},\ and\ \citenamefont
  {Goldfield}}]{stephen2012multifractal}%
  \BibitemOpen
  \bibfield  {author} {\bibinfo {author} {\bibfnamefont {D.~G.}\ \bibnamefont
  {Stephen}}, \bibinfo {author} {\bibfnamefont {W.-H.}\ \bibnamefont {Hsu}},
  \bibinfo {author} {\bibfnamefont {D.}~\bibnamefont {Young}}, \bibinfo
  {author} {\bibfnamefont {E.~L.}\ \bibnamefont {Saltzman}}, \bibinfo {author}
  {\bibfnamefont {K.~G.}\ \bibnamefont {Holt}}, \bibinfo {author}
  {\bibfnamefont {D.~J.}\ \bibnamefont {Newman}}, \bibinfo {author}
  {\bibfnamefont {M.}~\bibnamefont {Weinberg}}, \bibinfo {author}
  {\bibfnamefont {R.~J.}\ \bibnamefont {Wood}}, \bibinfo {author}
  {\bibfnamefont {R.}~\bibnamefont {Nagpal}},\ and\ \bibinfo {author}
  {\bibfnamefont {E.~C.}\ \bibnamefont {Goldfield}},\ }\bibfield  {title}
  {\bibinfo {title} {Multifractal fluctuations in joint angles during infant
  spontaneous kicking reveal multiplicativity-driven coordination},\
  }\href@noop {} {\bibfield  {journal} {\bibinfo  {journal} {Chaos, Solitons \&
  Fractals}\ }\textbf {\bibinfo {volume} {45}},\ \bibinfo {pages} {1201}
  (\bibinfo {year} {2012})}\BibitemShut {NoStop}%
\bibitem [{\citenamefont {Humeau}\ \emph {et~al.}(2008)\citenamefont {Humeau},
  \citenamefont {Chapeau-Blondeau}, \citenamefont {Rousseau}, \citenamefont
  {Rousseau}, \citenamefont {Trzepizur},\ and\ \citenamefont
  {Abraham}}]{humeau2008multifractality}%
  \BibitemOpen
  \bibfield  {author} {\bibinfo {author} {\bibfnamefont {A.}~\bibnamefont
  {Humeau}}, \bibinfo {author} {\bibfnamefont {F.}~\bibnamefont
  {Chapeau-Blondeau}}, \bibinfo {author} {\bibfnamefont {D.}~\bibnamefont
  {Rousseau}}, \bibinfo {author} {\bibfnamefont {P.}~\bibnamefont {Rousseau}},
  \bibinfo {author} {\bibfnamefont {W.}~\bibnamefont {Trzepizur}},\ and\
  \bibinfo {author} {\bibfnamefont {P.}~\bibnamefont {Abraham}},\ }\bibfield
  {title} {\bibinfo {title} {Multifractality, sample entropy, and wavelet
  analyses for age-related changes in the peripheral cardiovascular system:
  {P}reliminary results},\ }\href@noop {} {\bibfield  {journal} {\bibinfo
  {journal} {Medical Physics}\ }\textbf {\bibinfo {volume} {35}},\ \bibinfo
  {pages} {717} (\bibinfo {year} {2008})}\BibitemShut {NoStop}%
\bibitem [{\citenamefont {Goldberger}\ \emph {et~al.}(2002)\citenamefont
  {Goldberger}, \citenamefont {Amaral}, \citenamefont {Hausdorff},
  \citenamefont {Ivanov}, \citenamefont {Peng},\ and\ \citenamefont
  {Stanley}}]{goldberger2002fractal}%
  \BibitemOpen
  \bibfield  {author} {\bibinfo {author} {\bibfnamefont {A.~L.}\ \bibnamefont
  {Goldberger}}, \bibinfo {author} {\bibfnamefont {L.~A.}\ \bibnamefont
  {Amaral}}, \bibinfo {author} {\bibfnamefont {J.~M.}\ \bibnamefont
  {Hausdorff}}, \bibinfo {author} {\bibfnamefont {P.~C.}\ \bibnamefont
  {Ivanov}}, \bibinfo {author} {\bibfnamefont {C.-K.}\ \bibnamefont {Peng}},\
  and\ \bibinfo {author} {\bibfnamefont {H.~E.}\ \bibnamefont {Stanley}},\
  }\bibfield  {title} {\bibinfo {title} {Fractal dynamics in physiology:
  {A}lterations with disease and aging},\ }\href@noop {} {\bibfield  {journal}
  {\bibinfo  {journal} {Proceedings of the National Academy of Sciences}\
  }\textbf {\bibinfo {volume} {99}},\ \bibinfo {pages} {2466} (\bibinfo {year}
  {2002})}\BibitemShut {NoStop}%
\bibitem [{\citenamefont {Suckling}\ \emph {et~al.}(2008)\citenamefont
  {Suckling}, \citenamefont {Wink}, \citenamefont {Bernard}, \citenamefont
  {Barnes},\ and\ \citenamefont {Bullmore}}]{suckling2008endogenous}%
  \BibitemOpen
  \bibfield  {author} {\bibinfo {author} {\bibfnamefont {J.}~\bibnamefont
  {Suckling}}, \bibinfo {author} {\bibfnamefont {A.~M.}\ \bibnamefont {Wink}},
  \bibinfo {author} {\bibfnamefont {F.~A.}\ \bibnamefont {Bernard}}, \bibinfo
  {author} {\bibfnamefont {A.}~\bibnamefont {Barnes}},\ and\ \bibinfo {author}
  {\bibfnamefont {E.}~\bibnamefont {Bullmore}},\ }\bibfield  {title} {\bibinfo
  {title} {Endogenous multifractal brain dynamics are modulated by age,
  cholinergic blockade and cognitive performance},\ }\href@noop {} {\bibfield
  {journal} {\bibinfo  {journal} {Journal of Neuroscience Methods}\ }\textbf
  {\bibinfo {volume} {174}},\ \bibinfo {pages} {292} (\bibinfo {year}
  {2008})}\BibitemShut {NoStop}%
\bibitem [{\citenamefont {West}\ \emph {et~al.}(2003)\citenamefont {West},
  \citenamefont {Latka}, \citenamefont {Glaubic-Latka},\ and\ \citenamefont
  {Latka}}]{west2003multifractality}%
  \BibitemOpen
  \bibfield  {author} {\bibinfo {author} {\bibfnamefont {B.~J.}\ \bibnamefont
  {West}}, \bibinfo {author} {\bibfnamefont {M.}~\bibnamefont {Latka}},
  \bibinfo {author} {\bibfnamefont {M.}~\bibnamefont {Glaubic-Latka}},\ and\
  \bibinfo {author} {\bibfnamefont {D.}~\bibnamefont {Latka}},\ }\bibfield
  {title} {\bibinfo {title} {Multifractality of cerebral blood flow},\
  }\href@noop {} {\bibfield  {journal} {\bibinfo  {journal} {Physica A:
  Statistical Mechanics and its Applications}\ }\textbf {\bibinfo {volume}
  {318}},\ \bibinfo {pages} {453} (\bibinfo {year} {2003})}\BibitemShut
  {NoStop}%
\bibitem [{\citenamefont {Jeon}\ and\ \citenamefont
  {Metzler}(2010{\natexlab{b}})}]{jeon2010fractional}%
  \BibitemOpen
  \bibfield  {author} {\bibinfo {author} {\bibfnamefont {J.-H.}\ \bibnamefont
  {Jeon}}\ and\ \bibinfo {author} {\bibfnamefont {R.}~\bibnamefont {Metzler}},\
  }\bibfield  {title} {\bibinfo {title} {Fractional {B}rownian motion and
  motion governed by the fractional {L}angevin equation in confined
  geometries},\ }\href@noop {} {\bibfield  {journal} {\bibinfo  {journal}
  {Physical Review E}\ }\textbf {\bibinfo {volume} {81}},\ \bibinfo {pages}
  {021103} (\bibinfo {year} {2010}{\natexlab{b}})}\BibitemShut {NoStop}%
\bibitem [{\citenamefont {Kolmogorov}(1940)}]{kolmogorov1940wiener}%
  \BibitemOpen
  \bibfield  {author} {\bibinfo {author} {\bibfnamefont {A.~N.}\ \bibnamefont
  {Kolmogorov}},\ }\bibfield  {title} {\bibinfo {title} {The wiener spiral and
  some other interesting curves in hilbert space},\ }\href@noop {} {\bibfield
  {journal} {\bibinfo  {journal} {Doklady Akademii Nauk SSSR}\ }\textbf
  {\bibinfo {volume} {26}},\ \bibinfo {pages} {115} (\bibinfo {year}
  {1940})}\BibitemShut {NoStop}%
\bibitem [{\citenamefont {Mandelbrot}\ and\ \citenamefont
  {Van~Ness}(1968)}]{mandelbrot1968fractional}%
  \BibitemOpen
  \bibfield  {author} {\bibinfo {author} {\bibfnamefont {B.~B.}\ \bibnamefont
  {Mandelbrot}}\ and\ \bibinfo {author} {\bibfnamefont {J.~W.}\ \bibnamefont
  {Van~Ness}},\ }\bibfield  {title} {\bibinfo {title} {Fractional {B}rownian
  motions, fractional noises and applications},\ }\href@noop {} {\bibfield
  {journal} {\bibinfo  {journal} {SIAM Review}\ }\textbf {\bibinfo {volume}
  {10}},\ \bibinfo {pages} {422} (\bibinfo {year} {1968})}\BibitemShut
  {NoStop}%
\bibitem [{\citenamefont {Davies}\ and\ \citenamefont
  {Harte}(1987)}]{davies1987tests}%
  \BibitemOpen
  \bibfield  {author} {\bibinfo {author} {\bibfnamefont {R.~B.}\ \bibnamefont
  {Davies}}\ and\ \bibinfo {author} {\bibfnamefont {D.~S.}\ \bibnamefont
  {Harte}},\ }\bibfield  {title} {\bibinfo {title} {Tests for {H}urst effect},\
  }\href@noop {} {\bibfield  {journal} {\bibinfo  {journal} {Biometrika}\
  }\textbf {\bibinfo {volume} {74}},\ \bibinfo {pages} {95} (\bibinfo {year}
  {1987})}\BibitemShut {NoStop}%
\bibitem [{\citenamefont {Hosking}(1984)}]{hosking1984modeling}%
  \BibitemOpen
  \bibfield  {author} {\bibinfo {author} {\bibfnamefont {J.~R.}\ \bibnamefont
  {Hosking}},\ }\bibfield  {title} {\bibinfo {title} {Modeling persistence in
  hydrological time series using fractional differencing},\ }\href@noop {}
  {\bibfield  {journal} {\bibinfo  {journal} {Water Resources Research}\
  }\textbf {\bibinfo {volume} {20}},\ \bibinfo {pages} {1898} (\bibinfo {year}
  {1984})}\BibitemShut {NoStop}%
\bibitem [{\citenamefont {Wood}\ and\ \citenamefont
  {Chan}(1994)}]{wood1994simulation}%
  \BibitemOpen
  \bibfield  {author} {\bibinfo {author} {\bibfnamefont {A.~T.}\ \bibnamefont
  {Wood}}\ and\ \bibinfo {author} {\bibfnamefont {G.}~\bibnamefont {Chan}},\
  }\bibfield  {title} {\bibinfo {title} {Simulation of stationary {G}aussian
  processes in [0, 1] d},\ }\href@noop {} {\bibfield  {journal} {\bibinfo
  {journal} {Journal of Computational and Graphical Statistics}\ }\textbf
  {\bibinfo {volume} {3}},\ \bibinfo {pages} {409} (\bibinfo {year}
  {1994})}\BibitemShut {NoStop}%
\bibitem [{\citenamefont {Mu{\~n}oz-Gil}\ \emph {et~al.}(2022)\citenamefont
  {Mu{\~n}oz-Gil}, \citenamefont {Romero-Aristizabal}, \citenamefont {Mateos},
  \citenamefont {Campelo}, \citenamefont {de~Llobet~Cucalon}, \citenamefont
  {Beato}, \citenamefont {Lewenstein}, \citenamefont {Garcia-Parajo},\ and\
  \citenamefont {Torreno-Pina}}]{munoz2022stochastic}%
  \BibitemOpen
  \bibfield  {author} {\bibinfo {author} {\bibfnamefont {G.}~\bibnamefont
  {Mu{\~n}oz-Gil}}, \bibinfo {author} {\bibfnamefont {C.}~\bibnamefont
  {Romero-Aristizabal}}, \bibinfo {author} {\bibfnamefont {N.}~\bibnamefont
  {Mateos}}, \bibinfo {author} {\bibfnamefont {F.}~\bibnamefont {Campelo}},
  \bibinfo {author} {\bibfnamefont {L.~I.}\ \bibnamefont {de~Llobet~Cucalon}},
  \bibinfo {author} {\bibfnamefont {M.}~\bibnamefont {Beato}}, \bibinfo
  {author} {\bibfnamefont {M.}~\bibnamefont {Lewenstein}}, \bibinfo {author}
  {\bibfnamefont {M.~F.}\ \bibnamefont {Garcia-Parajo}},\ and\ \bibinfo
  {author} {\bibfnamefont {J.~A.}\ \bibnamefont {Torreno-Pina}},\ }\bibfield
  {title} {\bibinfo {title} {Stochastic particle unbinding modulates growth
  dynamics and size of transcription factor condensates in living cells},\
  }\href@noop {} {\bibfield  {journal} {\bibinfo  {journal} {Proceedings of the
  National Academy of Sciences}\ }\textbf {\bibinfo {volume} {119}},\ \bibinfo
  {pages} {e2200667119} (\bibinfo {year} {2022})}\BibitemShut {NoStop}%
\bibitem [{\citenamefont {Scher}\ and\ \citenamefont
  {Montroll}(1975)}]{scher1975anomalous}%
  \BibitemOpen
  \bibfield  {author} {\bibinfo {author} {\bibfnamefont {H.}~\bibnamefont
  {Scher}}\ and\ \bibinfo {author} {\bibfnamefont {E.~W.}\ \bibnamefont
  {Montroll}},\ }\bibfield  {title} {\bibinfo {title} {Anomalous transit-time
  dispersion in amorphous solids},\ }\href@noop {} {\bibfield  {journal}
  {\bibinfo  {journal} {Physical Review B}\ }\textbf {\bibinfo {volume} {12}},\
  \bibinfo {pages} {2455} (\bibinfo {year} {1975})}\BibitemShut {NoStop}%
\bibitem [{\citenamefont {Klafter}\ and\ \citenamefont
  {Sokolov}(2011)}]{klafter2011first}%
  \BibitemOpen
  \bibfield  {author} {\bibinfo {author} {\bibfnamefont {J.}~\bibnamefont
  {Klafter}}\ and\ \bibinfo {author} {\bibfnamefont {I.~M.}\ \bibnamefont
  {Sokolov}},\ }\href@noop {} {\emph {\bibinfo {title} {First {S}teps in
  {R}andom {W}alks: {F}rom {T}ools to {A}pplications}}}\ (\bibinfo  {publisher}
  {Oxford University Press, Oxford, UK},\ \bibinfo {year} {2011})\BibitemShut
  {NoStop}%
\bibitem [{\citenamefont {Massignan}\ \emph {et~al.}(2014)\citenamefont
  {Massignan}, \citenamefont {Manzo}, \citenamefont {Torreno-Pina},
  \citenamefont {Garc{\'\i}a-Parajo}, \citenamefont {Lewenstein},\ and\
  \citenamefont {Lapeyre~Jr}}]{massignan2014nonergodic}%
  \BibitemOpen
  \bibfield  {author} {\bibinfo {author} {\bibfnamefont {P.}~\bibnamefont
  {Massignan}}, \bibinfo {author} {\bibfnamefont {C.}~\bibnamefont {Manzo}},
  \bibinfo {author} {\bibfnamefont {J.}~\bibnamefont {Torreno-Pina}}, \bibinfo
  {author} {\bibfnamefont {M.}~\bibnamefont {Garc{\'\i}a-Parajo}}, \bibinfo
  {author} {\bibfnamefont {M.}~\bibnamefont {Lewenstein}},\ and\ \bibinfo
  {author} {\bibfnamefont {G.}~\bibnamefont {Lapeyre~Jr}},\ }\bibfield  {title}
  {\bibinfo {title} {Nonergodic subdiffusion from {B}rownian motion in an in
  homogeneous medium},\ }\href@noop {} {\bibfield  {journal} {\bibinfo
  {journal} {Physical Review Letters}\ }\textbf {\bibinfo {volume} {112}},\
  \bibinfo {pages} {150603} (\bibinfo {year} {2014})}\BibitemShut {NoStop}%
\bibitem [{\citenamefont {Klafter}\ and\ \citenamefont
  {Zumofen}(1994)}]{klafter1994levy}%
  \BibitemOpen
  \bibfield  {author} {\bibinfo {author} {\bibfnamefont {J.}~\bibnamefont
  {Klafter}}\ and\ \bibinfo {author} {\bibfnamefont {G.}~\bibnamefont
  {Zumofen}},\ }\bibfield  {title} {\bibinfo {title} {L{\'e}vy statistics in a
  hamiltonian system},\ }\href@noop {} {\bibfield  {journal} {\bibinfo
  {journal} {Physical Review E}\ }\textbf {\bibinfo {volume} {49}},\ \bibinfo
  {pages} {4873} (\bibinfo {year} {1994})}\BibitemShut {NoStop}%
\bibitem [{\citenamefont {Lanoisel{\'e}e}\ \emph {et~al.}(2018)\citenamefont
  {Lanoisel{\'e}e}, \citenamefont {Sikora}, \citenamefont {Grzesiek},
  \citenamefont {Grebenkov},\ and\ \citenamefont
  {Wy{\l}oma{\'n}ska}}]{lanoiselee2018optimal}%
  \BibitemOpen
  \bibfield  {author} {\bibinfo {author} {\bibfnamefont {Y.}~\bibnamefont
  {Lanoisel{\'e}e}}, \bibinfo {author} {\bibfnamefont {G.}~\bibnamefont
  {Sikora}}, \bibinfo {author} {\bibfnamefont {A.}~\bibnamefont {Grzesiek}},
  \bibinfo {author} {\bibfnamefont {D.~S.}\ \bibnamefont {Grebenkov}},\ and\
  \bibinfo {author} {\bibfnamefont {A.}~\bibnamefont {Wy{\l}oma{\'n}ska}},\
  }\bibfield  {title} {\bibinfo {title} {Optimal parameters for
  anomalous-diffusion-exponent estimation from noisy data},\ }\href@noop {}
  {\bibfield  {journal} {\bibinfo  {journal} {Physical Review E}\ }\textbf
  {\bibinfo {volume} {98}},\ \bibinfo {pages} {062139} (\bibinfo {year}
  {2018})}\BibitemShut {NoStop}%
\bibitem [{\citenamefont {Katz}\ and\ \citenamefont
  {George}(1985)}]{katz1985fractals}%
  \BibitemOpen
  \bibfield  {author} {\bibinfo {author} {\bibfnamefont {M.~J.}\ \bibnamefont
  {Katz}}\ and\ \bibinfo {author} {\bibfnamefont {E.~B.}\ \bibnamefont
  {George}},\ }\bibfield  {title} {\bibinfo {title} {Fractals and the analysis
  of growth paths},\ }\href@noop {} {\bibfield  {journal} {\bibinfo  {journal}
  {Bulletin of Mathematical Biology}\ }\textbf {\bibinfo {volume} {47}},\
  \bibinfo {pages} {273} (\bibinfo {year} {1985})}\BibitemShut {NoStop}%
\bibitem [{\citenamefont {Tejedor}\ \emph {et~al.}(2010)\citenamefont
  {Tejedor}, \citenamefont {B{\'e}nichou}, \citenamefont {Voituriez},
  \citenamefont {Jungmann}, \citenamefont {Simmel}, \citenamefont
  {Selhuber-Unkel}, \citenamefont {Oddershede},\ and\ \citenamefont
  {Metzler}}]{tejedor2010quantitative}%
  \BibitemOpen
  \bibfield  {author} {\bibinfo {author} {\bibfnamefont {V.}~\bibnamefont
  {Tejedor}}, \bibinfo {author} {\bibfnamefont {O.}~\bibnamefont
  {B{\'e}nichou}}, \bibinfo {author} {\bibfnamefont {R.}~\bibnamefont
  {Voituriez}}, \bibinfo {author} {\bibfnamefont {R.}~\bibnamefont {Jungmann}},
  \bibinfo {author} {\bibfnamefont {F.}~\bibnamefont {Simmel}}, \bibinfo
  {author} {\bibfnamefont {C.}~\bibnamefont {Selhuber-Unkel}}, \bibinfo
  {author} {\bibfnamefont {L.~B.}\ \bibnamefont {Oddershede}},\ and\ \bibinfo
  {author} {\bibfnamefont {R.}~\bibnamefont {Metzler}},\ }\bibfield  {title}
  {\bibinfo {title} {Quantitative analysis of single particle trajectories:
  {M}ean maximal excursion method},\ }\href@noop {} {\bibfield  {journal}
  {\bibinfo  {journal} {Biophysical Journal}\ }\textbf {\bibinfo {volume}
  {98}},\ \bibinfo {pages} {1364} (\bibinfo {year} {2010})}\BibitemShut
  {NoStop}%
\bibitem [{\citenamefont {Ernst}\ \emph {et~al.}(2014)\citenamefont {Ernst},
  \citenamefont {K{\"o}hler},\ and\ \citenamefont {Weiss}}]{ernst2014probing}%
  \BibitemOpen
  \bibfield  {author} {\bibinfo {author} {\bibfnamefont {D.}~\bibnamefont
  {Ernst}}, \bibinfo {author} {\bibfnamefont {J.}~\bibnamefont {K{\"o}hler}},\
  and\ \bibinfo {author} {\bibfnamefont {M.}~\bibnamefont {Weiss}},\ }\bibfield
   {title} {\bibinfo {title} {Probing the type of anomalous diffusion with
  single-particle tracking},\ }\href@noop {} {\bibfield  {journal} {\bibinfo
  {journal} {Physical Chemistry Chemical Physics}\ }\textbf {\bibinfo {volume}
  {16}},\ \bibinfo {pages} {7686} (\bibinfo {year} {2014})}\BibitemShut
  {NoStop}%
\bibitem [{\citenamefont {Helmuth}\ \emph {et~al.}(2007)\citenamefont
  {Helmuth}, \citenamefont {Burckhardt}, \citenamefont {Koumoutsakos},
  \citenamefont {Greber},\ and\ \citenamefont {Sbalzarini}}]{helmuth2007novel}%
  \BibitemOpen
  \bibfield  {author} {\bibinfo {author} {\bibfnamefont {J.~A.}\ \bibnamefont
  {Helmuth}}, \bibinfo {author} {\bibfnamefont {C.~J.}\ \bibnamefont
  {Burckhardt}}, \bibinfo {author} {\bibfnamefont {P.}~\bibnamefont
  {Koumoutsakos}}, \bibinfo {author} {\bibfnamefont {U.~F.}\ \bibnamefont
  {Greber}},\ and\ \bibinfo {author} {\bibfnamefont {I.~F.}\ \bibnamefont
  {Sbalzarini}},\ }\bibfield  {title} {\bibinfo {title} {A novel supervised
  trajectory segmentation algorithm identifies distinct types of human
  adenovirus motion in host cells},\ }\href@noop {} {\bibfield  {journal}
  {\bibinfo  {journal} {Journal of Structural Biology}\ }\textbf {\bibinfo
  {volume} {159}},\ \bibinfo {pages} {347} (\bibinfo {year}
  {2007})}\BibitemShut {NoStop}%
\bibitem [{\citenamefont {Burnecki}\ and\ \citenamefont
  {Weron}(2010)}]{burnecki2010fractional}%
  \BibitemOpen
  \bibfield  {author} {\bibinfo {author} {\bibfnamefont {K.}~\bibnamefont
  {Burnecki}}\ and\ \bibinfo {author} {\bibfnamefont {A.}~\bibnamefont
  {Weron}},\ }\bibfield  {title} {\bibinfo {title} {Fractional {L}{\'e}vy
  stable motion can model subdiffusive dynamics},\ }\href@noop {} {\bibfield
  {journal} {\bibinfo  {journal} {Physical Review E}\ }\textbf {\bibinfo
  {volume} {82}},\ \bibinfo {pages} {021130} (\bibinfo {year}
  {2010})}\BibitemShut {NoStop}%
\bibitem [{\citenamefont {Saxton}(1993)}]{saxton1993lateral}%
  \BibitemOpen
  \bibfield  {author} {\bibinfo {author} {\bibfnamefont {M.~J.}\ \bibnamefont
  {Saxton}},\ }\bibfield  {title} {\bibinfo {title} {Lateral diffusion in an
  archipelago. {S}ingle-particle diffusion},\ }\href@noop {} {\bibfield
  {journal} {\bibinfo  {journal} {Biophysical Journal}\ }\textbf {\bibinfo
  {volume} {64}},\ \bibinfo {pages} {1766} (\bibinfo {year}
  {1993})}\BibitemShut {NoStop}%
\bibitem [{\citenamefont {d'Agostino}\ and\ \citenamefont
  {Pearson}(1973)}]{d1973tests}%
  \BibitemOpen
  \bibfield  {author} {\bibinfo {author} {\bibfnamefont {R.}~\bibnamefont
  {d'Agostino}}\ and\ \bibinfo {author} {\bibfnamefont {E.~S.}\ \bibnamefont
  {Pearson}},\ }\bibfield  {title} {\bibinfo {title} {Tests for departure from
  normality. empirical results for the distributions of $b^{2}$ and
  $\sqrt{b^{1}}$},\ }\href@noop {} {\bibfield  {journal} {\bibinfo  {journal}
  {Biometrika}\ }\textbf {\bibinfo {volume} {60}},\ \bibinfo {pages} {613}
  (\bibinfo {year} {1973})}\BibitemShut {NoStop}%
\bibitem [{\citenamefont {Aghion}\ \emph {et~al.}(2021)\citenamefont {Aghion},
  \citenamefont {Meyer}, \citenamefont {Adlakha}, \citenamefont {Kantz},\ and\
  \citenamefont {Bassler}}]{aghion2021moses}%
  \BibitemOpen
  \bibfield  {author} {\bibinfo {author} {\bibfnamefont {E.}~\bibnamefont
  {Aghion}}, \bibinfo {author} {\bibfnamefont {P.~G.}\ \bibnamefont {Meyer}},
  \bibinfo {author} {\bibfnamefont {V.}~\bibnamefont {Adlakha}}, \bibinfo
  {author} {\bibfnamefont {H.}~\bibnamefont {Kantz}},\ and\ \bibinfo {author}
  {\bibfnamefont {K.~E.}\ \bibnamefont {Bassler}},\ }\bibfield  {title}
  {\bibinfo {title} {Moses, {N}oah and {J}oseph effects in {L}{\'e}vy walks},\
  }\href@noop {} {\bibfield  {journal} {\bibinfo  {journal} {New Journal of
  Physics}\ }\textbf {\bibinfo {volume} {23}},\ \bibinfo {pages} {023002}
  (\bibinfo {year} {2021})}\BibitemShut {NoStop}%
\bibitem [{\citenamefont {Mandelbrot}\ and\ \citenamefont
  {Wallis}(1968)}]{mandelbrot1968noah}%
  \BibitemOpen
  \bibfield  {author} {\bibinfo {author} {\bibfnamefont {B.~B.}\ \bibnamefont
  {Mandelbrot}}\ and\ \bibinfo {author} {\bibfnamefont {J.~R.}\ \bibnamefont
  {Wallis}},\ }\bibfield  {title} {\bibinfo {title} {Noah, {J}oseph, and
  operational hydrology},\ }\href@noop {} {\bibfield  {journal} {\bibinfo
  {journal} {Water Resources Research}\ }\textbf {\bibinfo {volume} {4}},\
  \bibinfo {pages} {909} (\bibinfo {year} {1968})}\BibitemShut {NoStop}%
\bibitem [{\citenamefont {Chen}\ \emph {et~al.}(2017)\citenamefont {Chen},
  \citenamefont {Bassler}, \citenamefont {McCauley},\ and\ \citenamefont
  {Gunaratne}}]{chen2017anomalous}%
  \BibitemOpen
  \bibfield  {author} {\bibinfo {author} {\bibfnamefont {L.}~\bibnamefont
  {Chen}}, \bibinfo {author} {\bibfnamefont {K.~E.}\ \bibnamefont {Bassler}},
  \bibinfo {author} {\bibfnamefont {J.~L.}\ \bibnamefont {McCauley}},\ and\
  \bibinfo {author} {\bibfnamefont {G.~H.}\ \bibnamefont {Gunaratne}},\
  }\bibfield  {title} {\bibinfo {title} {Anomalous scaling of stochastic
  processes and the {M}oses effect},\ }\href@noop {} {\bibfield  {journal}
  {\bibinfo  {journal} {Physical Review E}\ }\textbf {\bibinfo {volume} {95}},\
  \bibinfo {pages} {042141} (\bibinfo {year} {2017})}\BibitemShut {NoStop}%
\bibitem [{\citenamefont {Balcerek}\ \emph {et~al.}(2021)\citenamefont
  {Balcerek}, \citenamefont {Burnecki}, \citenamefont {Sikora},\ and\
  \citenamefont {Wy{\l}oma{\'n}ska}}]{balcerek2021discriminating}%
  \BibitemOpen
  \bibfield  {author} {\bibinfo {author} {\bibfnamefont {M.}~\bibnamefont
  {Balcerek}}, \bibinfo {author} {\bibfnamefont {K.}~\bibnamefont {Burnecki}},
  \bibinfo {author} {\bibfnamefont {G.}~\bibnamefont {Sikora}},\ and\ \bibinfo
  {author} {\bibfnamefont {A.}~\bibnamefont {Wy{\l}oma{\'n}ska}},\ }\bibfield
  {title} {\bibinfo {title} {Discriminating {G}aussian processes via quadratic
  form statistics},\ }\href@noop {} {\bibfield  {journal} {\bibinfo  {journal}
  {Chaos: An Interdisciplinary Journal of Nonlinear Science}\ }\textbf
  {\bibinfo {volume} {31}} (\bibinfo {year} {2021})}\BibitemShut {NoStop}%
\end{thebibliography}%

\end{document}